\documentclass[mathpazo]{cicp}
\usepackage{amssymb}

\usepackage{stmaryrd,amsmath,amssymb}
\usepackage{tabularx,epsfig,color,tabularx}
\usepackage{algorithm,algorithmic,multirow}
\usepackage{graphicx,caption,subcaption}
\usepackage{mathtools,epstopdf,hyperref}

\newcommand\bx{\mathbf{x}}

\newcommand{\p}{\partial}

\newcommand{\fl}[2]{\frac{#1}{#2}}
\newcommand{\dt}{\delta}
\newcommand{\nn}{\nonumber}
\newcommand{\ap}{\alpha}
\newcommand{\bt}{\beta}

\newcommand{\vep}{\varepsilon}

\newcommand{\be}{\begin{equation}}
\newcommand{\ee}{\end{equation}}
\newcommand{\ba}{\begin{array}}
\newcommand{\ea}{\end{array}}
\def\bea{\begin{eqnarray}}
\def\eea{\end{eqnarray}}
\def \beas{\begin{eqnarray*}}
\def \eeas{\end{eqnarray*}}

\begin{document}

%
\title{Mathematical models and numerical methods for spinor Bose-Einstein condensates}

 \author[Weizhu Bao et.~al.]{Weizhu Bao\affil{1},
          Yongyong Cai\affil{2}\comma\corrauth}
\address{\affilnum{1}\ Department of Mathematics, National University of
Singapore, Singapore 119076\\
\affilnum{2}\  Beijing Computational Science Research Center,
No. 10 East Xibeiwang Road,
Beijing 100193,  P. R. China}
\emails{{\tt matbaowz@nus.edu.sg} (W.~Bao),
{\tt yongyong.cai@csrc.ac.cn} (Y.~Cai)}

\begin{abstract}
In this paper, we systematically review mathematical models, theories and
numerical methods for ground states and dynamics of spinor Bose-Einstein condensates (BECs)
based on the coupled Gross-Pitaevskii equations (GPEs). We start with a pseudo spin-1/2 BEC system with/without an internal atomic Josephson junction and spin-orbit coupling including (i) existence and uniqueness
as well as non-existence of ground states under different parameter regimes, (ii) ground state structures under different limiting parameter
regimes,  (iii) dynamical properties, and (iv) efficient and accurate numerical methods for computing ground states and dynamics.
Then we extend these results to spin-1 BEC and spin-2 BEC.
Finally, extensions to dipolar spinor systems and/or 
general spin-$F$ ($F\ge3$) BEC are discussed.
\end{abstract}

\


\maketitle{}

\tableofcontents

\section{Introduction}
%
%
The remarkable experimental achievement of Bose-Einstein condensation (BEC) of dilute alkali gases
in 1995  \cite{Anderson,Bradley,Davis} reached a milestone  in atomic, molecular and optical (AMO) physics
and quantum optics, and it provided a unique opportunity
to observe the  mysterious quantum world directly in laboratory.
The  BEC phenomenon was predicted by Einstein in 1924  \cite{Einstein1,Einstein2} when he generalized
the studies of Bose  \cite{Bose} concerning photons to atoms which assume the same statistical rule.
Based on the derived Bose-Einstein statistics,
Einstein figured that, there exits a critical temperature, below which a finite fraction
of all the particles ``condense'' into the same quantum state.

Einstein's  prediction was for a system of noninteracting bosons and did not receive much
attention until the observation of superfluidity in liquid $^4$He
below the $\lambda$ temperature ($2.17$K) in 1938, when London \cite{London}
suggested that despite the strong interatomic interactions,
part of the system is in the BEC state resulting in its superfluidity.
Over the years, the major difficulty to realize BEC state in laboratory is that
almost all the substances become solid or liquid (strong interatomic interactions) at low
temperature where the BEC phase transition occurs.
With the development of magnetic trapping and laser cooling techniques, BEC was finally
achieved in the system of weakly interacting dilute alkali gases \cite{Anderson,Bradley,Davis} in 1995.
The key is to bring down the temperature of the gas before  its relaxation to solid state.
In  most BEC experiments, the system reaches quantum degeneracy between $50$ nK and $2$ $\mu$K, at densities
between $10^{11}$ and $10^{15}$ cm$^{-3}$. The largest condensates are of
$100$ million atoms for sodium, and a billion for hydrogen; the smallest are just a
few hundred atoms. Depending on the magnetic trap, the shape of the condensate is
either approximately round, with a diameter of $10$--$15$ $\mu$m, or cigar-shaped with about
$15$ $\mu$m in diameter and $300$ $\mu$m in length. The full cooling cycle that
produces a condensate may take from a few seconds to as long as several minutes \cite{Cornell,Ketterle}.
For better understanding of the long history towards the BEC and its physics study,
we refer to the Nobel lectures \cite{Cornell,Ketterle} and
several review papers \cite{Bar,BlochDZ,Dalfovo,Fetter,Lahaye,Leggett} as well as the two books \cite{Pethick,PitaevskiiStringari} in physics.

The pioneering experiments \cite{Anderson,Bradley,Davis} were conducted for single species of atoms, which can be theoretically described by
a scalar order parameter (or wavefunction) satisfying
the Gross-Pitaevskii equation (GPE)
(or the nonlinear Schr\"{o}dinger equation (NLSE) with cubic nonlinearity)
\cite{Dalfovo,Fetter,Leggett,Pethick,PitaevskiiStringari}.
For the mathematical models and numerical methods
of single-component BEC based on the GPE, we refer to
\cite{AftalionDu,ABB2013,Bao0,Baocai2013,Dion,Huang,Erd,
LiebSeiringerPra2000,Johnson} and references therein.
 A natural generalization is to explore the multi-component  BEC system, where inter-species interactions
 lead to more interesting phases and involve vector order parameters.  In 1996, one year after the major breakthrough, an overlapping two component BEC was
produced with $|F=2,m=2\rangle$ and $|F=1,m=-1\rangle$ spin
states of $^{87}$Rb \cite{Myatt}, by employing a double  magneto-optic trap. During the process,
two condensates were cooled together and the interaction between different components was observed.
Later, it was proposed that the binary BEC system can generate coherent matter wave
(also called atom laser) analogous to the coherent light emitted from a laser.  In view of such potential applications,
multi-component BEC systems have attracted numerous research interests \cite{Jaksch,Bao,Pethick,PitaevskiiStringari}.

In the early experiments, magnetic traps were  used  and the spin degrees of the atoms were then frozen.
In 1998, by using an optical dipole trap, a spinor BEC was first produced with spin-1 $^{23}$Na gases \cite{StmaperAndrews}, where the internal spin degrees of freedom were activated.
In the optical trap, particles with different hyperfine states allow different angular momentum in space, resulting in a rich variety of spin texture.
Therefore, degenerate quantum spinor gases maintain both magnetism and superfluidity, and are quite promising for   many fields, such as topological quantum structure, fractional quantum Hall effect \cite{Ueda,Ueda2014,StamperUeda}. For a spin-$F$  Bose condensate, there are $2F+1$ hyperfine states and  the spinor condensate can be described by a $2F+1$ component vector wavefunction \cite{Ho1,Ueda,Pethick,PitaevskiiStringari,Ohmi,StamperUeda}.

Up to now, various spinor condensates  including spin-1/2 $^{87}$Rb condensate (pseudo spin-1/2)\cite{Myatt}, spin-1 $^{23}$Na condensate\cite{StmaperAndrews}, spin-1 $^{87}$Rb condensate \cite{Barrett} and spin-2 $^{87}$Rb condensate \cite{Chang},  have been achieved in experiments. For the experimental and theoretical studies of spinor
BEC, we refer to the two recent review papers in physics
\cite{Ueda,StamperUeda} and references therein.
In this  growing research direction,   mathematical models and analysis as well as
numerical simulation have been playing an important role  in  understanding
the theoretical part of spinor BEC and predicting and guiding the experiments.
The goal of this review paper is to offer a short survey on mathematical models and theories as well as
numerical methods for spinor BEC based on the coupled Gross-Pitaevskii equations (CGPEs)
\cite{Baocai2013,Gross,Ueda,Pitaevskii,PitaevskiiStringari,StamperUeda}.

The paper is organized as follows. In section 2, we present the results on
the ground states and the dynamics for pseudo spin-1/2 BEC system with/without Josephson junction based on the CGPEs, including the semi-classical limit and the Bogoliubov excitation. Both theoretical and numerical results will be shown.
As a generalization, a spin-1/2 BEC with spin-orbit-coupling is then discussed in section 3.
Section 4 is devoted to the study of spin-1 system, and spin-2 system is considered in section 5.
 Some perspectives on
spin-3 system and spinor dipolar BEC system are discussed in section 6.

\section{Pseudo-spin-1/2 system} \label{sec:2}
In this section, we consider a two-component (pseudo spin-1/2) BEC system with/without Josephson junction \cite{Williams,Ashhab} and discuss
its ground state and dynamics based on the mean-field theory \cite{BaoCai0}. In the derivation of the mean-field Gross-Pitaevskii (GP) theory \cite{Baocai2013,LiebSeiringerPra2000,LieSolo,Pethick,PitaevskiiStringari}, the many body
Hamiltonian of the system  with two-body interaction is approximated by a single particle Hamiltonian (mean field approximation), leading to the
time dependent GPE in Heisenberg picture and the associated Gross-Pitaevskii (GP) energy functional. We refer to \cite{Baocai2013,LiebSeiringerPra2000,LieSolo,Pethick,PitaevskiiStringari,Ueda}
and references therein for the derivation of GPE in single component and two component BECs.

\subsection{Coupled Gross-Pitaevskii equations}
At temperature $T$ much smaller than the BEC critical temperature $T_c$, a pseudo spin-1/2 BEC with Josephson junction can be well described by the following coupled
Gross-Pitaevskii equations (CGPEs) in three dimensions (3D) \cite{Bao,Baocai2013,BaoCai0,Jungel,Pitaevskii,PitaevskiiStringari,Anton}:
\begin{equation}\label{eq:cgpe:sec2}
\begin{split} &i\hbar\partial_t \psi_{\uparrow}=\left[-\frac{\hbar^2} {2m}\nabla^2
+\tilde{V}_\uparrow(\bx)+\frac{\hbar\tilde{\delta}}{2}
+(g_{\uparrow\uparrow}|\psi_\uparrow|^2+
g_{\uparrow\downarrow}|\psi_\downarrow|^2)\right]
\psi_\uparrow+\frac{\hbar\tilde{\Omega}}{2}
\psi_\downarrow,\quad \bx\in\mathbb{R}^3,\\
&i\hbar\partial_t \psi_{\downarrow}=\left[-\frac{\hbar^2} {2m}\nabla^2
+\tilde{V}_\downarrow(\bx)-\frac{\hbar\tilde{\delta}}{2}
+(g_{\downarrow\uparrow}|\psi_\uparrow|^2+
g_{\downarrow\downarrow}|\psi_\downarrow|^2)\right]
\psi_\downarrow+\frac{\hbar\tilde{\Omega}}{2}
\psi_\uparrow,
\quad \bx\in\mathbb{R}^3.\end{split} \end{equation}
 Here, $t$ is time,
$\bx=(x,y,z)^T\in \mathbb{R}^3$ is the Cartesian coordinate
vector,  $\Psi(\bx,t):=(\psi_{\uparrow}(\bx,t),\psi_{\downarrow}(\bx,t))^T$ is the
complex-valued macroscopic wave function corresponding to the spin-up and spin-down components,  $\nabla^2=\Delta$ is the Laplace operator,
  $\tilde{\Omega}$ is the effective
Rabi frequency to realize the internal atomic
Josephson junction  by a Raman transition, $\tilde{\delta}$ is the Raman transition constant, and $g_{jl}=\frac{4\pi\hbar^2}{m}a_{jl}$ with
$a_{jl}=a_{lj}$ ($j,l=\uparrow,\downarrow$) being the $s$-wave scattering
lengths between the $j$th and $l$th component (positive for
repulsive interaction and negative for attractive interaction), $m$ is the mass of the particle and $\hbar$ is the reduced Planck constant.
$\tilde{V}_j(\bx)$ ($j=\uparrow,\downarrow$) are the external trapping potentials and may vary in different applications, and the most commonly used
ones in experiments are the following harmonic potentials
\be\label{eq:hot:sec2}
\tilde{V}_j(\bx)=\frac{m}{2}\left[\omega_x^2(x-\tilde{x}_j)^2+\omega_y^2y^2+
\omega_z^2z^2\right],
 \qquad j=\uparrow,\downarrow,\quad\bx=(x,y,z)^T\in {\mathbb R}^3,
\ee
with $\omega_{x}$, $\omega_y$ and $\omega_z$ being the trapping frequencies in $x$-, $y$- and $z$-directions, respectively, and
$\tilde{x}_j$ ($j=\uparrow,\downarrow$) are the shifts in the $x$-direction
of $\tilde{V}_j(\bx)$ from the origin.

 The wavefunction $\Psi$ is normalized
as
\begin{equation}\label{eq:N:sec2}
\|\Psi(\cdot,t)\|^2:=\int_{{\mathbb R}^3}
\left[|\psi_\uparrow(\bx,t)|^2+|\psi_\downarrow(\bx,t)|^2\right]\,d\bx=N,
\end{equation}
where $N$ is the total number of particles in the condensate.

{\it Nondimensionalization and dimension reduction.} To nondimensionalize \eqref{eq:cgpe:sec2}, introduce
\begin{equation}\label{eq:scale:sec2}
\tilde{t}=\frac{t}{t_s},\qquad \tilde{\bx}=\frac{\bx}{x_s},\qquad \tilde{\Psi}(\tilde{\bx},\tilde{t})=\frac{\Psi(\bx,t)}{x_s^{-3/2}N^{1/2}},
\end{equation}
where $t_s=1/\omega_s$ and $x_s=\sqrt{\hbar/m\omega_s}$ with $\omega_s=\min\{\omega_x,\omega_y,\omega_z\}$ are the time and length units, respectively. Plugging \eqref{eq:scale:sec2} into \eqref{eq:cgpe:sec2},
multiplying by $t_s^2/mx_s^{1/2}N^{1/2}$ and then removing all $\tilde{}$, we obtain the following dimensionless CGPEs for $\Psi=(\psi_\uparrow,\psi_\downarrow)^T$ as
\begin{equation}\label{eq:cgpen:sec20}
\begin{split} &i\partial_t \psi_{\uparrow}=\left[-\frac{1} {2}\nabla^2
+V_\uparrow(\bx)+\frac{\delta}{2}
+(\kappa_{\uparrow\uparrow}|\psi_\uparrow|^2+
\kappa_{\uparrow\downarrow}|\psi_\downarrow|^2)\right]
\psi_\uparrow+\frac{\Omega}{2}
\psi_\downarrow,\quad \quad\bx\in {\mathbb R}^3,\\
&i\partial_t \psi_{\downarrow}=\left[-\frac{1} {2}\nabla^2
+V_\downarrow(\bx)-\frac{\delta}{2}
+(\kappa_{\downarrow\uparrow}|\psi_\uparrow|^2+
\kappa_{\downarrow\downarrow}|\psi_\downarrow|^2)\right]
\psi_\downarrow+\frac{\Omega}{2}
\psi_\uparrow,\quad \quad\bx\in {\mathbb R}^3,
\end{split}
\end{equation}
where
$\kappa_{jl}=\frac{4\pi N a_{jl}}{x_s}$ ($j,l=\uparrow,\downarrow$), $\Omega=\frac{\tilde \Omega}{\omega_s}$, $\delta=\frac{\tilde\delta}{\omega_s}$ and the trapping potentials are given as
\be\label{eq:hotn:sec2}
V_j(\bx)=\frac{1}{2}(\gamma_x^2(x-x_j)^2+\gamma_y^2y^2+\gamma_z^2z^2),\quad j=\uparrow,\downarrow,\quad \bx=(x,y,z)^T\in\mathbb{R}^3,
\ee
with $\gamma_x=\omega_x/\omega_s$, $\gamma_y=\omega_y/\omega_s$,
$\gamma_z=\omega_z/\omega_s$ and $x_j=\tilde x_j/x_s$ ($j=\uparrow,\downarrow$).
The normalization for \eqref{eq:cgpen:sec20} becomes
 \be \label{eq:normn:sec20}
\|\Psi\|^2:=
\|\psi_\uparrow(\cdot,t)\|^2+\|\psi_\downarrow(\cdot,t)\|^2:=\int_{{\mathbb R}^3}
\left[|\psi_\uparrow(\bx,t)|^2+|\psi_\downarrow(\bx,t)|^2\right]\,d\bx=1. \ee

 In practice, when the harmonic traps \eqref{eq:hotn:sec2} are strongly anisotropic, e.g. when $\gamma_x=O(1)$, $\gamma_y=O(1)$ and $\gamma_z\gg1$, following the dimension reduction process for GPE in \cite{Baocai2013,bamsw}, the 3D CGPEs \eqref{eq:cgpen:sec2} can be reduced to a system in two dimensions (2D)
under effective trapping potentials $V_{j}(x,y)=\frac{1}{2}(\gamma_x^2(x-x_j)^2+\gamma_y^2y^2)$ ($j=\uparrow,\downarrow$) and
effective interaction strengths
$\beta_{jl}=\frac{\sqrt{\gamma_z}}{\sqrt{2\pi}}\kappa_{jl}$  ($j,l=\uparrow,\downarrow$); and respectively,
when $\gamma_x=O(1)$, $\gamma_y\gg1$ and $\gamma_z\gg1$,
the 3D CGPEs \eqref{eq:cgpen:sec2} can be reduced to a system in one dimension (1D) under effective trapping potentials $V_{j}(x)=\frac{1}{2}\gamma_x^2(x-x_j)^2$ ($j=\uparrow,\downarrow$) and
effective interaction strengths
$\beta_{jl}=\frac{\sqrt{\gamma_y\gamma_z}}{2\pi}\kappa_{jl}$ ($j,l=\uparrow,\downarrow$).

In fact, the CGPEs \eqref{eq:cgpen:sec20} in 3D and the corresponding CGPEs
in 2D and 1D obtained from \eqref{eq:cgpen:sec20} by dimension reduction
under strongly anisotropic trapping potentials can be
written in a unified form in $d$-dimensions ($d=3,2,1$) as
\begin{equation}\label{eq:cgpen:sec2}
\begin{split} &i\partial_t \psi_{\uparrow}=\left[-\frac{1} {2}\nabla^2
+V_\uparrow(\bx)+\frac{\delta}{2}
+(\beta_{\uparrow\uparrow}|\psi_\uparrow|^2+
\beta_{\uparrow\downarrow}|\psi_\downarrow|^2)\right]
\psi_\uparrow+\frac{\Omega}{2}
\psi_\downarrow,\quad \quad\bx\in {\mathbb R}^d,\\
&i\partial_t \psi_{\downarrow}=\left[-\frac{1} {2}\nabla^2
+V_\downarrow(\bx)-\frac{\delta}{2}
+(\beta_{\downarrow\uparrow}|\psi_\uparrow|^2+
\beta_{\downarrow\downarrow}|\psi_\downarrow|^2)\right]
\psi_\downarrow+\frac{\Omega}{2}
\psi_\uparrow,\quad \quad\bx\in {\mathbb R}^d,
\end{split}
\end{equation}
where the interaction strengths and harmonic trapping potentials  are given as
\begin{equation}\label{eq:hopot:sec2}
\beta_{jl}=\begin{cases}
\kappa_{jl}=\frac{4\pi N a_{jl}}{x_s},\\
\frac{\sqrt{\gamma_z}}{\sqrt{2\pi}}\kappa_{jl},\\
\frac{\sqrt{\gamma_y\gamma_z}}{2\pi}\kappa_{jl}, \\
\end{cases}
V_j(\bx)=\begin{cases}\frac{1}{2}(\gamma_x^2(x-x_j)^2+\gamma_y^2y^2+\gamma_z^2
z^2),& d=3,\\
\frac{1}{2}(\gamma_x^2(x-x_j)^2+\gamma_y^2y^2),&d=2,\\
\frac{1}{2}\gamma_x^2(x-x_j)^2,&d=1,
\end{cases}\quad j,l=\uparrow,\downarrow,
\end{equation}
with $\bx=(x,y,z)^T$ in 3D, $\bx=(x,y)^T $ in 2D, and $\bx=x$ in 1D.
The normalization for \eqref{eq:cgpen:sec2} becomes
 \be \label{eq:normn:sec2}
\|\Psi\|^2:=
\|\psi_\uparrow(\cdot,t)\|^2+\|\psi_\downarrow(\cdot,t)\|^2:=\int_{{\mathbb R}^d}
\left[|\psi_\uparrow(\bx,t)|^2+|\psi_\downarrow(\bx,t)|^2\right]\,d\bx=1. \ee
Without loss of generality and for mathematical convenience,
we shall assume $\Omega$, $\delta$ and $\beta_{jl}$ satisfying $\beta_{jl}=\beta_{lj}$ ($j,l=\uparrow,\downarrow$) are given real constants,
and $V_j(\bx)$ ($j=\uparrow,\downarrow$) are given non-negative real functions.

Despite the normalization (or mass conservation) \eqref{eq:normn:sec2}, the CGPEs \eqref{eq:cgpen:sec2} possess another important conserved quantity, i.e. {\sl energy per particle},
\begin{align} E(\Psi)=&\int_{\mathbb{R}^d}\biggl[
 \sum_{j=\uparrow,\downarrow}\left(\frac12|\nabla\psi_{j}|^2+V_j(\bx)|\psi_j|^2\right)+\frac{\delta}{2}
\left(|\psi_\uparrow|^2-|\psi_\downarrow|^2\right)+\frac 12 \beta_{\uparrow\uparrow}|\psi_\uparrow|^4+\frac
12\beta_{\downarrow\downarrow}|\psi_\downarrow|^4\nonumber\\
&\qquad\qquad
+\beta_{\uparrow\downarrow}|\psi_\uparrow|^2|\psi_\downarrow|^2+\Omega\;\text{Re}
(\psi_\uparrow\overline{\psi_\downarrow})\biggl]d\bx,\label{eq:energy:sec2}
\end{align}
where $\bar{f}$ and Re$(f)$ denote the conjugate and real parts of a
function $f$, respectively.

\subsection{Ground states}
The ground state $\Phi_g:=\Phi_g(\bx)=(\phi_\uparrow^g(\bx),\phi_\downarrow^g(\bx))^T$ of
the pseudo spin-1/2 BEC with an internal atomic Josephson junction governed by
\eqref{eq:cgpen:sec2} is defined as the minimizer of the following nonconvex
minimization problem:
 \begin{quote}
  Find $\left(\Phi_g \in S\right)$, such that
\end{quote}
  \begin{equation}\label{eq:minimize:sec2}
    E_g := E\left(\Phi_g\right) = \min_{\Phi \in S}
    E\left(\Phi\right),
  \end{equation}
where $S$ is a nonconvex set defined as
\be\label{eq:nonconset:sec2}
S:=\left\{\Phi=(\phi_\uparrow,\phi_\downarrow)^T \ | \ \|\Phi\|^2=\int_{{\mathbb
R}^d}\left(|\phi_\uparrow(\bx)|^2+|\phi_\downarrow(\bx)|^2\right)d\bx=1,\
E(\Phi)<\infty \right\}.\ee
It is easy to see that the ground state
$\Phi_g$  satisfies the following Euler-Lagrange equations \be
\label{eq:grd1:sec2}
\begin{split} &\mu\,\phi_\uparrow=\left[-\frac 12\nabla^2
+V_\uparrow(\bx)+\frac{\delta}{2}
+(\beta_{\uparrow\uparrow}|\phi_\uparrow|^2+
\beta_{\uparrow\downarrow}|\phi_\downarrow|^2)\right]
\phi_\uparrow+\frac{\Omega}{2}
\phi_\downarrow,\qquad \bx\in {\mathbb R}^d,\\
&\mu \,\phi_\downarrow=\left[-\frac 12\nabla^2
+V_\downarrow(\bx)-\frac{\delta}{2}+(\beta_{\downarrow\uparrow}
|\phi_\uparrow|^2+\beta_{\downarrow\downarrow}|\phi_\downarrow|^2)\right]
\phi_\downarrow+\frac{\Omega}{2}
\phi_\uparrow,\qquad \bx\in {\mathbb R}^d,
\end{split}
\ee under the constraint \be \label{eq:norm1:sec2}
\|\Phi\|^2:=\|\Phi\|_2^2=\int_{{\mathbb R}^d}
\left[|\phi_\uparrow(\bx)|^2+|\phi_\downarrow(\bx)|^2\right]\,d\bx=1, \ee
 with the eigenvalue  $\mu$ being the Lagrange multiplier
 (or chemical potential in physics literatures)  corresponding to the constraint \eqref{eq:norm1:sec2}, which can be
computed as
\be\mu=\mu(\Phi)=E(\Phi)+\int_{{\mathbb R}^d}\left[
 \frac{\beta_{\uparrow\uparrow}}{2}|\phi_\uparrow|^4
+\frac{\beta_{\downarrow\downarrow}}{2}|\phi_\downarrow|^4
+\beta_{\uparrow\downarrow}|\phi_\uparrow|^2|\phi_\downarrow|^2\right]\,d\bx. \label{eq:chemp3:sec2}
\ee
  In
fact, the above time-independent CGPEs \eqref{eq:grd1:sec2} can also be
obtained from the CGPEs \eqref{eq:cgpen:sec2} by substituting the ansatz
\begin{equation} \label{eq:anst1:sec2} \psi_\uparrow(\bx,t)=e^{-i\mu t}\phi_\uparrow(\bx),\qquad
\psi_\downarrow(\bx,t)=e^{-i\mu t}\phi_\downarrow(\bx), \qquad \bx\in {\mathbb R}^d.
\end{equation}
The eigenfunctions of  the nonlinear eigenvalue problem \eqref{eq:grd1:sec2}
under the normalization \eqref{eq:norm1:sec2} are usually called as
stationary states of the two-component BEC \eqref{eq:cgpen:sec2} \cite{LieSolo,LinWei,LuLiu}. Among
them, the  eigenfunction with the  minimum energy is the ground state and
those whose energy are larger than that of the ground state are
usually called as excited states.

\subsubsection{Mathematical theories}
Before presenting mathematical theories on ground states, some notations are introduced below.
Define the function $I(\bx)$ as
\be\label{eq:Ibx:sec2}
I(\bx)=\left(V_\uparrow(\bx)-V_\downarrow(\bx)+\delta\right)^2 +(\beta_{\uparrow\uparrow}-\beta_{\uparrow\downarrow})^2+
(\beta_{\uparrow\downarrow}-\beta_{\downarrow\downarrow})^2,\quad \bx\in {\mathbb R}^d,
\ee
where $I(\bx)\equiv0$ means that the spin-1/2 BEC with $\Omega=0$ is essentially one component;
denote the interaction matrix as
\be\label{eq:Bforl1:sec}B=\begin{pmatrix}\beta_{\uparrow\uparrow}
&\beta_{\uparrow\downarrow}\\
\beta_{\uparrow\downarrow}&\beta_{\downarrow\downarrow}\end{pmatrix},
\ee and we say $B$ is positive semi-definite iff $\beta_{\uparrow\uparrow}\ge0$ and
$\beta_{\uparrow\uparrow}\beta_{\downarrow\downarrow}-\beta_{\uparrow\downarrow}^2\ge0$; and $B$ is nonnegative iff
$\beta_{\uparrow\uparrow}\ge0$, $\beta_{\uparrow\downarrow}\ge0$ and $\beta_{\downarrow\downarrow}\ge0$.
 In 2D, i.e. $d=2$, let $C_b$ be the
best constant as \cite{Weinstein}
\be\label{eq:bestc:sec2}C_b:=\inf_{0\ne f\in H^1({\mathbb R}^2)} \frac{\|\nabla
f\|_{L^2(\mathbb R^2)}^2\|f\|_{L^2(\mathbb R^2)}^2}{\|f\|_{L^4(\mathbb R^2)}^4}=\pi\cdot (1.86225\ldots).\ee

For the ground state of \eqref{eq:minimize:sec2}, we have \cite{BaoCai0,Baocai2013}

\begin{theorem} [existence and uniqueness of \eqref{eq:minimize:sec2} \cite{BaoCai0}] \label{thm:con1:sec2}
Suppose  $V_j(\bx)\ge 0$ ($j=\uparrow,\downarrow$) satisfying
$\lim\limits_{|\bx|\to\infty}V_j(\bx)=+\infty$ and at least one of the
following conditions holds
\begin{enumerate}\renewcommand{\labelenumi}{(\roman{enumi})}
 \item $d=1$;
\item  $d=2$ and  $\beta_{\uparrow\uparrow}
> -C_{b}$\,,  $\beta_{\downarrow\downarrow}> -C_b$\,, and $\beta_{\uparrow\downarrow}\ge -C_b-
\sqrt{C_b+\beta_{\uparrow\uparrow}}\sqrt{C_b+\beta_{\downarrow\downarrow}}$;
\item $d=3$ and  $B$ is
either positive semi-definite or nonnegative;
\end{enumerate}
there exists a
ground state $\Phi_g=(\phi_\uparrow^g,\phi_\downarrow^g)^T$ of \eqref{eq:minimize:sec2}. In
addition,
 $\widetilde{\Phi}_g:=
 (e^{i\theta_\uparrow}|\phi_\uparrow^g|,e^{i\theta_\downarrow}
 |\phi_\downarrow^g|)$ is also a
ground state of \eqref{eq:minimize:sec2} with two
constants$\theta_\uparrow$, $\theta_\downarrow\in [0,2\pi)$ satisfying $\theta_\uparrow-\theta_\downarrow=\pm\pi$ when $\Omega>0$ and
$\theta_\uparrow-\theta_\downarrow=0$ when $\Omega<0$, respectively. Furthermore,
if the matrix $B$ is positive semi-definite, $\Omega\ne0$ and  $I(\bx)\neq0$,  the ground state
$(|\phi_\uparrow^g|,-\text{sign}(\Omega)|\phi_\downarrow^g|)^T$ is unique.
 In contrast, if one of the following conditions holds,
\begin{enumerate}\renewcommand{\labelenumi}{(\roman{enumi})$^\prime$}
\item  $d=2$ and $\beta_{\uparrow\uparrow}\leq-C_b$ or $\beta_{\downarrow\downarrow}\leq-C_b$ or $\beta_{\uparrow\downarrow}<
-C_b-\sqrt{C_b+\beta_{\uparrow\uparrow}}\sqrt{C_b+\beta_{\downarrow\downarrow}}$ ;
\item $d=3$ and $\beta_{\uparrow\uparrow}<0$ or $\beta_{\downarrow\downarrow}<0$ or $\beta_{\uparrow\downarrow}<0$ with
$\beta_{\uparrow\downarrow}^2> \beta_{\uparrow\uparrow}\beta_{\downarrow\downarrow}$;
\end{enumerate}
there exists no ground state of \eqref{eq:minimize:sec2}, i.e. $\inf\limits_{\Phi\in S}E(\Phi)=-\infty$.
\end{theorem}


\begin{theorem}[limiting behavior when $|\Omega|\to+\infty$  \cite{BaoCai0}]\label{thm:omg:sec2}
Suppose $V_j(\bx)\ge 0$ ($j=\uparrow,\downarrow$) satisfying
$\lim\limits_{|\bx|\to\infty}V_j(\bx)=+\infty$ and $B$ is
either positive semi-definite or nonnegative. For fixed
$V_j(\bx)$ ($j=\uparrow,\downarrow$), $B$  and $\delta$, let
$\Phi^\Omega=(\phi_\uparrow^\Omega,\phi_\downarrow^\Omega)^T$ be a ground state of
\eqref{eq:minimize:sec2}. Then when $|\Omega|\to
+\infty$, we have   \be \|\ |\phi_j^\Omega|-\phi^g\|\to 0, \quad
j=\uparrow,\downarrow, \qquad \quad E(\Phi^\Omega)\approx 2 E_1(\phi^g)-|\Omega|/2,\ee where
$\phi^g$ is the unique positive minimizer \cite{LiebSeiringerPra2000} of
\be\label{eq:oglim:sec2} E_1(\phi)=\int_{\mathbb{R}^d}\left[\frac 12|\nabla
\phi|^2+V(\bx) |\phi|^2+\frac{\beta}{2} |\phi|^4\right]d\bx
\ee under the constraint \be
\|\phi\|^2=\int_{\mathbb{R}^d}|\phi(\bx)|^2\,d\bx=\frac
12, \ee with  $\beta=\frac{\beta_{\uparrow\uparrow}+\beta_{\downarrow\downarrow}+2\beta_{\uparrow\downarrow}}{2}$ and $V(\bx)=\frac{1}{2}(V_\uparrow(\bx)+V_\downarrow(\bx))$.
\end{theorem}

\begin{theorem}[limiting behavior when $\delta\to\pm \infty$ \cite{BaoCai0}] \label{thm:lim2:sec2}
Suppose $V_j(\bx)\ge 0$ ($j=\uparrow,\downarrow$) satisfying
$\lim\limits_{|\bx|\to\infty}V_j(\bx)=+\infty$ and $B$ is
either positive semi-definite or nonnegative. For fixed
$V_j(\bx)$ ($j=\uparrow,\downarrow$), $B$  and $\Omega$, let
$\Phi^\delta=(\phi_\uparrow^\dt,\phi_\downarrow^\delta)^T$ be a ground state of
\eqref{eq:minimize:sec2} with respect to $\delta$. Then when $\delta\to +\infty$,
we have \be \|\phi_\uparrow^\delta\|\to 0,\quad \| |\phi_\downarrow^\delta|-\phi^g\|\to 0, \qquad \quad E(\Phi^\delta)\approx
E_2(\phi^g)-\frac{\delta}{2},
 \ee and resp.; when $\delta\to-\infty$, we have
\be \|\ |\phi_\uparrow^\delta|-\phi^g\|\to 0, \quad
\|\phi_\downarrow^\delta\|\to 0, \qquad \quad E(\Phi^\delta)\approx
E_2(\phi^g)+\frac{\delta}{2},\ee where $\phi^g$ is the unique positive minimizer \cite{LiebSeiringerPra2000}
of \be
E_2(\phi)=\int_{\mathbb{R}^d}\left[\frac 12|\nabla
\phi|^2+V_\ast(\bold{x})|\phi|^2+\frac{\beta_\ast}{2}|\phi|^4\right]d\bold{x}
\ee under the constraint \be
\|\phi\|^2=\int_{\mathbb{R}^d}|\phi(\bx)|^2\,d\bold{x}=1,
\ee with $\beta_\ast=\beta_{\downarrow\downarrow}$ and $V_\ast(\bx)=V_\downarrow(\bx)$ when $\delta>0$, and resp., $\beta_\ast=\beta_{\uparrow\uparrow}$, $V_\ast(\bx)=V_\uparrow(\bx)$
when $\delta<0$.
\end{theorem}

\subsubsection{Numerical methods and results}\label{numeric-gs:sec2}
In order to compute the ground state  \eqref{eq:minimize:sec2}, we construct
the following continuous normalized gradient flow (CNGF) for $\Phi(\bx,t)=(\phi_\uparrow(\bx,t),\phi_\downarrow(\bx,t))^T$ \cite{BaoCai0}:
\bea\label{eq:cgf1:sec2}\begin{split} &\qquad
\frac{\partial\phi_\uparrow(\bx,t)}{\partial t}=\left[\frac 1 2\nabla^2
-V_\uparrow(\bx)-\frac{\delta}{2}-(\beta_{\uparrow\uparrow}
|\phi_\uparrow|^2+\beta_{\uparrow\downarrow}|\phi_\downarrow|^2)
\right]\phi_\uparrow-\frac{\Omega}{2}
\phi_\downarrow+\mu_\Phi(t)\phi_\uparrow,  \\
&\qquad \frac{\partial\phi_\downarrow(\bx,t)}{\partial t}=\left[\frac
12\nabla^2-V_\downarrow(\bx)+\frac{\delta}{2}-(\beta_{\uparrow
\downarrow}|\phi_\uparrow|^2+\beta_{\downarrow\downarrow}|
\phi_\downarrow|^2)\right]\phi_\downarrow-\frac{\Omega}{2}
\phi_\uparrow+\mu_\Phi(t)\phi_\downarrow,
\end{split}
\eea
with a prescribed initial data $\Phi(\bx,0)=\Phi_0(\bx)=(\phi_{\uparrow}^0(\bx),\phi_{\downarrow}^0(\bx))^T$ satisfying
$\|\Phi_0\|=1$, where
$\mu_\Phi(t)$ is chosen such that the above CNGF is mass (or
normalization) conservative and energy diminishing.
By taking $\mu_\Phi(t)=\frac{\mu(\Phi(\cdot,t))}{\|\Phi(\cdot,t)\|^2}$ with $\mu(\Phi)$ given in (\ref{eq:chemp3:sec2}) \cite{BaoCai0}, it is readily to check that the CNGF \eqref{eq:cgf1:sec2}
conserves the mass and is energy diminishing \cite{BaoCai0}. Therefore, one can compute ground sates of \eqref{eq:minimize:sec2} by discretizing the CNGF \eqref{eq:cgf1:sec2}.

In practical computation, an efficient way to discretize  the CNGF \eqref{eq:cgf1:sec2} is
through the construction of the following gradient flow with
discrete normalization (GFDN) via a time-splitting approach:
let $\tau>0$ be a chosen time step and denote
 $t_n=n\tau$ for $n\ge0$. One solves
\be\label{eq:dngf1:sec2} \begin{split} &\qquad
\frac{\partial\phi_\uparrow}{\partial t}=\left[\frac 1 2\nabla^2
-V_\uparrow(\bx)-\frac{\delta}{2}-(\beta_{\uparrow\uparrow}
|\phi_\uparrow|^2+\beta_{\uparrow\downarrow}|\phi_\downarrow|^2)
\right]\phi_\uparrow
-\frac{\Omega}{2}\phi_\downarrow,  \\
&\qquad \frac{\partial\phi_\downarrow}{\partial t}=\left[\frac
12\nabla^2-V_\downarrow(\bx)+\frac{\delta}{2}-(\beta_{\uparrow\downarrow}|\phi_\uparrow|^2+\beta_{\downarrow\downarrow}|\phi_\downarrow|^2)\right]\phi_\downarrow-\frac{\Omega}{2}
\phi_\uparrow, \end{split} \quad  t_{n}\le t<t_{n+1},\ee
followed by a
projection step as
\begin{eqnarray}\label{eq:dnproj:sec2}
\phi_l(\bx,t_{n+1}):=
\phi_l(\bx,t_{n+1}^+)=\sigma_l^{n+1}\;\phi_l(\bx\, ,t_{n+1}^-), \qquad
l=\uparrow,\downarrow, \quad n\ge0,
\end{eqnarray}
where $\phi_l(\bx,t_{n+1}^{\pm})=\lim\limits_{t\to
t_{n+1}^{\pm}}\phi_l(\bx,t)$  and the projection constants $\sigma_l^{n+1}$ ($l=\uparrow,\downarrow$)
are chosen such that \be \label{eq:pcont3:sec2}
 \|\Phi(\bx,t_{n+1})\|^2=\|\phi_\uparrow(\bx,t_{n+1})\|^2+
 \|\phi_\downarrow(\bx,t_{n+1})\|^2=1, \qquad n\ge0.
\ee
Since there are two projection constants to be determined,
i.e. $\sigma_\uparrow^{n+1}$ and $\sigma_\downarrow^{n+1}$ in \eqref{eq:dnproj:sec2}, and there is only one equation, i.e. \eqref{eq:pcont3:sec2}, to fix them, we need to find another condition
so that the two projection constants are uniquely determined. In fact,
the above GFDN \eqref{eq:dngf1:sec2}-\eqref{eq:dnproj:sec2} can be viewed as
applying the first-order splitting method to the CNGF \eqref{eq:cgf1:sec2}
and the projection step \eqref{eq:dnproj:sec2} is equivalent to solving the
following ordinary differential equations (ODEs) \be
\frac{\partial\phi_\uparrow(\bx,t)}{\partial t}= \mu_\Phi(t)\phi_\uparrow, \qquad
\frac{\partial\phi_\downarrow(\bx,t)}{\partial t}= \mu_\Phi(t)\phi_\downarrow, \qquad
\quad t_n\leq t\leq t_{n+1}, \ee which immediately suggests that the
projection constants in \eqref{eq:dnproj:sec2} could  be chosen as \cite{BaoCai0} \be
\label{eq:pcont1:sec2}\sigma_\uparrow^{n+1}=\sigma_{\downarrow}^{n+1}, \qquad n\ge0. \ee
Plugging \eqref{eq:pcont1:sec2} and \eqref{eq:dnproj:sec2} into
\eqref{eq:pcont3:sec2}, we get
  \be
\sigma_\uparrow^{n+1}=\sigma_\downarrow^{n+1}=\frac{1}{\|\Phi(\cdot,t_{n+1}^-)\|}=
\frac{1}{\sqrt{\|\phi_\uparrow(\cdot,t_{n+1}^-)\|^2
+\|\phi_\downarrow(\cdot,t_{n+1}^-)\|^2}}, \qquad n\ge0. \ee
In fact, the
gradient flow \eqref{eq:dngf1:sec2} can be viewed as applying the steepest
decent method to the energy functional $E(\Phi)$ in \eqref{eq:minimize:sec2}
without the constraint, and  then projecting the solution back to
the unit sphere $S$ in \eqref{eq:dnproj:sec2}. In addition, \eqref{eq:dngf1:sec2} can also be obtained
from the CGPEs \eqref{eq:cgpen:sec2} by the change of variable $t\to -i\;
t$, and thus this kind of algorithm is usually called as the
imaginary time method in the physics literature \cite{Lahaye,Leggett,Myatt,StamperUeda,WenLiuCai}.

To fully discretize the GFDN \eqref{eq:dngf1:sec2},  it is highly recommended to use backward Euler scheme in temporal discretization \cite{BaoDu,BaoCai0,Baocai2013} and adopt one's favorite numerical method, such as
finite difference method, spectral method and finite element, for spatial discretization. In practical computation,
the GFDN \eqref{eq:dngf1:sec2} (or  CNGF \eqref{eq:cgf1:sec2})
is usually truncated on a bounded domain
with either homogeneous Dirichlet boundary conditions
or periodic boundary conditions or homogeneous Neumann boundary conditions due to the trapping potentials $V_j(\bx)$ ($j=\uparrow,\downarrow$)  which ensure the wave function decays exponentially fast at far field.
For the convenience of readers and simplification of notations,
here we only present a modified
backward Euler finite difference (BEFD)  discretization of the GFDN \eqref{eq:dngf1:sec2}
in 1D, which is truncated on a bounded domain
$U=(a,b)$ with homogeneous Dirichlet boundary condition
\be
\label{eq:bc:sec2}
\Phi(a,t)=\Phi(b,t)={\bf 0},\quad t\ge0.
\ee
Choose a mesh size $h=(b-a)/L$ with $L$ being a positive integer and denote $x_j=a+jh$ ($0\leq j\leq L$)
as the grid points, then the BEFD discretization reads
\bea  \label{eq:befd1:sec21}
\frac{\phi^{(1)}_{\uparrow,j}-\phi_{\uparrow,j}^n}{\tau}&=&\frac{1}{2h^2}\left
[\phi_{\uparrow,j+1}^{(1)}-2\phi_{\uparrow,j}^{(1)}+\phi_{\uparrow,j-1}^{(1)}
\right]
-\left[V_\uparrow(x_j)+\frac{\delta}{2}+\alpha_\ast\right]\phi_{\uparrow,j}^{(1)}
-\frac{\Omega}{2}\phi_{\downarrow,j}^{(1)} \nonumber\\
&&-\left(\beta_{\uparrow\uparrow}|\phi_{\uparrow,j}^n|^2+
\beta_{\uparrow\downarrow}|\phi_{\downarrow,j}^n|^2\right)
\phi_{\uparrow,j}^{(1)}+\alpha_\ast\phi_{\uparrow,j}^n,\quad
1\le j\le L-1,\\
\label{eq:befd1:sec22}
\frac{\phi^{(1)}_{\downarrow,j}-\phi_{\downarrow,j}^n}{\tau}&=&\frac{1}{2h^2}
\left[\phi_{\downarrow,j+1}^{(1)}-2\phi_{\downarrow,j}^{(1)}+
\phi_{\downarrow,j-1}^{(1)}\right]
-\left[V_\downarrow(x_j)-\frac{\delta}{2}+\alpha_\ast\right]
\phi_{\uparrow,j}^{(1)}-\frac{\Omega}{2}\phi_{\uparrow,j}^{(1)} \nonumber\\
&&-\left(\beta_{\uparrow\downarrow}|\phi_{\uparrow,j}^n|^2+
\beta_{\downarrow\downarrow}|\phi_{\downarrow,j}^n|^2\right)
\phi_{\downarrow,j}^{(1)}+\alpha_\ast\phi_{\downarrow,j}^n,\quad 1\le j\le L-1,\\
\label{eq:befd1:sec23}
\phi_{l,j}^{n+1}&=&\frac{\phi^{(1)}_{l,j}}{\|\Phi^{(1)}\|_h},\qquad
j=0,1,\ldots,L,\quad n\ge0, \quad l=\uparrow,\downarrow;
\eea where $\alpha_\ast\ge0$ is a stabilization parameter \cite{BaoLim} chosen in such a way that
the time step $\tau$ is independent of the effective Rabi frequency
$\Omega$ and \be \|\Phi^{(1)}\|_h:=\sqrt{h\sum\limits_{j=0}^{L-1}
\left[|\phi_{\uparrow,j}^{(1)}|^2+|\phi_{\downarrow,j}^{(1)}|^2\right]}.\ee The
initial and boundary conditions are discretized as
\be\label{initd1d}
\phi_{l,j}^{0}=\phi_{l}^0(x_j), \quad j=0,1,\ldots,L;\qquad
\phi_{l,0}^{n}=\phi_{l,L}^{n}=0,\quad n\ge0; \qquad l=\uparrow,\downarrow.
\ee

We remark here that many other numerical methods proposed
in the literatures for computing the ground state of single-component
BEC \cite{Adhikari-2000,Ant,Ant1,BaoCaiWang,BaoChernLim,BaoT,BaoWangP,Can,CC2,
Chang1,Cerim,Dana,Dana2017,Edw-Bur-1995,Ming,Ruan,Wu} can be extended to computing numerically the ground state
of pseudo spin-$1/2$ BEC \cite{Bao,Baocai2015}.

\begin{figure}[t!]
\centerline{
\includegraphics[height=5cm,width=7cm]{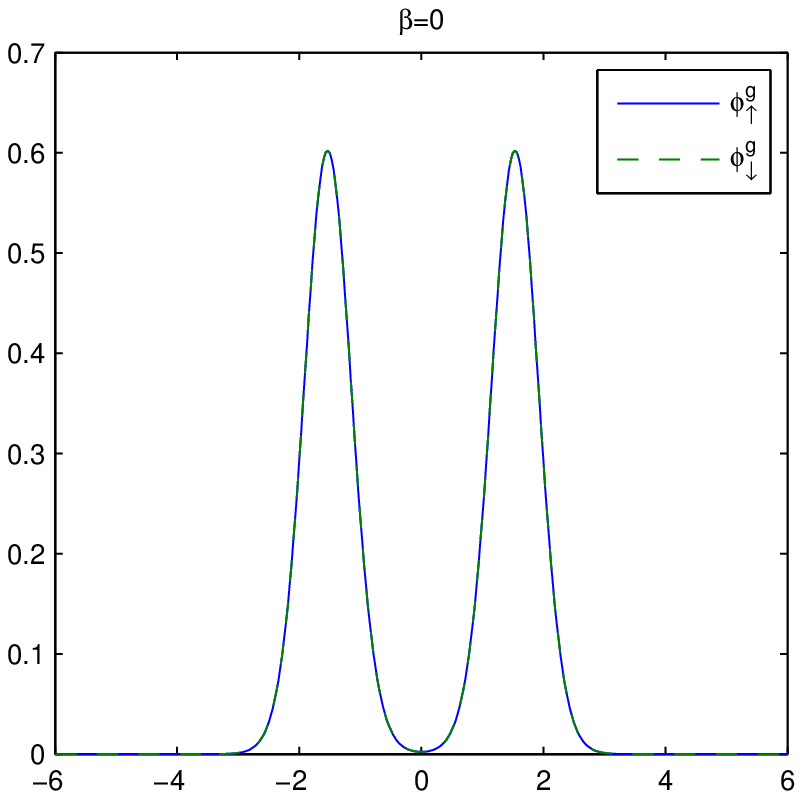} \qquad
\includegraphics[height=5cm,width=7cm]{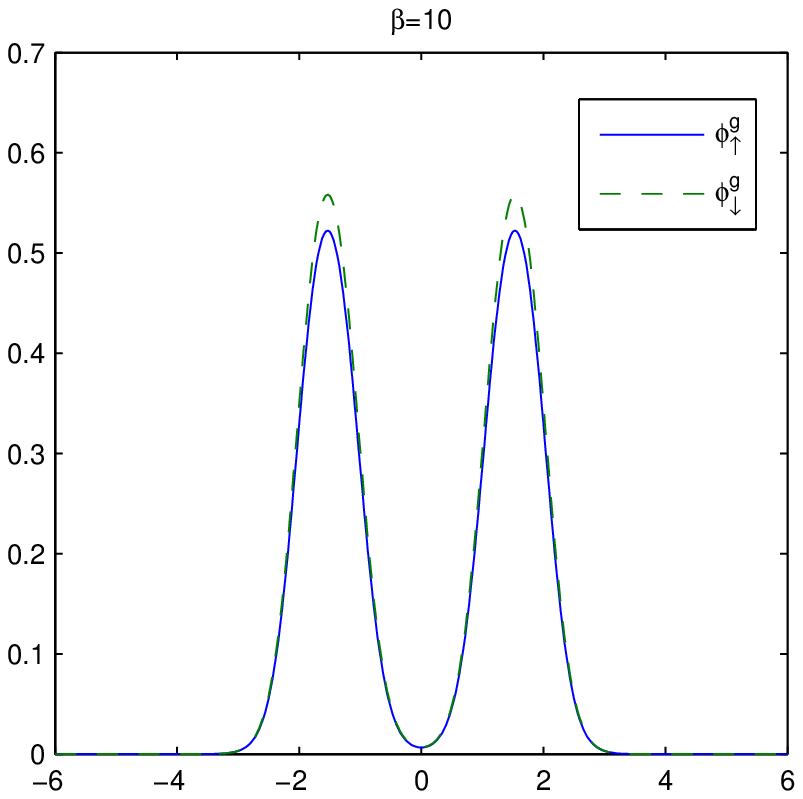}}
\centerline{
\includegraphics[height=5cm,width=7cm]{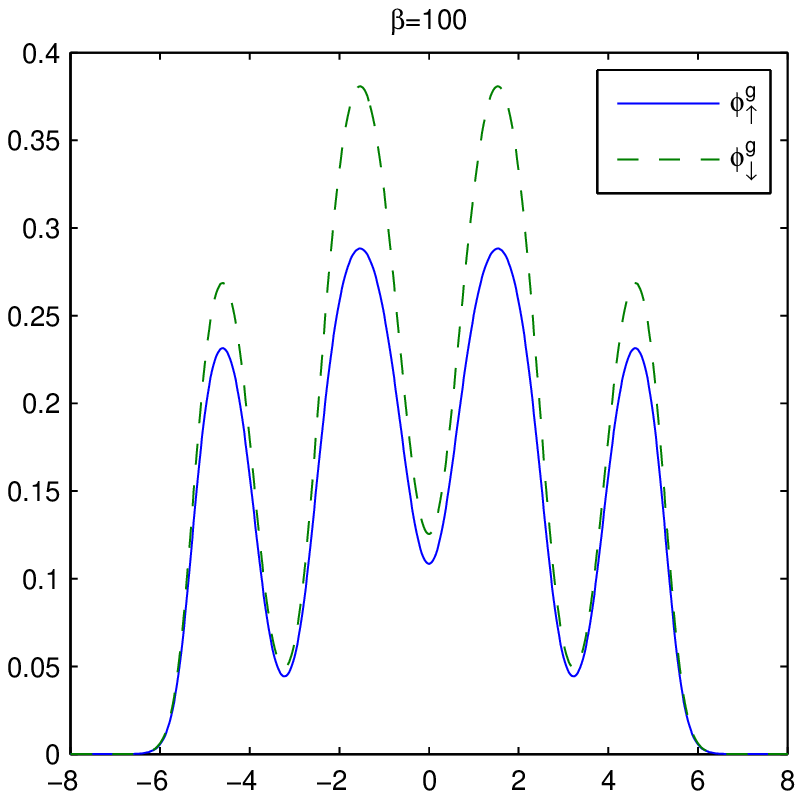} \qquad
\includegraphics[height=5cm,width=7cm]{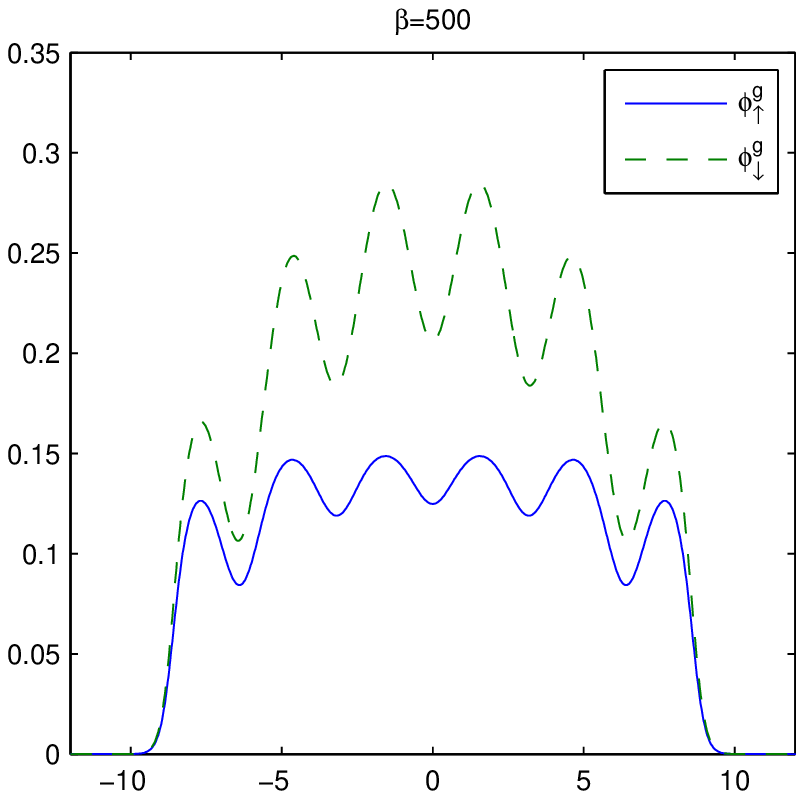}}

\caption{Ground states $\Phi_g=(\phi_\uparrow^g,\phi_\downarrow^g)^T$ in Example 2.1 with $\delta=0$ and $\Omega=-2$ for different $\beta$. } \label{fig:sec2}
\end{figure}

\begin{figure}[t!]
\centerline{
\includegraphics[height=5cm,width=7cm]{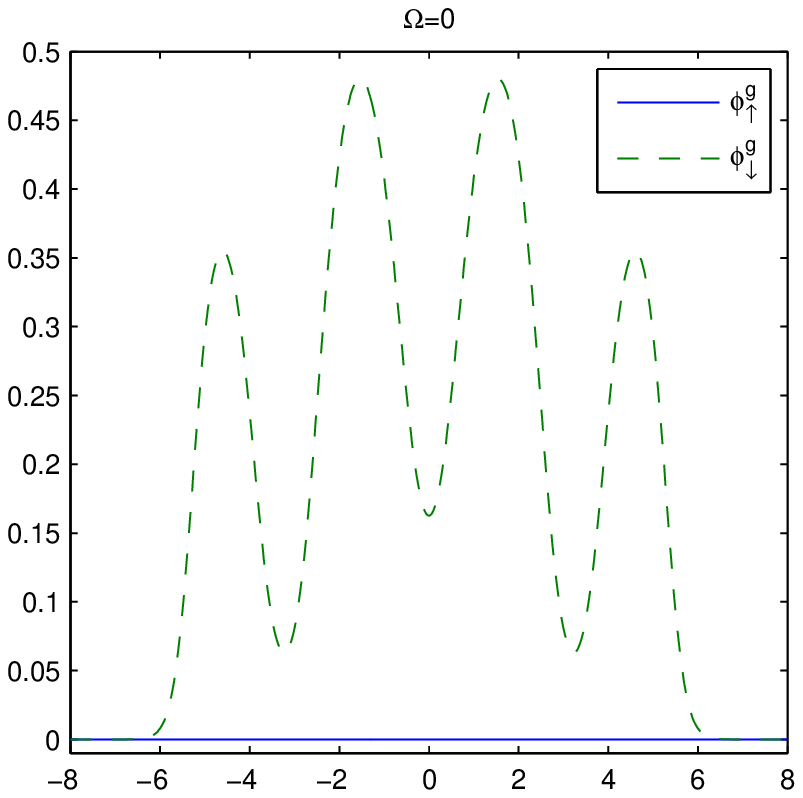} \qquad
\includegraphics[height=5cm,width=7cm]{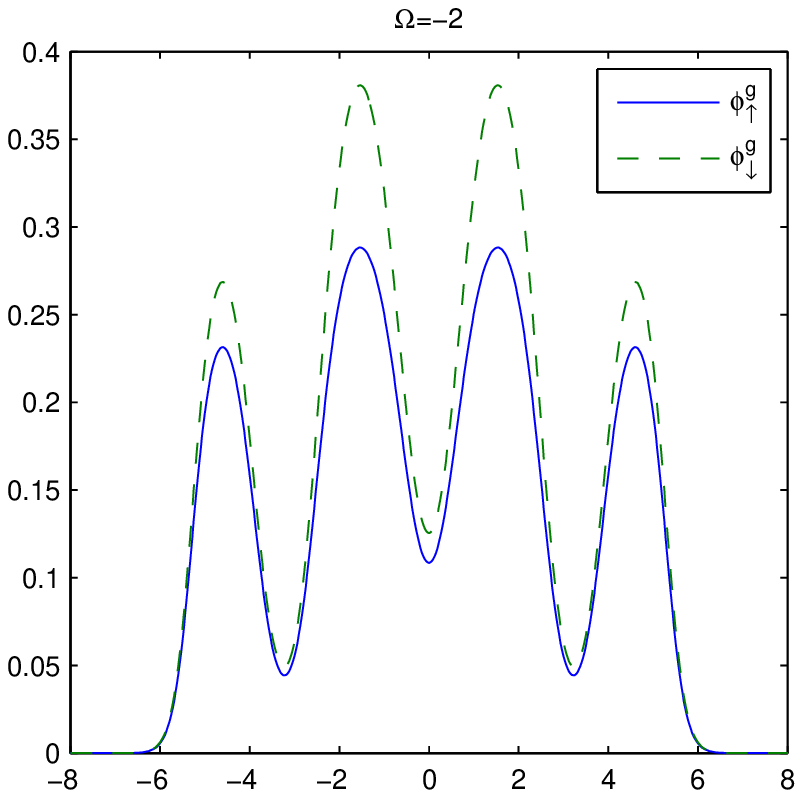}}
\centerline{
\includegraphics[height=5cm,width=7cm]{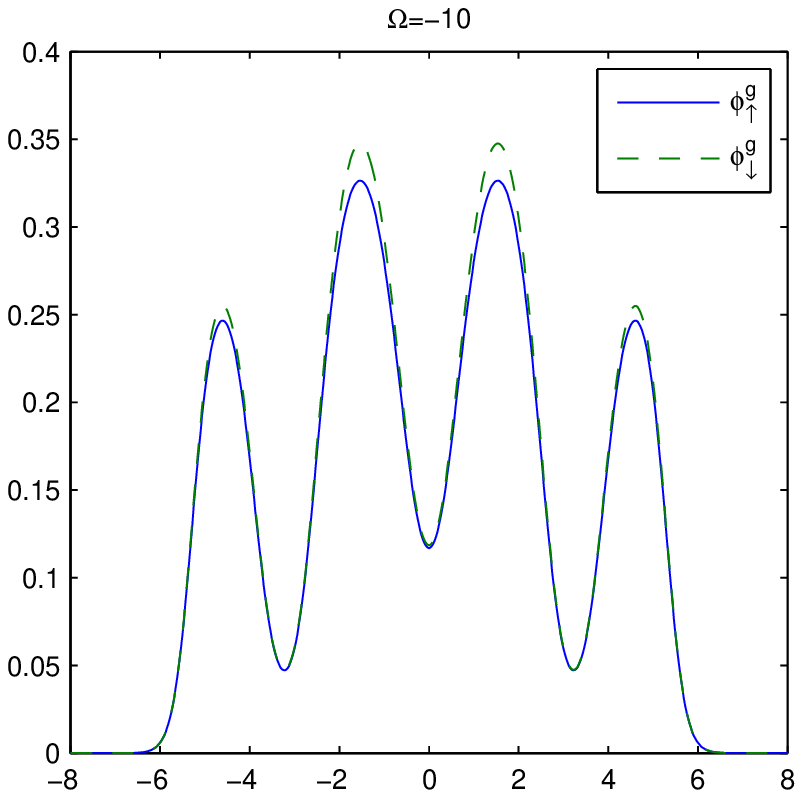} \qquad
\includegraphics[height=5cm,width=7cm]{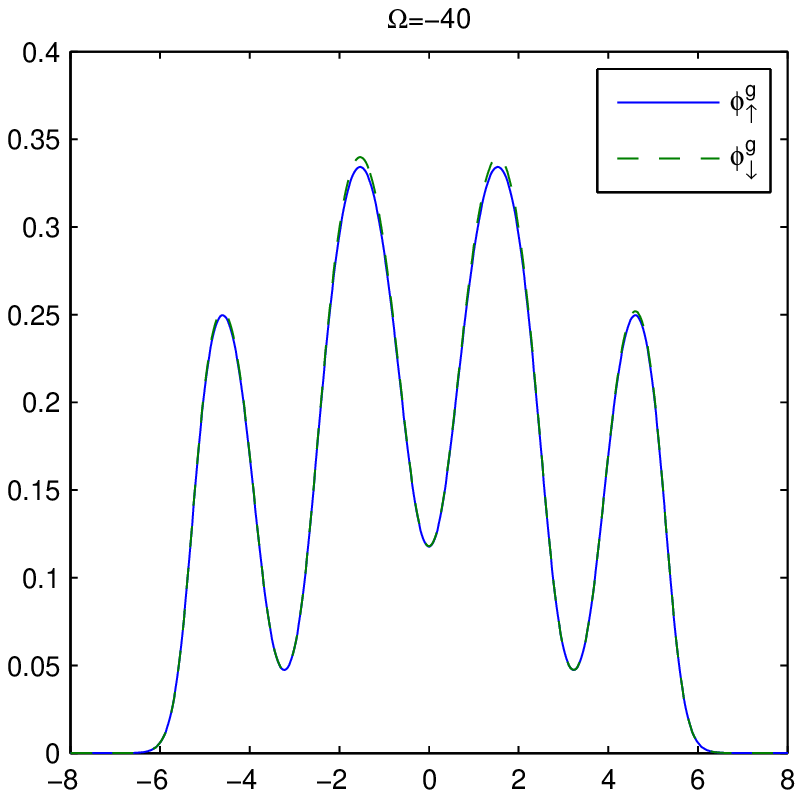}}

\caption{Ground states $\Phi_g=(\phi_\uparrow^g,\phi_\downarrow^g)^T$  in Example 2.1 with $\delta=0$ and $\beta=100$ for different $\Omega$. } \label{fig2:sec2}
\end{figure}

{\it Example 2.1}. To demonstrate the efficiency of the BEFD method
\eqref{eq:befd1:sec21}-\eqref{initd1d} for computing the ground state
of \eqref{eq:minimize:sec2},
we take $d=1$,  $V_\uparrow(x)=V_\downarrow(x)=\frac 12 x^2+24\cos^2(x)$ and
$\beta_{\uparrow\uparrow}:\beta_{\uparrow\downarrow}:
\beta_{\downarrow\downarrow}=(1.03:1:0.97)\beta$ in
\eqref{eq:energy:sec2} with $\beta$ being a real constant.  The
 computational domain is $U=[-16,16]$ with mesh size
$h=\frac{1}{32}$ and time step $\tau=0.1$. Figure \ref{fig:sec2} plots the ground states $\Phi_g$ when $\delta=0$ and
$\Omega=-2$ for different $\beta$, and Figure \ref{fig2:sec2} depicts
similar results when $\delta=0$ and $\beta=100$ for different $\Omega$.

\subsubsection{Another type ground state without Josephson junction}

If there is no internal atomic Josephson junction, i.e. $\Omega=0$ in \eqref{eq:cgpen:sec2}, then the mass of each component is also conserved \cite{BaoCai0}, i.e.
\be \label{normud964}
\|\psi_l(\cdot,t)\|^2:=\int_{{\mathbb R}^d}
|\psi_l(\bx,t)|^2\,d\bx\equiv \int_{{\mathbb R}^d}
|\psi_l(\bx,0)|^2\,d\bx, \quad t\ge0, \qquad l=\uparrow,\downarrow.
\ee
Without loss of generality, we can assume $\delta=0$.
In this case, for any given $\nu\in[0,1]$, one can consider
another type ground state
$\Phi_g^\nu(\bx)=(\phi_\uparrow^\nu(\bx),\phi_\downarrow^\nu(\bx))^T$ of the
spin-1/2 BEC, which  is defined as the minimizer of the following
nonconvex minimization problem:
 \begin{quote}
  Find $\left(\Phi_g^\nu \in S_\nu\right)$, such that
\end{quote}
  \begin{equation}\label{eq:minim2:sec2}
    E_g^\nu := E_0\left(\Phi_g^\nu\right) = \min_{\Phi \in S_\nu}
    E_0\left(\Phi\right),
  \end{equation}
where  $E_0(\cdot)$ is the energy in \eqref{eq:energy:sec2} with $\Omega=\delta=0$, i.e.
\be\label{eq:energy0:sec2}
E_0(\Phi):=\int_{\mathbb{R}^d}\left[
 \sum_{j=\uparrow,\downarrow}\left(\frac12|\nabla\phi_{j}|^2+V_j(\bx)
 |\phi_j|^2\right)+\frac 12 \beta_{\uparrow\uparrow}|\phi_\uparrow|^4+\frac
12\beta_{\downarrow\downarrow}|\phi_\downarrow|^4
+\beta_{\uparrow\downarrow}|\phi_\uparrow|^2|\phi_\downarrow|^2\right]d\bx,
\ee
and
$S_\nu$ is a nonconvex set defined as \be\label{eq:nonconset2:sec2}
S_\nu:=\left\{\Phi=(\phi_\uparrow,\phi_\downarrow)^T \ | \ \|\phi_\uparrow\|^2=\nu,\
\|\phi_\downarrow\|^2=1-\nu,\  E_0(\Phi)<\infty \right\}.\ee
 Again, it is
easy to see that the ground state $\Phi_g^\nu$  satisfies the
following Euler-Lagrange equations \be \label{eq:grd2:sec2} \begin{split}
&\mu_\uparrow\,\phi_\uparrow=\left[-\frac 12\nabla^2 +V_\uparrow(\bx)
+(\beta_{\uparrow\uparrow}|\phi_\uparrow|^2+\beta_{\uparrow\downarrow}|\phi_\downarrow|^2)\right]\phi_\uparrow,\qquad\bx\in {\mathbb R}^d,\\
&\mu_\downarrow \,\phi_\downarrow=\left[-\frac 12\nabla^2
+V_\downarrow(\bx)+(\beta_{\uparrow\downarrow}|\phi_\uparrow|^2+\beta_{\downarrow\downarrow}|\phi_\downarrow|^2)\right]\phi_\downarrow,\qquad
\bx\in {\mathbb R}^d,
\end{split}
\ee under the two constraints \be \label{eq:norm2:sec2}
\|\phi_\uparrow\|^2:=\int_{{\mathbb R}^d} |\phi_\uparrow(\bx)|^2\,d\bx=\nu,\qquad
\|\phi_\downarrow\|^2:=\int_{{\mathbb R}^d} |\phi_\downarrow(\bx)|^2\,d\bx=1-\nu, \ee
with $\mu_\uparrow$ and $\mu_\downarrow$ being the Lagrange multipliers or chemical
potentials corresponding to the two constraints in (\ref{eq:norm2:sec2}).
Again, the above time-independent CGPEs (\ref{eq:grd2:sec2}) can also be
obtained from the CGPEs \eqref{eq:cgpen:sec2} with $\Omega=\delta=0$ by substituting
the ansatz
\begin{equation}\label{eq:anst2:sec2} \psi_\uparrow(\bx,t)=e^{-i\mu_\uparrow t}\phi_\uparrow(\bx),\qquad
\psi_\downarrow(\bx,t)=e^{-i\mu_\downarrow t}\phi_\downarrow(\bx),\qquad
\bx\in {\mathbb R}^d.
\end{equation}
We remark here that, when $\Omega=\delta=0$ in \eqref{eq:energy:sec2},
the ground state $\Phi_g$ defined in \eqref{eq:minimize:sec2} can be computed from the ground states $\Phi_g^\nu$ ($0\le \nu\le 1$) in \eqref{eq:energy0:sec2} as
\be
E_g:=E(\Phi_g)=\min_{\Phi\in S} E_0(\Phi)=\min_{0\le \nu\le 1} E_g^\nu=\min_{0\le \nu\le 1} E_0(\Phi_g^\nu)=\min_{0\le \nu\le 1}\;\min_{\Phi\in S_\nu} E_0(\Phi).
\ee

If $\nu=0\text{ or } 1$ in
the nonconvex minimization problem \eqref{eq:minim2:sec2}, it reduces to
the ground state of single-component BEC, which
has been well studied in the literature \cite{LiebSeiringerPra2000,Baocai2013}.
Thus here we assume $\nu\in(0,1)$ and  denote
$$
\beta_{\uparrow\uparrow}^\prime:=\nu\beta_{\uparrow\uparrow},\quad \beta_{\downarrow\downarrow}^\prime
=(1-\nu)\beta_{\downarrow\downarrow},\quad
\beta_{\uparrow\downarrow}^\prime=\sqrt{\nu(1-\nu)} \beta_{\uparrow\downarrow},\quad
\nu^\prime=\nu(1-\nu),
$$
and we have the  following conclusions \cite{BaoCai0}.

\begin{theorem}[existence and uniqueness of \eqref{eq:minim2:sec2} \cite{BaoCai0}] \label{eq:thmnojj:sec2}
Suppose $V_j(\bx)\ge 0$ ($j=\uparrow,\downarrow$) satisfying
$\lim\limits_{|\bx|\to\infty}V_j(\bx)=+\infty$ and at least one of the
following conditions holds
\begin{enumerate}\renewcommand{\labelenumi}{(\roman{enumi})}
 \item $d=1$;
\item  $d=2$ and  $\beta^\prime_{\uparrow\uparrow}
> -C_{b}$\,, $\beta^\prime_{\downarrow\downarrow}> -C_b$\,, and  $\beta^\prime_{\uparrow\downarrow}
\ge-\sqrt{(C_b+\beta^\prime_{\uparrow\uparrow})(C_b+
\beta_{\downarrow\downarrow}^\prime)}$;
\item $d=3$ and  $B$ is
either positive semi-definite or nonnegative;
\end{enumerate}
 there exists a ground state $\Phi_g^\nu=(\phi_\uparrow^\nu,\phi_\downarrow^\nu)^T$ of
\eqref{eq:minim2:sec2} for any given $\nu\in(0,1)$. In addition,
 $\widetilde{\Phi}_g^\nu:=
 (e^{i\theta_\uparrow}|\phi_\uparrow^\nu|,e^{i\theta_\downarrow}
 |\phi_\downarrow^\nu|)$ is also a
ground state of \eqref{eq:minim2:sec2} with two real phase constants $\theta_\uparrow$ and
$\theta_\downarrow$. Furthermore, if the matrix $B$ is positive
semi-definite,  the ground state $(|\phi_\uparrow^\nu|,|\phi_\downarrow^\nu|)^T$ of
\eqref{eq:minim2:sec2} is unique.
 In contrast, if one of the following conditions holds
\begin{enumerate}\renewcommand{\labelenumi}{(\roman{enumi})$^\prime$}
\item  $d=2$ and $\beta_{\uparrow\uparrow}^\prime\leq-C_b$ or $\beta_{\downarrow\downarrow}^\prime\leq-C_b$
or $\beta^\prime_{\uparrow\downarrow}<-\frac{1}{2\sqrt{\nu^\prime}}
\left(\nu\beta^\prime_{\uparrow\uparrow}+
(1-\nu)\beta_{\downarrow\downarrow}^\prime+C_b\right)$;
\item $d=3$ and $\beta_{\uparrow\uparrow}<0$ or $\beta_{\downarrow\downarrow}<0$ or $\beta_{\uparrow\downarrow}<-\frac{1}
{2\nu^\prime}(\nu^2\beta_{\uparrow\uparrow}+(1-\nu)^2
\beta_{\downarrow\downarrow})$,
\end{enumerate}
there exists no ground state of \eqref{eq:minim2:sec2}.
\end{theorem}

  Similarly, the BEFD method for computing the ground state of
\eqref{eq:minimize:sec2} can be directly extended to compute
the groud state of \eqref{eq:minim2:sec2} by replacing the projection
step \eqref{eq:befd1:sec23} by
\be
\phi_{\uparrow,j}^{n+1}= \frac{\nu \;\phi^{(1)}_{\uparrow,j}}{\|\phi_\uparrow^{(1)}\|_h}, \quad
\phi_{\downarrow,j}^{n+1}= \frac{(1-\nu )\;\phi^{(1)}_{\downarrow,j}}{\|\phi_\downarrow^{(1)}\|_h},
\quad j=0,1,\ldots, L, \quad n\ge0,
\ee
where $\|\phi_l^{(1)}\|_h:=\sqrt{h\sum_{j=0}^{L-1} |\phi_{l,j}^{(1)}|^2}$
for $l=\uparrow, \downarrow$.

{\it Phase separation}. From Theorem \ref{eq:thmnojj:sec2}, we know the positive ground state of \eqref{eq:minim2:sec2} is unique when $B$ is positive semi-definite,  i.e.
the inter-component interaction strength $|\beta_{\uparrow\downarrow}|\leq \sqrt{\beta_{\uparrow\uparrow}\beta_{\downarrow\downarrow}}$. When $\beta_{\uparrow\downarrow}>\sqrt{\beta_{\uparrow\uparrow}
\beta_{\downarrow\downarrow}}$,
the large inter-component interaction will drive the BEC to a segregated phase where the two components $\phi_\uparrow$ and $\phi_\downarrow$ tend to be separated \cite{BaoCai0,CalSqua}, especially when $\beta_{\uparrow\downarrow}\to+\infty$, the two components tend to be
completely separated.

For simplicity of notations,
we take $\nu=1/2$, $\beta_{\uparrow\uparrow}=\beta_{\downarrow\downarrow}=\beta>0$ and $\Omega=\delta=0$ in \eqref{eq:minim2:sec2} with $\beta_{\uparrow\downarrow}\ge0$ as a parameter, i.e.
each component has the same mass $\frac12$ with $\|\phi_\uparrow^{1/2}\|^2=
\|\phi_\downarrow^{1/2}\|^2=\frac{1}{2}$. From Theorem \ref{eq:thmnojj:sec2}, we know for $0\leq\beta_{\uparrow\downarrow}<\sqrt{\beta_{\uparrow\uparrow}
\beta_{\downarrow\downarrow}}=\beta$, the positive ground state $\Phi_g^{1/2}$ is unique and by symmetry there must hold
$\phi_\uparrow^{1/2}=\phi_\downarrow^{1/2}$.
To measure phase separation for different inter-component interaction $\beta_{\uparrow\downarrow}\ge0$, we define the mixing factor for the positive ground state
$\Phi_g^{1/2}=(\phi_\uparrow^{1/2},\phi_\downarrow^{1/2})^T\in S_{1/2}$ as
\begin{equation}\label{eq:factor:sec2}
0\le \eta:=2\int_{\mathbb{R}^d}\phi_\uparrow^{1/2}\phi_\downarrow^{1/2}\,d\bx\le 2 \|\phi_\uparrow^{1/2}\|\times \|\phi_\downarrow^{1/2}\|\le 2\times \frac{1}{\sqrt{2}} \times \frac{1}{\sqrt{2}}= 1.
\end{equation}
In fact, when $\eta=1$, it means that
the two components are totally mixed, i.e.
$\phi_\uparrow^{1/2}\equiv \phi_\downarrow^{1/2}$,
 and resp., when $\eta=0$, it
indicates that the two components are totally separated, i.e. $\phi_\uparrow^{1/2}\phi_\downarrow^{1/2}\equiv 0$.
In this scenario,
for a uniform spin-1/2 BEC system without kinetic energy terms, i.e.
the problem \eqref{eq:minim2:sec2} is defined on a bounded domain $U\subset\mathbb{R}^d$ with periodic boundary condition and
$V_j(\bx)\equiv 0$ ($j=\uparrow,\downarrow$),  then the ground state $\phi_\uparrow^{1/2}$ and $\phi_\downarrow^{1/2}$ are constants.  In this case, $0\le \beta_{\uparrow\downarrow}\leq =\beta$ is a sharp criteria for the phase separation, i.e. $\eta=1$ when $0\le \beta_{\uparrow\downarrow}\leq \beta$, and
resp., $0\le \eta<1$ when $\beta_{\uparrow\downarrow}> \beta$ with $\eta\to 0$ when $\beta_{\uparrow\downarrow}\to +\infty$ \cite{WenLiuCai}. However, as observed and proved in \cite{WenLiuCai},
when the BEC system is no longer uniform in the presence of the external confinement, i.e. there exists kinetic energy, the  phase
separation will be affected by the kinetic energy \cite{WenLiuCai}. More specifically, we consider a box potential $V_{\uparrow}(\bx)=V_{\downarrow}(\bx)=V(\bx)$ taken as
\begin{equation}\label{eq:boxpot:sec2}
V(\bx)=\begin{cases}0,&\bx\in U\subset\mathbb{R}^d,\\
+\infty,&\text{otherwise}.
\end{cases}
\end{equation}
In this case, there exists a constant $\beta_{\uparrow\downarrow}^{\rm cr}>\beta$, which depends on $\beta$ and $U$,  such that the total  mixing still holds, i.e. $\eta=1$  for $0\le \beta_{\uparrow\downarrow}\le \beta_{\uparrow\downarrow}^{\rm cr}$,
and resp., $0\le \eta<1$ when $\beta_{\uparrow\downarrow}> \beta_{\uparrow\downarrow}^{\rm cr}$ with $\eta\to 0$ as $\beta_{\uparrow\downarrow}\to +\infty$. In other words,
the phase separation position is shifted from
$\beta_{\uparrow\downarrow}=\beta$ in the uniform case to
$\beta_{\uparrow\downarrow}= \beta_{\uparrow\downarrow}^{\rm cr}>\beta$
in the nonuniform case due to the appearance of the kinetic energy,
which is illustrated in Figure 2.3.

\begin{figure}[t!]
\centerline{
\includegraphics[height=6cm,width=10cm]{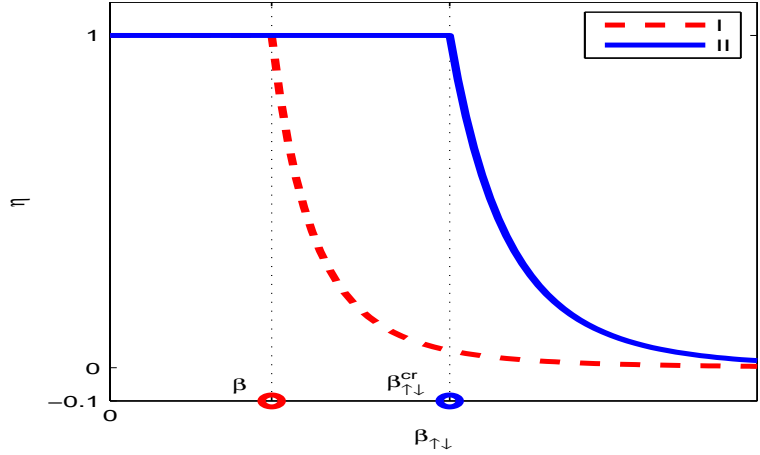}}
\caption{Illustration of the mixing factor $\eta$ defined in \eqref{eq:factor:sec2} v.s. the inter-component interaction $\beta_{\uparrow\downarrow}\ge0$ for a pseudo spin-$1/2$ BEC with $\beta_{\uparrow\uparrow}=\beta_{\downarrow\downarrow}=\beta>0$ and $\Omega=\delta=0$ under a uniform potential (dashed line I) and a
 box potential \eqref{eq:boxpot:sec2} (solid line II). } \label{figbeta:sec2}
\end{figure}

In addition, the following result on phase separation
was established in \cite{WenLiuCai}.

\begin{theorem}[phase separation of \eqref{eq:minim2:sec2} \cite{WenLiuCai}]\label{thm:phase:sec2} Suppose $\beta_{\uparrow\uparrow}=\beta_{\downarrow\downarrow}=\beta\ge0$, $\Omega=\delta=0$ and $V_{j}(\bx)=V(\bx)$ ($j=\uparrow,\downarrow$) in \eqref{eq:boxpot:sec2} with $U$ sufficiently smooth, then there exists a constant $\beta_{\uparrow\downarrow}^{\rm cr}>\beta$ which depends on $\beta$ and $U$, such that the mixing factor $\eta$ defined  in \eqref{eq:factor:sec2} for the positive ground state
$\Phi_g^{1/2}=(\phi_\uparrow^{1/2},\phi_\downarrow^{1/2})^T\in S_{1/2}$ of \eqref{eq:minim2:sec2} with $\nu=1/2$ satisfies $\eta=1$ when $0\leq\beta_{\uparrow\downarrow}\leq\beta_{\uparrow\downarrow}^{\rm cr}$,
and resp., $0\le \eta<1$ when $\beta_{\uparrow\downarrow}>\beta_{\uparrow\downarrow}^{\rm cr}$
with $\eta\to0$ as $\beta_{\uparrow\downarrow}\to +\infty$
(cf. Fig~\ref{figbeta:sec2} dashed line).  In other words, the box potential confinement $V(\bx)$ (by adjusting the size of $U$) can control phase separation of the underlying pseudo spin-1/2 BEC system without the internal atomic Josephson Junction.
\end{theorem}

\begin{remark} Theorem \ref{thm:phase:sec2} can be extended to the pseudo spin-1/2 BEC system in the whole space with the harmonic trapping potentials by using  the fundamental gap result of the Schr\"{o}dinger operator,
which was established in \cite{Andrews}.
\end{remark}

\subsection{Dynamics}
Here we discuss dynamical properties of a spin-1/2 BEC system described by the CGPEs \eqref{eq:cgpen:sec2}, including the center-of-mass (COM) motion
and the spin dynamics (or mass transfer). For the study of dynamics,
the initial condition for \eqref{eq:cgpen:sec2}  is usually given as
\be\label{init:sec2}
\Psi(\bx,t=0)=\Psi_0(\bx)=(\psi_\uparrow^0(\bx),\psi_\downarrow^0(\bx))^T,
\quad \bx\in {\mathbb R}^d \qquad \hbox{with} \quad \|\Psi_0\|=1.
\ee

\subsubsection{Dynamical properties}
Let $\Psi(\bx,t)=(\psi_\uparrow(\bx,t),\psi_\downarrow(\bx,t))^T$ be a solution of the CGPEs \eqref{eq:cgpen:sec2} with \eqref{init:sec2}, and the total mass is defined as
\be
N(t)=\|\Psi(\cdot,t)\|^2=N_\uparrow(t)+N_\downarrow(t), \qquad t\ge0,
\ee where the mass of each component and their difference are defined as
\be\label{eq:density:sec2}
\quad N_j(t)=\|\psi_j(\cdot,t)\|^2:=\int_{\mathbb{R}^d}
|\psi_j(\bx,t)|^2d{\bx},\quad j=\uparrow,\downarrow; \qquad \delta_N(t)=N_\uparrow(t)-N_\downarrow(t),\quad t\ge0.
\ee
Since $N(t)=N_\uparrow(t)+N_\downarrow(t)$ is conserved, it suffices to consider the difference $\delta_N(t)$.
The condensate width is defined as
\bea
\label{eq:def_sigma:sec2}
\sigma_{\alpha}(t) = \sqrt{\delta_{\alpha}(t)},\quad\mbox{where}\quad
\delta_\alpha(t) = \langle  \alpha^2\rangle(t) =
\int_{\mathbb{R}^d}\alpha^2|\Psi(\bx,t)|^2d{\bx},\qquad t\ge0,
\eea
with $\alpha$ being either $x,y$ or $z$;
the center-of-mass is given as
\be\label{eq:centerofm:sec2}
{\bx}_c(t)=\int_{\Bbb R^d}\bx |\Psi(\bx,t)|^2\,d\bx,\quad t\ge0;
\ee
and the momentum  is defined as
\be\label{eq:momentum:sec2}
{\bf P}(t)=\int_{\mathbb R^d}\sum\limits_{j=\uparrow,\downarrow}\text{Im}(\overline{\psi_j(\bx,t)}
\nabla\psi_j(\bx,t))\,d\bx,\qquad t\ge0,
\ee
where $\mathrm{Im}(f)$ denotes the imaginary part of $f$. Then we could obtain the following results.

\begin{lemma}[mass difference \cite{BaoCai0}] \label{6LEMMA1}
Suppose $\Psi(\bx,t)=(\psi_\uparrow(\bx,t),\psi_\downarrow(\bx,t))^T$ is a sufficiently regular solution of the
CGPEs \eqref{eq:cgpen:sec2} with \eqref{init:sec2}, then we have
\begin{align} \label{eq:ode11:sec2}
\ddot{\delta}_N(t)=&2\Omega\;\mathrm{Re}\int_{\mathbb R^d}\bigg[\big(V_\uparrow(\bx)-V_\downarrow(\bx)+\delta+
(\beta_{\uparrow\uparrow}-\beta_{\uparrow\downarrow})|\psi_\uparrow|^2
+(\beta_{\uparrow\downarrow}
-\beta_{\downarrow\downarrow})|\psi_\downarrow|^2\big)
\overline{\psi_\uparrow}\psi_\downarrow\bigg]\,d\bx\nonumber\\
&-\Omega^2\,\delta_N(t), \qquad t\ge0,
\end{align}
with initial conditions
\begin{equation*}
\delta_N(0)=\delta^{(0)}=\|\psi_\uparrow(\cdot,0)\|^2-
\|\psi_\downarrow(\cdot,0)\|^2,\quad\dot \delta_N(0)=\delta^{(1)}=2\Omega\, \mathrm{Im}\int_{\mathbb R^d}\overline{\psi_\uparrow(\bx,0)}\psi_\downarrow(\bx,0)\,d\bx.\end{equation*}
Therefore, if $\Omega\ne 0$ and
\be
V_\uparrow({\bx})-
V_\downarrow({\bx})\equiv -\delta,\quad \bx\in{\mathbb R}^d, \qquad\quad \beta_{\uparrow\uparrow} =
\beta_{\uparrow\downarrow} = \bt_{\downarrow\downarrow},\ee
 we have  \be
\delta_N(t)=\delta^{(0)}\;\cos(\Omega t)+\frac{\delta^{(1)}}{\Omega}\sin(\Omega t),
\qquad t\ge0,
 \ee
which implies the mass of each component is a periodic
function with period $T = \frac{2\pi}{|\Omega|}$ depending only on $\Omega$.
\end{lemma}

\begin{lemma}[center-of-mass motion \cite{BaoCai0}] \label{lem2:sec2}
Assume $V_\uparrow(\bx)=V_\downarrow(\bx)$ are harmonic potentials in \eqref{eq:hopot:sec2} with $x_\uparrow=x_\downarrow=0$, and  $\Psi(\bx,t)$ is a sufficiently regular solution of the
CGPEs \eqref{eq:cgpen:sec2} with \eqref{init:sec2}, then we have
\begin{align}
\dot{\bx}_c(t)={\bf P}(t),\quad \dot{\bf P}(t)=-\Lambda \bx_c(t),\qquad t\ge0,
\end{align}
with the initial conditions
\be
\bx_c(0)=\int_{\mathbb R^d}\bx|\Psi(\bx,0)|^2\,d\bx, \qquad
\dot{\bx}_c(0)={\bf P}(0)=\int_{\mathbb R^d}\sum\limits_{j=\uparrow,\downarrow}\mathrm{Im}(\overline{\psi_j(\bx,0)}
\nabla\psi_j(\bx,0))\,d\bx,
\ee
where $\Lambda=\gamma_x^2$ in 1D, $\Lambda=\mathrm{diag}(\gamma_x^2,\gamma_y^2)$ in 2D and $\Lambda=\mathrm{diag}(\gamma_x^2,\gamma_y^2,\gamma_z^2)$ in 3D.
\end{lemma}

\begin{lemma}[condensate width \cite{BaoCai0}] Assume  $\Psi(\bx,t)=(\psi_\uparrow(\bx,t),\psi_\downarrow(\bx,t))^T$ is a sufficiently regular solution of the
CGPEs \eqref{eq:cgpen:sec2} with \eqref{init:sec2}, then we have
\begin{align} &\ddot\delta_{\ap}(t) = \int_{{\Bbb
R}^d} \sum_{j=\uparrow,\downarrow}\left[2|\partial_\alpha\psi_j|^2
-2\alpha|\psi_j|^2\partial_\alpha(V_j(\bx))
+|\psi_j|^2\sum_{l=\uparrow,\downarrow} \bt_{jl}|\psi_l|^2\right]d{\bx},\quad t\geq 0,\\
&\delta_{\alpha}(0) = \delta_{\alpha}^{(0)} = \int_{\Bbb{R}^d}
\alpha^2\left(|\psi_\uparrow^0({\bx})|^2+|\psi_\downarrow^0({\bx})|^2\right)
d{\bx},\qquad \ap = x, y, z,\\
&\dot\delta_{\alpha}(0) = \dt_{\ap}^{(1)} = 2\sum_{j=\uparrow,\downarrow}
\int_{\Bbb{R}^d}\alpha \left[ {\rm Im}
\left((\psi_j^0)^*\p_\ap\psi_j^0\right)\right]\; d{\bx}.
\end{align}
\end{lemma}
\begin{remark} The above results can be generalized to the case where an angular momentum rotating term is added in the CGPEs \eqref{eq:cgpen:sec2}, see \cite{ZhangBaoLi} for more details.
\end{remark}

\subsubsection{Numerical methods and results}\label{dynanum:sec2}
In order to solve the CGPEs \eqref{eq:cgpen:sec2} with \eqref{init:sec2} numerically, similar to the ground state case,  the CGPEs \eqref{eq:cgpen:sec2} with \eqref{init:sec2} are  truncated
 onto a bounded computational
domain $U\subset\mathbb{R}^d$ with homogeneous Dirichlet boundary conditions:
 \begin{align}
\label{eq:DGPEs:sec2} &i\frac{\p\psi_\uparrow}{\p
t}=\left[-\frac{1}{2}\nabla^2+V_{\uparrow}({\bx})+\frac{\delta}{2}+(\beta_{\uparrow\uparrow}|\psi_\uparrow|^2+\beta_{\uparrow\downarrow}|\psi_\downarrow|^2)\right]\psi_\uparrow
+\frac{\Omega}{2}\psi_{\downarrow},\quad
{\bx}\in U,\quad t> 0,\\
\label{eq:DGPEs2:sec2} &i\frac{\p\psi_\downarrow}{\p
t}=\left[-\frac{1}{2}\nabla^2+V_{\downarrow}({\bx})-\frac{\delta}{2}+(\beta_{\uparrow\downarrow}|\psi_\uparrow|^2+\beta_{\downarrow\downarrow}|\psi_\downarrow|^2)\right]\psi_\downarrow
+\frac{\Omega}{2}\psi_{\uparrow},\quad
{\bx}\in U,\quad t> 0,\\
\label{eq:DGPEs6:sec2} &\psi_j({\bx},t) = 0, \quad
{\bx}\in\p U,\quad j=\uparrow,\downarrow,\quad
t\geq 0,\\
\label{eq:DGPEs3:sec2} & \psi_j({\bx}, 0) = \psi_j^0({\bx}), \quad
{\bx}\in\overline{U},\quad j=\uparrow,\downarrow.
\end{align}
In practical computation, a large bounded computational
domain $U$ is usually taken such that the truncation error can be
neglected due to that the  homogeneous Dirichlet boundary
conditions \eqref{eq:DGPEs6:sec2} are adopted.
Different numerical methods have been proposed for
discretizing the problem \eqref{eq:DGPEs:sec2}-\eqref{eq:DGPEs3:sec2}
in the literature \cite{ABB2013,Baocai2013,Baocai2015,BaoCai0,ZhangBaoLi}.
Here we only present
one of the most efficient and accurate time splitting spectral method (TSSP)
\cite{BaoJakschP,BaoLiShen,BaoShen,BaoWang}.

{\it Time-splitting procedure.} For $n\ge0$,  from time
$t=t_n=n \tau$ to $t=t_{n+1}=t_n+\tau$, the CGPEs \eqref{eq:DGPEs:sec2}-\eqref{eq:DGPEs2:sec2}
are solved in two splitting steps. One first
solves
\be \label{eq:ODE1:sec2} \begin{split}
i\frac{\p\psi_\uparrow(\bx,t)}{\p t} = -\frac{1}{2}\nabla^2\psi_\uparrow(\bx,t)+\frac{\Omega}{2}\psi_{\downarrow}(\bx,t), \\
i\frac{\p\psi_\downarrow(\bx,t)}{\p t} = -\frac{1}{2}\nabla^2\psi_\downarrow(\bx,t)+\frac{\Omega}{2}\psi_{\uparrow}(\bx,t),
\end{split}\ee
for the time step of length $\tau$, followed by solving \be
\label{eq:ODE3:sec2} \begin{split}i\frac{\p\psi_\uparrow(\bx,t)}{\p t} =
V_\uparrow({\bx})\psi_\uparrow(\bx,t)+\frac{\delta}{2}+
(\beta_{\uparrow\uparrow}|\psi_\uparrow(\bx,t)|^2+\beta_{\uparrow\downarrow}
|\psi_\downarrow(\bx,t)|^2)\psi_\uparrow(\bx,t),\\
i\frac{\p\psi_\downarrow(\bx,t)}{\p t} =
V_\downarrow({\bx})\psi_\downarrow(\bx,t)-\frac{\delta}{2}+
(\beta_{\uparrow\downarrow}|\psi_\uparrow(\bx,t)|^2+\beta_{\downarrow\downarrow}
|\psi_\downarrow(\bx,t)|^2)\psi_\downarrow(\bx,t),
\end{split}
\ee for the same time step. For time $t\in[t_n,t_{n+1}]$,
the ODE system \eqref{eq:ODE3:sec2} leaves $|\psi_\uparrow({\bx},t)|$ and
$|\psi_\downarrow({\bx},t)|$
 invariant in $t$, and thus it
can be integrated {\sl exactly} as \cite{BaoCai0}
 \bea \label{eq:solution3:sec2}\begin{split}
\psi_\uparrow({\bx},t) = \psi_\uparrow(\bx,t_n)\exp\left[-i\left(V_\uparrow({\bx})+\frac{\delta}{2}+
\beta_{\uparrow\uparrow}\left|\psi_\uparrow(\bx,t_n)\right|^2+
\beta_{\uparrow\downarrow}\left|\psi_\downarrow(\bx,t_n)\right|^2\right)
(t-t_n)\right],\\
\psi_\downarrow({\bx},t) = \psi_\downarrow(\bx,t_n)\exp\left[-i\left(V_\downarrow({\bx})-\frac{\delta}{2}+
\beta_{\uparrow\downarrow}\left|\psi_\uparrow(\bx,t_n)\right|^2+
\beta_{\downarrow\downarrow}\left|\psi_\downarrow(\bx,t_n)\right|^2\right)
(t-t_n)\right]. \end{split}\quad \eea
For \eqref{eq:ODE1:sec2}, it can be discretized in space by the
sine spectral method and then integrated (in phase space or Fourier space) in time
{\sl analytically}. For details,  we refer the readers to \cite{Bao,BaoDuZhang,Baocai2013,ABB2013} and references therein.

For the convenience of readers and simplicity of notations, here we present the method in 1D.
Extensions to 2D and 3D are straightforward. In 1D, let $h=\Delta x=(b-a)/L$
($L$ a positive integer), $x_j=a+jh$ ($j=0,\ldots,L$),
$\Psi_j^n=(\psi_{\uparrow,j}^n,\psi_{\downarrow,j}^n)^T$ be the numerical approximation
of $\Psi(x_j,t_n)=(\psi_\uparrow(x_j,t_n),\psi_{\downarrow}(x_j,t_n))^T$,
and for each fixed $l=\uparrow,\downarrow$, denote $\Psi_{l}^n$  to be the vector consisting of
$\psi_{l,j}^n$ for $j=0,1,\ldots,L-1$.
From time $t=t_n$ to $t=t_{n+1}$,  a second-order time-splitting sine pseudospectral (TSSP)
method for the CGPEs \eqref{eq:DGPEs:sec2}-\eqref{eq:DGPEs3:sec2} in 1D reads
\begin{equation}\label{TSSP1d}
\begin{split}
&\Psi^{(1)}_j
 =\sum\limits_{k=1}^{L-1}\sin(\lambda_k(x_j-a))\,Q_0^T\,
 e^{-\frac{i\tau}{4} U_k}\;Q_0(\widetilde{\Psi^n})_k,\\
&\Psi_j^{(2)}=e^{-i\tau P_j^{(1)}}\;\Psi_j^{(1)},  \qquad \qquad j=0,1,\ldots, L, \\
&\Psi^{n+1}_j
 =\sum\limits_{k=1}^{L-1}\sin(\lambda_k(x_j-a))\,Q_0^T\,
 e^{-\frac{i\tau}{4} U_k}\;Q_0(\widetilde{\Psi^{(2)}})_k,
\end{split}
 \end{equation}
where  $\lambda_k=\frac{k\pi}{b-a}$,
$(\widetilde{\Psi}^n)_k=((\widetilde{\psi_{\uparrow}^n})_k,
(\widetilde{\psi}_{\downarrow}^n)_k)^T$
with $(\widetilde{\psi_l^n})_k=\frac{2}{ L}\sum_{j=1}^{L-1}(\psi_l^n)_j\sin(\pi jk/L)$  ($k=1,2,\ldots,L-1$) being the discrete sine transform
coefficients of $\psi_l^n$ ($l=\uparrow,\downarrow$), $U_k=\text{diag}\left(\lambda_k^2+\Omega, \lambda_k^2-\Omega\right)$
is a diagonal matrix, $P_j^{(1)}=\text{diag}\big(V_\uparrow(x_j)+\sum\limits_
{l=\uparrow,\downarrow}\beta_{\uparrow l}|\psi_{l,j}^{(1)}|^2,
V_\downarrow(x_j)+\sum\limits_{l=\uparrow,\downarrow}\beta_{\downarrow l}|\psi_{l,j}^{(1)}|^2\big)$ for $j=0,1\ldots,L$, and
\[
 Q_0=\begin{pmatrix}\frac{1}{\sqrt{2}}&\frac{1}
 {\sqrt{2}}\\
  -\frac{1}{\sqrt{2}}
&\frac{1}{\sqrt{2}}\end{pmatrix}.
\]

We remark here again that many other numerical methods proposed
in the literatures for computing the dynamics of single-component
BEC \cite{ABB2013,Baocai2013,BaoCai2,BaoCaiWang,BaoDuZhang,
BaoJakschP,BaoJ,BaoJP1,BaoLiShen,Bao2013,BaoShen,BaoTX,BaoTZ,BaoWang,Cerim,Dion,
Edw-Bur-1995,Huang,Jiang,Min,Ming,Thalhammer,Xiong} can be extended to computing numerically the dynamics
of pseudo spin-$1/2$ BEC, i.e. the problem \eqref{eq:DGPEs:sec2}-\eqref{eq:DGPEs3:sec2}.

\begin{figure}[htb]
\centerline{
\includegraphics[height=5cm,width=7cm]{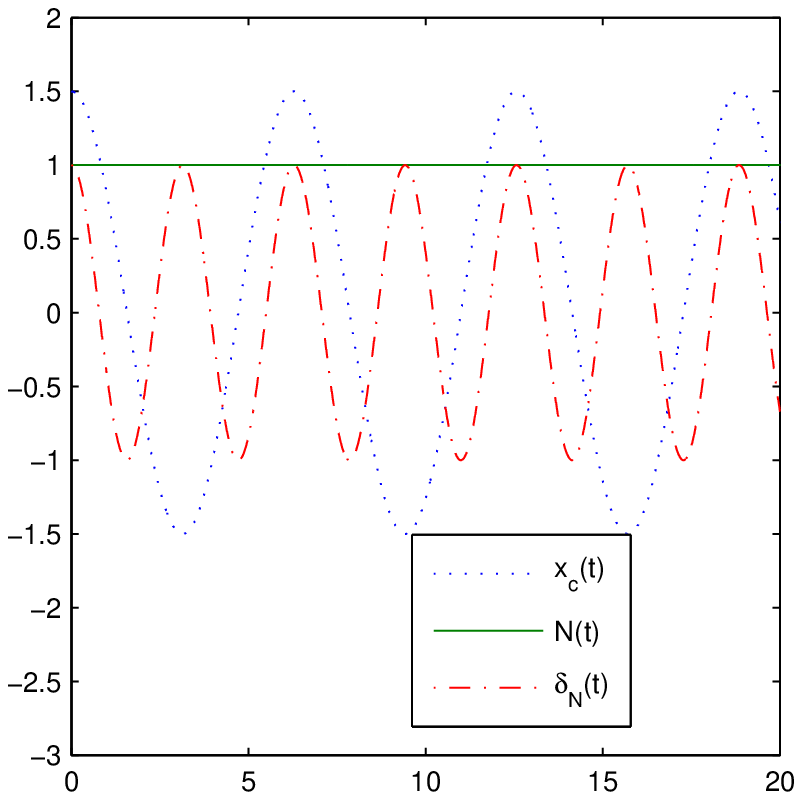} \quad
\includegraphics[height=5cm,width=7cm]{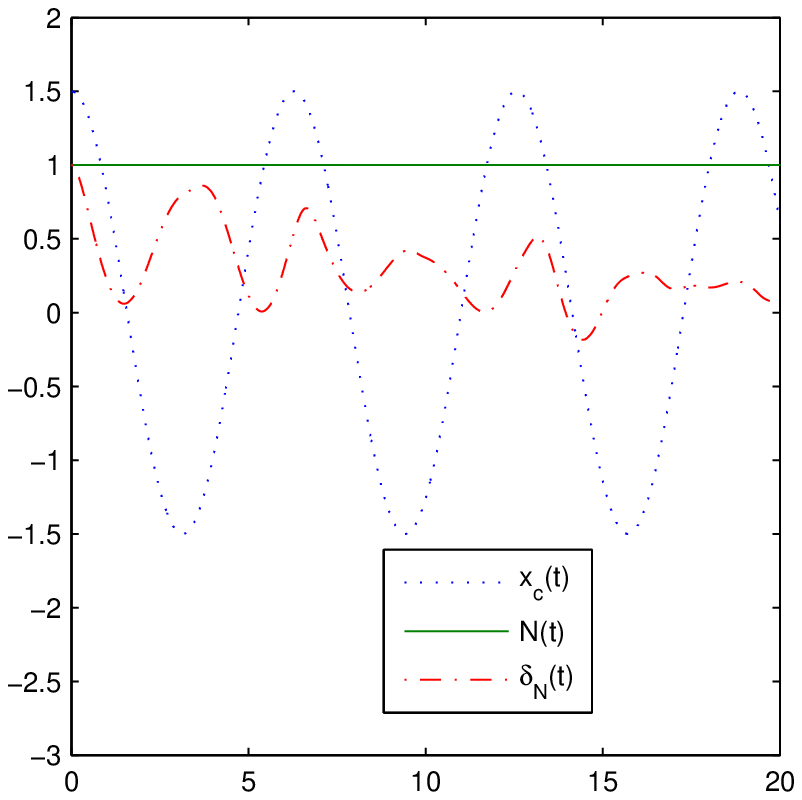}}
\caption{Dynamics of $x_c(t)$, $N(t)$ and $\delta_N(t)$ in Example 2.2
for $\beta_{\uparrow\uparrow}=
\beta_{\uparrow\downarrow}=\beta_{\downarrow\downarrow}=20$ (left)
and $\beta_{\uparrow\uparrow}=20$, $\beta_{\uparrow\downarrow}=8$, $\beta_{\downarrow\downarrow}=6$ (right). }
\label{fig:dy:sec2}
\end{figure}

{\it Example 2.2}. To demonstrate the efficiency of the TSSP method
\eqref{TSSP1d} for computing the dynamics
of \eqref{eq:cgpen:sec2} with \eqref{init:sec2},
we take $d=1$, $\Omega=2$, $\delta=0$ and
$V_\uparrow(x)=V_\downarrow(x)=\frac12x^2$ in \eqref{eq:cgpen:sec2}
and the initial data in \eqref{init:sec2} as
\be\psi_\uparrow^0(x)=\frac{1}{\pi^{1/4}}e^{-(x-1)^2/2},\quad \psi_\downarrow^0(x)=0, \qquad x\in {\mathbb R}.\ee
  The
 computational domain is $U=[-16,16]$ with mesh size
$h=\frac{1}{32}$ and time step $\tau=10^{-4}$. Figure \ref{fig:dy:sec2}  shows  time evolution of the center-of-mass $x_c(t)$, the total mass $N(t)$ and the mass difference $\delta_N(t)$ for different interaction strengths, which confirms the conclusions in Lemmas \ref{6LEMMA1}\&\ref{lem2:sec2}.

\subsection{Bogoliubov excitation}\label{sec:bdg:2}
 In the mean field description of a BEC system, the many body effects are absent in the mean field ground states. However, due to the interaction of the atoms,
 there are excitations in the system even in the lowest energy state, which is a result of many body effect. Such excitations could be regarded as quasi particles and
 are known as Bogoliubov excitations (or collective excitation or linear response) \cite{BlochDZ}.

 To determine the Bogoliubov excitation spectrum, we consider small perturbations
around the ground state of the CGPEs \eqref{eq:cgpen:sec2} with $\Omega\neq0$.  Assume $\Phi_g(\bx)=(\phi_\uparrow^g(\bx),\phi_\downarrow^g(\bx))^T$ is a ground state of the CGPEs \eqref{eq:cgpen:sec2} with chemical potential $\mu_g$,
we write the perturbed wave function $\Psi(\bx,t)$ as \cite{Bao0,PitaevskiiStringari}
\be\label{eq:bogo:sec2}
\Psi(\bx,t)=e^{-i\mu_g t}\left[\Phi_g(\bx)+u(\bx)\,e^{-i\omega t}+\overline{v(\bx)}\,e^{i\overline{\omega} t}\right],
\ee
where $\omega$ is the frequency of perturbation and $u(\bx)=(u_\uparrow,u_\downarrow)^T$ and $v(\bx)=(v_\uparrow,v_\downarrow)^T$ are the two vector amplitude functions.
Plugging \eqref{eq:bogo:sec2} into the CGPEs \eqref{eq:cgpen:sec2} and keep only the linear terms (w.r.t. $u$ and $v$), separating the $e^{-i(\mu_g-\omega)t}$ and $e^{-i(\mu_g+\overline{\omega})t}$ parts, we could find
\begin{equation}\label{eq:Bdg:sec2}
\begin{pmatrix}
\mathcal{L}_1&\beta_{\uparrow\uparrow}(\phi_\uparrow^g)^2
&\beta_{\uparrow\downarrow}\overline{\phi_\downarrow^g}\phi_\uparrow^g+
\frac{\Omega}{2}&
\beta_{\uparrow\downarrow}\phi_\downarrow^g\phi_\uparrow^g\\
-\beta_{\uparrow\uparrow}(\overline{\phi_\uparrow^g})^2&-\mathcal{L}_1
&-\beta_{\uparrow\downarrow}\overline{\phi_\downarrow^g}\overline{\phi_\uparrow^g}
&-\beta_{\uparrow\downarrow}\phi_\downarrow^g\overline{\phi_\uparrow^g}
-\frac{\Omega}{2}\\
\beta_{\uparrow\downarrow}\phi_\downarrow^g\overline{\phi_\uparrow^g}
+\frac{\Omega}{2}&\beta_{\uparrow\downarrow}\phi_\uparrow^g\phi_\downarrow^g
&
\mathcal{L}_2&\beta_{\downarrow\downarrow}(\phi_\downarrow^g)^2\\
-\beta_{\uparrow\downarrow}\overline{\phi_\uparrow^g}\overline{\phi_\downarrow^g}
&-\beta_{\uparrow\downarrow}\overline{\phi_\downarrow^g}\phi_\uparrow^g-
\frac{\Omega}{2}&
-\beta_{\downarrow\downarrow}(\overline{\phi_\downarrow^g})^2&-\mathcal{L}_2
\end{pmatrix}\begin{pmatrix}u_{\uparrow}\\ v_{\uparrow} \\ u_{\downarrow} \\ v_{\downarrow}\end{pmatrix}=\omega\begin{pmatrix}u_{\uparrow}\\ v_{\uparrow} \\ u_{\downarrow} \\ v_{\downarrow}\end{pmatrix},
\end{equation}
where
\begin{align}
\mathcal{L}_1=-\frac{1}{2}\nabla^2+V_\uparrow+\frac{\delta}{2}
+2\beta_{\uparrow\uparrow}|\phi_\uparrow^g|^2+\beta_{\uparrow\downarrow}
|\phi_\downarrow^g|^2-\mu_g,\\
\mathcal{L}_2=-\frac{1}{2}\nabla^2+V_\downarrow-\frac{\delta}{2}
+\beta_{\uparrow\downarrow}|\phi_\uparrow^g|^2+2\beta_{\downarrow\downarrow}
|\phi_\downarrow^g|^2-\mu_g.
\end{align}
The Bogoliubov-de Gennes (BdG) equations \eqref{eq:Bdg:sec2}, which can be numerically solved, determine the spectrum of the quasi-particle excitations. The solution $(\omega,u(\bx), v(\bx))$ is crucial in understanding the collective excitations in the BEC system.

\subsection{Semiclassical scaling and limits}
Let $\beta_{\rm max}=\max\{\beta_{\uparrow\uparrow}, \bt_{\uparrow\downarrow}, \bt_{\downarrow\downarrow}\}$. If
$\beta_{\rm max}\gg 1$, i.e. in the strong repulsive interaction
regime or there are many particles in the condensate,
under the normalization \eqref{eq:normn:sec2}, the semiclassical
scaling for the CGPEs \eqref{eq:cgpen:sec2} with harmonic trapping potentials \eqref{eq:hopot:sec2} is also very useful in
practice by choosing
\be \label{eq:6SCALE1:sec2} \tilde {\bx} =
\vep^{-1/2}{\bx}, \qquad \Psi^{\vep} = \vep^{d/4}\Psi, \qquad
\varepsilon = \beta_{\rm max}^{-2/(d+2)}.\ee Substituting
\eqref{eq:6SCALE1:sec2} into \eqref{eq:cgpen:sec2} and then
remove all $\tilde{ }$, we get the CGPEs in the
semiclassical (or Thomas-Fermi) scaling under the normalization \eqref{eq:normn:sec2} with $\Psi=\Psi^\vep$: \be\label{eq:cgpesm:sec2}
\begin{split} &i\vep\partial_t \psi_{\uparrow}^\vep=\biggl[-\frac{\vep^2}{2}\nabla^2
+V_\uparrow(\bx)+\frac{\vep\delta}{2}
+\sum_{j=\uparrow,\downarrow}\beta_{\uparrow j}^\vep|\psi^\vep_{j}|^2\biggr]\psi^\vep_{\uparrow}+\frac{\vep\Omega}{2}
\psi^\vep_{\downarrow}, \\
&i\vep\partial _t \psi^\vep_{\downarrow}=\biggl[-\frac{\vep^2}{2}\nabla^2
+V_\downarrow(\bx)-\frac{\vep\delta}{2}+\sum_{j=\uparrow,\downarrow}\beta_{\downarrow j}^\vep|\psi^\vep_{j}|^2\biggr]\psi^\vep_{\downarrow}
+\frac{\vep\Omega}{2}
\psi^\vep_{\uparrow},\end{split} \ee
where $\beta_{jl}^\vep=\beta_{jl}/\beta_{\rm max}$ with $\beta_{jl}^\vep\to\beta_{jl}^{0}$ as $\vep\to0^+$.

If $\Omega=0$ and $0<\vep\ll 1$ in \eqref{eq:cgpesm:sec2}, take the WKB ansatz \cite{GMMP,Carles01,JinMS}
 \be\label{WKBsec2}
 \psi_j^{\vep}({\bx},t) =
\sqrt{\rho_j^{\vep}({\bx}, t)}\exp\left(\fl{i}{\vep}
S_j^{\vep}({\bx}, t)\right), \qquad j = \uparrow, \downarrow,\ee where
$\rho_j^{\vep}=|\psi_j^{\vep}|^2$ and $S_j^{\vep}=\vep\; {\rm
arg}\left(\psi_j^{\vep}\right)$ are the position density and phase
of the wave function $\psi_j^{\vep}$ of $j$-component ($j=\uparrow,\downarrow$),
respectively. Then the coupled transport equations for the densities
$\rho_j^{\vep}$ and the Hamilton-Jacobi equations for the phases
$S_j^{\vep}$ ($j=\uparrow,\downarrow$) can be written as:
\begin{align*} &\p_t\rho_j^{\vep}+{\rm
div}\left(\rho_j^{\vep}\nabla
S_j^{\vep}\right) = 0, \\
 &\p_t S_j^{\vep}+\frac{1}{2}|\nabla
S_j^{\vep}|^2+V_j({\bx})+\frac{\vep\delta_j}{2}+\sum_{l=\uparrow,\downarrow}\beta^\vep_{j l}\rho_l^{\vep} =
\frac{{\vep}^2}{2\sqrt{\rho_j^{\vep}}}\nabla^2\sqrt{\rho_j^{\vep}},\qquad
j = \uparrow, \downarrow,\end{align*}
where $\delta_\uparrow=\delta$ and $\delta_\downarrow=-\delta$. As $\vep\to0^+$, by formally dropping the $\vep$ terms, we have
\begin{align*} &\p_t\rho_j^{0}+{\rm
div}\left(\rho_j^{0}\nabla
S_j^{0}\right) = 0, \\
 &\p_t S_j^{0}+\fl{1}{2}|\nabla
S_j^{0}|^2+V_j({\bx})+\sum_{l=\uparrow,\downarrow}\beta^0_{j l}\rho_l^{0} =0,\qquad j
= \uparrow, \downarrow,\end{align*}
 with $\rho^0_j = \lim_{\vep\to0^+}\rho_j^{\vep}$ and
$S_j^0 = \lim_{\vep\to0^+}S_j^{\vep}$.

Introduce  the
current densities \be {\bf J}_j^{\vep}(\bx,
t)=\rho_j^{\vep}\nabla S_j^{\vep} = \vep\,{\rm
Im}\left[\overline{\psi_j^{\vep}}\nabla\psi_j^{\vep}\right],
\qquad j = \uparrow,\downarrow, \ee
we can obtain an Euler system for the densities $\rho_j$ and ${\bf J}_j$ and the details are omitted here. A rigorous proof can be found in \cite{Lee}.
 We remark here that it is a tough problem
to study the semiclassical limit of the CGPEs \eqref{eq:cgpesm:sec2} when $\Omega\ne0$
since, in general,  the ansatz \eqref{WKBsec2} is no longer valid. Wigner transform is another widely used
tool in semiclassical analysis, and will be discussed in section \ref{sec:sem-so} for a different system.

\section{Spin-orbit-coupled BEC}
\setcounter{equation}{0}
\setcounter{figure}{0}
 Spin-orbit  (SO) coupling  is the interaction
between the spin and  motion of a particle, and is crucial for understanding
many physical phenomenon, such as quantum Hall effects and topological insulators.
However, the SO coupling is only for fermions
in solid state matters.
In a recent experiment \cite{Lin2011}, Lin et al. have
created a spin-orbit-coupled BEC with two spin states of
$^{85}$Rb: $\left|\uparrow\rangle\right.=|F=1,\,m_f=0\rangle$ and
$\left|\downarrow\rangle\right.=|F=1,\,m_f=-1\rangle$.  It is then desirable to study the SO coupling in the context of
BEC.
\subsection{The mathematical model}
We focus on the experimental case, where
the SO-coupled BEC is described by
the macroscopic wave function $\Psi:=\Psi(\bx,t)=(\psi_{\uparrow}(\bx,t),\psi_{\downarrow}(\bx,t))^T:=(\psi_\uparrow,\psi_\downarrow)^T$
 governed by the CGPEs in 3D \cite{Lin2011,Li2012,Zhai,Zhai1}
\be \label{eq:cgpe:sec3}
\begin{split}
&i\hbar\partial_t \psi_{\uparrow}=\biggl[-\frac{\hbar^2}{2m}\nabla^2
+\tilde{V}_\uparrow(\bx)+\frac{i\hbar^2 \tilde{k}_0}{m}\p_x+\frac{\hbar\tilde{\delta}}{2}
+\sum_{l=\uparrow,\downarrow} \tilde{g}_{l\uparrow}|\psi_{l}|^2\biggr]\psi_{\uparrow}+\frac{\hbar\tilde{\Omega}}{2}
\psi_{\downarrow}, \\
&i\hbar\partial _t \psi_{\downarrow}=\biggl[-\frac{\hbar^2}{2m}\nabla^2
+\tilde{V}_\downarrow(\bx)-\frac{i\hbar^2\tilde{k}_0}{m} \p_x-\frac{\hbar\tilde{\delta}}{2}+
\sum_{l=\uparrow,\downarrow} \tilde{g}_{l\downarrow}|\psi_{l}|^2\biggr]\psi_{\downarrow}
+\frac{\hbar\tilde{\Omega}}{2}
\psi_{\uparrow},\end{split} \ee
where $\tilde{k}_0$
is the wave number of Raman lasers representing the SO coupling strength, and all the other parameters are the same as those in
psuedo spin-1/2 BEC system \eqref{eq:cgpe:sec2}. Again, here the wave function $\Psi$ is normalized according to \eqref{eq:N:sec2}.

Similar to the nondimensionalization and dimension reduction of \eqref{eq:cgpe:sec2}, by introducing the same scaling as \eqref{eq:scale:sec2} and performing necessary dimension reduction process
from 3D to 1D or 2D,
we can obtain the CGPEs for $\Psi=(\psi_\uparrow,\psi_\downarrow)^T$ in $d$ ($d=1,2,3$) dimensions as
\be\label{eq:cgpen:sec3}
\begin{split} &i\partial_t \psi_{\uparrow}=\left[-\frac{1}{2}\nabla^2
+V_\uparrow(\bx)+ik_0\p_x+\frac{\delta}{2}
+(\beta_{\uparrow\uparrow}|\psi_{\uparrow}|^2+\beta_{\uparrow\downarrow}|\psi_{\downarrow}|^2)\right]\psi_{\uparrow}+\frac{\Omega}{2}
\psi_{\downarrow}, \\
&i\partial _t \psi_{\downarrow}=\left[-\frac{1}{2}\nabla^2
+V_\downarrow(\bx)-ik_0 \p_x-\frac{\delta}{2}+(\beta_{\uparrow\downarrow}|\psi_{\uparrow}|^2+\beta_{\downarrow\downarrow}|\psi_{\downarrow}|^2)\right]\psi_{\downarrow}
+\frac{\Omega}{2}
\psi_{\uparrow},\end{split} \ee
where $k_0=\tilde{k}_0\sqrt{\hbar/m\omega_0}$ and all the rest parameters are the same as those in \eqref{eq:cgpen:sec2}. The normalization condition for $\Psi$ becomes
\eqref{eq:normn:sec2}. The CGPEs \eqref{eq:cgpen:sec3} conserve the energy
\begin{align} E(\Psi)=&\int_{\mathbb{R}^d}\biggl[
 \sum_{j=\uparrow,\downarrow}\left(\frac12|\nabla\psi_{j}|^2+V_j(\bx)|\psi_j|^2\right)+\frac{\delta}{2}
\left(|\psi_\uparrow|^2-|\psi_\downarrow|^2\right)+\frac 12 \beta_{\uparrow\uparrow}|\psi_\uparrow|^4+\frac
12\beta_{\downarrow\downarrow}|\psi_\downarrow|^4\nonumber\\
&\qquad\qquad
+\beta_{\uparrow\downarrow}|\psi_\uparrow|^2|\psi_\downarrow|^2+ik_0\left(\overline{\psi}_\uparrow\p_x\psi_\uparrow-\overline{\psi}_\downarrow\p_x\psi_\downarrow\right)+\Omega\cdot\text{Re}
(\psi_\uparrow\overline{\psi_\downarrow})\biggl]d\bx. \label{eq:energy:sec3}
\end{align}

Finally, by introducing the following change of variables
\be\label{eq:changv:sec3}
\psi_\uparrow(\bx,t)=\tilde{\psi}_\uparrow(\bx,t)
 e^{i(\omega t+k_0x)}, \quad \psi_\downarrow(\bx,t)=\tilde{\psi}_\downarrow(\bx,t)
 e^{i(\omega t-k_0x)},\qquad \bx\in{\mathbb R}^d,
 \ee
with $\omega=\frac{-k_0^2}{2}$
in the CGPEs \eqref{eq:cgpen:sec3}, we obtain for $\bx\in{\mathbb R}^d$ and $t>0$
\begin{equation}\label{eq:cgpe199:sec3}
\begin{split}
&i\partial_t \tilde{\psi}_\uparrow=\left[-\frac{1}{2}\nabla^2
+V_\uparrow(\bx)+\frac{\delta}{2}
+\beta_{\uparrow\uparrow}|\tilde{\psi}_\uparrow|^2+\beta_{\uparrow\downarrow}|\tilde{\psi}_\downarrow|^2\right]\tilde{\psi}_\uparrow+\frac{\Omega}{2}
e^{-i2k_0x}\tilde{\psi}_\downarrow, \\
&i\partial _t \tilde{\psi}_\downarrow=\left[-\frac{1}{2}\nabla^2
+V_\downarrow(\bx)-\frac{\delta}{2}+\beta_{\uparrow\downarrow}|\tilde{\psi}_\uparrow|^2+\beta_{\downarrow\downarrow}|\tilde{\psi}_\downarrow|^2\right]
\tilde{\psi}_\downarrow+\frac{\Omega}{2}e^{i2k_0x}
\tilde{\psi}_\uparrow.
\end{split}
\end{equation}

If $\Omega=0$,  \eqref{eq:energy:sec3} is equivalent to a pseudo-spin 1/2 BEC system without Josephson junction through transformation \eqref{eq:changv:sec3}, which has been
studied in section 2.   Therefore, we will assume $\Omega\neq0$ through out this section.
\subsection{Ground states}
The ground state $\Phi_g:=\Phi_g(\bx)=(\phi_\uparrow^g(\bx),\phi_\downarrow^g(\bx))^T$ of
a two-component SO-coupled BEC based on \eqref{eq:cgpen:sec3}
is defined as the minimizer of the energy functional \eqref{eq:energy:sec3} under the
constraint \eqref{eq:norm1:sec2}, i.e.

  Find $\Phi_g \in S$, such that
\begin{equation}\label{eq:minimize:sec3}
    E_g := E\left(\Phi_g\right) = \min_{\Phi \in S}
    E\left(\Phi\right),
  \end{equation}
where $S$ is  defined in \eqref{eq:nonconset:sec2}. The ground state $\Phi_g$ is a solution of the following nonlinear
eigenvalue problem, i.e. Euler-Lagrange equation of the problem \eqref{eq:minimize:sec3}
\be\label{eq:e-l:1:sec3} \begin{split}
&\mu\, \phi_{\uparrow}=\left[-\frac{1}{2}\nabla^2
+V_\uparrow(\bx)+ik_0\p_x+\frac{\delta}{2}
+(\beta_{\uparrow\uparrow}|\phi_{\uparrow}|^2+\beta_{\uparrow\downarrow}|\phi_{\downarrow}|^2)\right]\phi_{\uparrow}+\frac{\Omega}{2}
\phi_{\downarrow}, \\
&\mu \,\phi_{\downarrow}=\left[-\frac{1}{2}\nabla^2
+V_\downarrow(\bx)-ik_0\p_x-\frac{\delta}{2}+(\beta_{\uparrow\downarrow}|\phi_{\uparrow}|^2+\beta_{\downarrow\downarrow}|\phi_{\downarrow}|^2)\right]\phi_{\downarrow}
+\frac{\Omega}{2}
\phi_{\uparrow},\end{split} \ee
under the normalization constraint $\Phi\in S$. For an eigenfunction
 $\Phi=(\phi_\uparrow,\phi_\downarrow)^T$ of \eqref{eq:e-l:1:sec3}, its corresponding eigenvalue (or chemical potential in the physics
literature)  $\mu:=\mu(\Phi)=\mu(\phi_\uparrow,\phi_\downarrow)$ can be computed as
\be
\mu=E(\Phi)+\int_{\mathbb R^d}\left(\frac{\beta_{\uparrow\uparrow}}{2}|\phi_\uparrow|^4+\frac{\beta_{\downarrow\downarrow}}{2}|\phi_\downarrow|^4
+\beta_{\uparrow\downarrow}|\phi_\uparrow|^2|\phi_\downarrow|^2\right)\,d\bx.
\ee
\subsubsection{Mathematical theories}
For the existence and uniqueness concerning the ground state, we have the following results \cite{Baocai2015}.

\begin{theorem}[existence and uniqueness \cite{Baocai2015}]\label{thm:mres:sec3}
Suppose  $V_j(\bx)\ge 0$ ($j=\uparrow,\downarrow$) satisfying
$\lim\limits_{|\bx|\to\infty}V_j(\bx)=+\infty$, then  there exists a
minimizer
 $\Phi_g=(\phi_\uparrow^g,\phi_\downarrow^g)^T\in S$ of
 \eqref{eq:minimize:sec3} if one of the following conditions holds
 \begin{enumerate}\renewcommand{\labelenumi}{(\roman{enumi})}
 \item $d=1$;
 \item $d=2$, $\beta_{\uparrow\uparrow}>-C_b$, $\beta_{\downarrow\downarrow}>-C_b$ and $\beta_{\uparrow\downarrow}\ge-C_b-\sqrt{(C_b+
     \beta_{\uparrow\uparrow})(C_b+\beta_{\downarrow\downarrow})}$;
\item $d=3$ and  the matrix $B$ \eqref{eq:Bforl1:sec}  is
either positive semi-definite or nonnegative.
\end{enumerate}
In addition, $e^{i\theta_0}\Phi_g$ is also a ground state
of \eqref{eq:minimize:sec3} for any $\theta_0\in\mathbb [0,2\pi)$.
In particular, when $k_0=0$ or $\Omega=0$, the ground state is unique  up to a  phase factor if the
matrix $B$ is positive semi-definite
and $I(\bx)\not\equiv0$ in \eqref{eq:Ibx:sec2}.
In contrast, there exists no ground state of \eqref{eq:minimize:sec3} if one of the following holds
 \begin{enumerate}\renewcommand{\labelenumi}{(\roman{enumi})$^\prime$}
 \item $d=2$, $\beta_{\uparrow\uparrow}\leq-C_b$ or $\beta_{\downarrow\downarrow}\leq-C_b$ or $\beta_{\uparrow\downarrow}<-C_b-\sqrt{(C_b+\beta_{\uparrow\uparrow})(C_b+\beta_{\downarrow\downarrow})}$;
\item $d=3$, $\beta_{\uparrow\uparrow}<0$ or $\beta_{\downarrow\downarrow}<0$ or $\beta_{\uparrow\downarrow}<0$ with $\beta_{\uparrow\downarrow}^2>\beta_{\uparrow\uparrow}\beta_{\downarrow\downarrow}$.
\end{enumerate}
\end{theorem}
As observed in \eqref{eq:cgpe199:sec3}, the SO coupling $k_0$ is competing with the Raman transition $\Omega$. Indeed, when letting either $|\Omega|$ or $|k_0|$
tend to infinity, the asymptotic profile of the ground state can be classified.
Introducing an auxiliary  energy functional $\tilde{E}_0(\tilde{\Phi})$ for $\tilde{\Phi}=(\tilde{\phi}_\uparrow,\tilde{\phi}_\downarrow)^T$
\begin{eqnarray}\label{eq:mini1:sec3} \tilde{E}_0(\tilde{\Phi})&=&\int_{{\mathbb R}^d}\biggl[
\sum\limits_{j=\uparrow,\downarrow}\left(\frac12|\nabla\tilde{\phi}_j|^2+V_j(\bx)|\tilde{\phi}_j|^2\right)+\frac{\delta}{2}
(|\tilde{\phi}_\uparrow|^2-|\tilde{\phi}_\downarrow|^2)+\frac{\beta_{\uparrow\uparrow}}{2}|\tilde{\phi}_\uparrow|^4+\frac{\beta_{\downarrow\downarrow}}{2}|\tilde{\phi}_\downarrow|^4
\nonumber\\
&&\qquad
+\beta_{\uparrow\downarrow}|\tilde{\phi}_\uparrow|^2|\tilde{\phi}_\downarrow|^2\biggl]d\bx=\tilde{E}(\tilde{\Phi})-\Omega \int_{{\mathbb R}^d}\text{Re}
(e^{i2k_0x}\tilde{\phi}_\uparrow\overline{\tilde{\phi}}_\downarrow)d\bx,
\end{eqnarray}
 we know that the nonconvex minimization problem
\be \label{eq:min987:sec3}
\tilde E_g^{(0)}:=\tilde{E}_0(\tilde\Phi_g^{(0)})=\min_{\tilde\Phi\in S}\tilde{E}_0(\tilde\Phi),
\ee
admits a unique positive minimizer $\tilde\Phi_g^{(0)}=(\tilde{\phi}_\uparrow^{g,0},
\tilde{\phi}_\downarrow^{g,0})^T\in S$  if the matrix $B$
is positive semi-definite and $I(\bx)\not\equiv0$ in \eqref{eq:Ibx:sec2}.  For a given $k_0\in {\mathbb R}$, let
$\tilde{\Phi}^{k_0}=(\tilde{\phi}_\uparrow^{k_0},\tilde{\phi}_\downarrow^{k_0})^T\in S$ be a ground state of
\eqref{eq:minimize:sec3} when all other parameters are fixed, then we can draw the conclusions as follows.

\begin{theorem}[large $k_0$ limit \cite{Baocai2015}]\label{thm:k0change1:sec3}  Suppose the matrix $B$ is positive semi-definite
and $I(\bx)\not\equiv0$ in \eqref{eq:Ibx:sec2}, and $\Phi^{k_0}=(\phi_\uparrow^{k_0},\phi_\downarrow^{k_0})^T$  is a ground state of \eqref{eq:minimize:sec3}.
 When $k_0\to\infty$,  let
$\tilde{\Phi}^{k_0}=(\tilde{\phi}_\uparrow^{k_0},\tilde{\phi}_\downarrow^{k_0})^T=(e^{-ik_0x}\phi_\uparrow^{k_0},e^{ik_0x}\phi_\downarrow^{k_0})^T$, then
 $\tilde{\Phi}^{k_0}$ converges to
a ground state of \eqref{eq:min987:sec3}
in  $L^{p_1}\times L^{p_2}$ sense with
$p_1,p_2$ satisfying (i) $p_1,p_2\in[2,6)$ when $d=3$, (ii)
$p_1,p_2\in[2,\infty)$ when $d=2$, and (iii)
$p_1,p_2\in [2,\infty]$ when  $d=1$. Equivalently speaking, there exist constants $\theta_{k_0}\in [0,2\pi)$
such that $e^{i\theta_{k_0}}(\tilde{\phi}_\uparrow^{k_0},\tilde{\phi}_\downarrow^{k_0})^T$ converges to the unique positive ground state
$\tilde\Phi_g^{(0)}$ of \eqref{eq:min987:sec3}.  In other words,
large $k_0$ in the CGPEs \eqref{eq:cgpen:sec3} will remove the effect of Raman coupling $\Omega$, i.e. large $k_0$ limit
is effectively letting $\Omega\to0$.
\end{theorem}
When either $\Omega$ or $\delta$ tends to infinity, similar results as Theorems \ref{thm:omg:sec2}\&\ref{thm:lim2:sec2} hold and they are omitted here for brevity.
Indeed,  large Raman coupling $\Omega$ will remove the effect of SO coupling  $k_0$
in  the asymptotic profile of the ground states of \eqref{eq:minimize:sec3}
and the reverse is true, i.e.
there is a competition between these two parameters.

\begin{theorem}[ground states property \cite{Baocai2015}]\label{thm:order:sec3} Suppose $\lim\limits_{|\bx|\to\infty}V_j(\bx)=+\infty$ ($j=\uparrow,\downarrow$),
 the matrix $B$ is either positive semi-definite or nonnegative, then
we have

(i) If $|\Omega|/|k_0|^2\gg1$, $|\Omega|\to+\infty$, the ground state $\Phi_g=(\phi^g_1,\phi^g_2)^T$
of \eqref{eq:minimize:sec3} for the CGPEs \eqref{eq:cgpen:sec3} converges
to a state $(\phi_g,\mathrm{sgn}(-\Omega)\phi_g)^T$, where $\phi_g$
minimizes the energy \eqref{eq:oglim:sec2} under the constraint $\|\phi_g\|=1/\sqrt{2}$.

(ii) If $|\Omega|/|k_0|\ll1$, $|k_0|\to+\infty$, the ground state $\Phi_g=(\phi^g_\uparrow,\phi^g_\downarrow)^T$ of
\eqref{eq:minimize:sec3} for the CGPEs \eqref{eq:cgpen:sec3} converges
to a state $(e^{-ik_0x}\tilde{\phi}_\uparrow^{g,0},e^{ik_0x}\tilde{\phi}_\downarrow^{g,0})^T$,
where $\widetilde{\Phi}_g^{(0)}=(\tilde{\phi}_\uparrow^{g,0},\tilde{\phi}_\downarrow^{g,0})^T$
is a ground state of  \eqref{eq:min987:sec3} for the energy $\tilde{E}_0(\cdot)$ in \eqref{eq:mini1:sec3}.

(iii) If $|k_0|\ll|\Omega|\ll |k_0|^2$ and $|k_0|\to+\infty$, the
leading order of the ground state energy $E_g:=E(\Phi_g)$ of \eqref{eq:minimize:sec3} for the CGPEs
\eqref{eq:cgpen:sec3} is given by
 $E_g=-\frac{k_0^2}{2}-C_0\frac{|\Omega|^2}{|k_0|^2}+o\left(\frac{|\Omega|^2}{|k_0|^2}\right)$,
 where $C_0>0$ is a generic constant.
\end{theorem}
\begin{remark}\label{rmk:order:sec3}
For $|k_0|\ll|\Omega|\ll |k_0|^2$, the ground state of \eqref{eq:minimize:sec3} is
very complicated. The  ground state energy expansion indicates that $-k_0^2/2$ is the leading order term and is much
larger than the next order term. In such situation,
the above theorem shows that  the ground state $\Phi_g\approx (e^{ik_0x}|\phi_1^g|,e^{-ik_0x}|\phi_2^g|)^T$,
and oscillation of ground state densities $|\phi_j^g|^2$ ($j=\uparrow,\downarrow$)
may occur at the order of  $O(|\Omega|/|k_0|^2)$ in amplitude
and $k_0$ in frequency. Such density oscillation is predicted in the physics literature
\cite{Li2012}, known as the density modulation.
\end{remark}
\subsubsection{Numerical methods and results}
Similar to the pseudo spin-1/2 case in section \ref{numeric-gs:sec2}, we  construct a GFDN
to compute the ground state $\Phi_g=(\phi_\uparrow^g,\phi_\downarrow^g)^T$ of \eqref{eq:minimize:sec3} for a SO-coupled BEC.
 Let $t_n=n\tau$ ($n=0,1,2,\ldots$) be the time steps with $\tau>0$ as the time step size and we evolve an initial state $\Phi_0:=(\phi_\uparrow^{(0)},\phi_\downarrow^{(0)})^T$
 through the following GFDN
\begin{equation}\label{eq:gfdn:sec3}
\begin{split}
&\partial_t \phi_\uparrow=\left[\frac{1}{2}\nabla^2
-V_\uparrow(\bx)-ik_0\p_x-\frac{\delta}{2}
-\sum_{l=\uparrow,\downarrow}\beta_{\uparrow l}|\phi_l|^2\right]\phi_\uparrow-\frac{\Omega}{2}
\phi_\downarrow, \quad t\in[t_n, t_{n+1}),\\
&\partial _t \phi_\downarrow=\left[\frac{1}{2}\nabla^2
-V_\downarrow(\bx)+ik_0\p_x+\frac{\delta}{2}-\sum_{l=\uparrow,\downarrow}^2\beta_{\downarrow l}|\phi_l|^2\right]
\phi_\downarrow-\frac{\Omega }{2}
\phi_\uparrow,\quad t\in[t_{n},t_{n+1}),\\
&\phi_\uparrow(\bx,t_{n+1})=\frac{\phi_\uparrow(\bx,t_{n+1}^-)}{\|\Phi(\cdot,t_{n+1}^-)\|},
\quad \phi_\downarrow(\bx,t_{n+1})=\frac{\phi_\downarrow(\bx,t_{n+1}^-)}{\|\Phi(\cdot,t_{n+1}^-)\|},\quad \bx\in\mathbb R^d,\\
&\phi_\uparrow(\bx,0)=\phi_\uparrow^{(0)}(\bx),\quad \phi_\downarrow(\bx,0)=\phi_\downarrow^{(0)}(\bx),\quad \bx\in\mathbb R^d.
\end{split}
\end{equation}
The above GFDN \eqref{eq:gfdn:sec3}
is then truncated on a bounded large computational domain $U$, e.g.
an interval $[a,b]$ in 1D, a rectangle $[a,b]\times[c,d]$ in 2D and
a box $[a,b]\times[c,d]\times[e,f]$ in 3D, with periodic boundary conditions.
The GFDN on $U$ can be further discretized in space via the pseudospectral
method with the Fourier basis or second-order central finite difference method and in time via backward Euler scheme as discussed in section \ref{numeric-gs:sec2}.
For more details, we refer to \cite{Bao2013,BaoCai0,BaoDu,BaoCaiWang,BaoCai2,BaoChernLim} and references therein.
\begin{remark} If the box potential \eqref{eq:boxpot:sec2}
is used in the CGPEs \eqref{eq:cgpen:sec3}
instead of the harmonic potentials \eqref{eq:hopot:sec2},
due to the appearance of the SO coupling, in order to compute the ground state,
it is better to construct the GFDN based on CGPEs \eqref{eq:cgpe199:sec3} (imaginary time) and then discretize it
via the backward Euler sine pseudospectral (BESP) method due to that the homogeneous Dirichlet
boundary conditions on $\partial U$ must be used in this case. Again, for details, we refer to
\cite{Bao2013,BaoCai0,BaoDu,BaoCaiWang} and references therein.
\end{remark}

\begin{figure}[htb]
\centerline{\includegraphics[height=5.2cm,width=8cm]{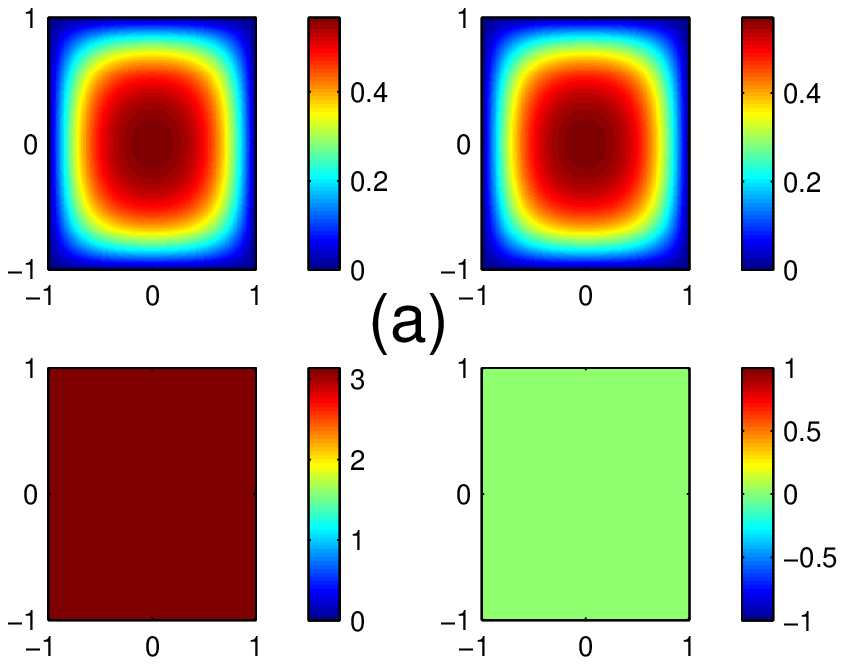}
\includegraphics[height=5.2cm,width=8cm]{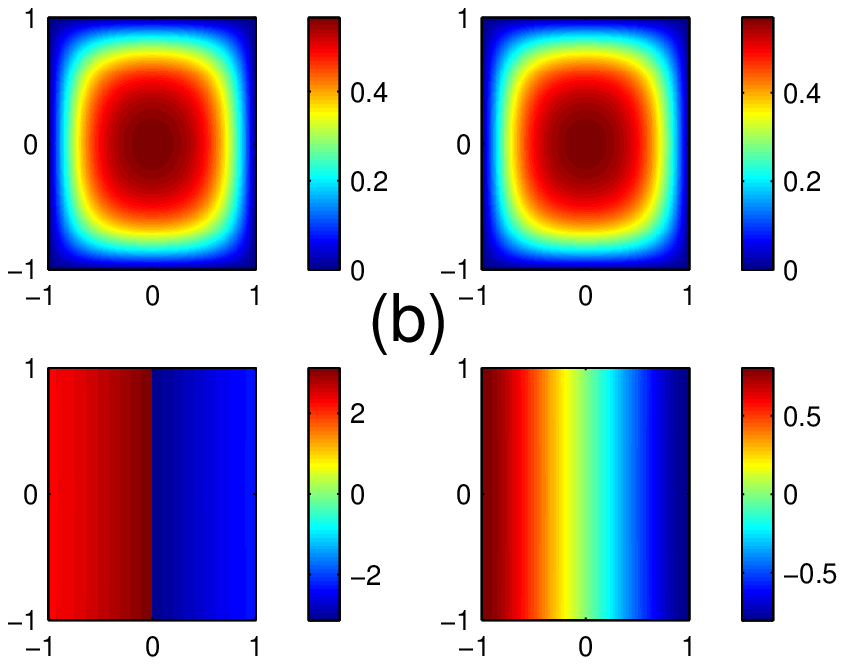}}
\centerline{\includegraphics[height=5.2cm,width=8cm]{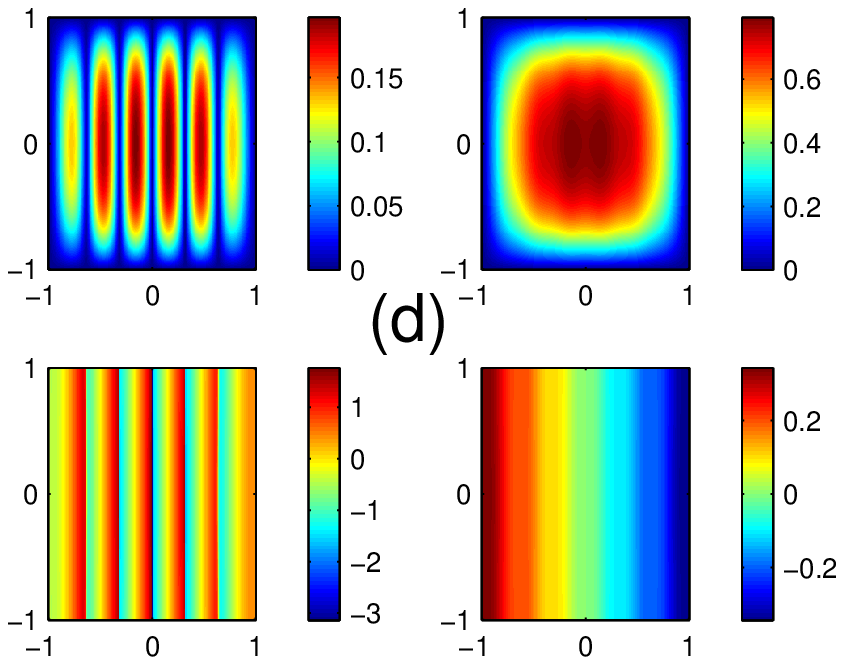}
\includegraphics[height=5.2cm,width=8cm]{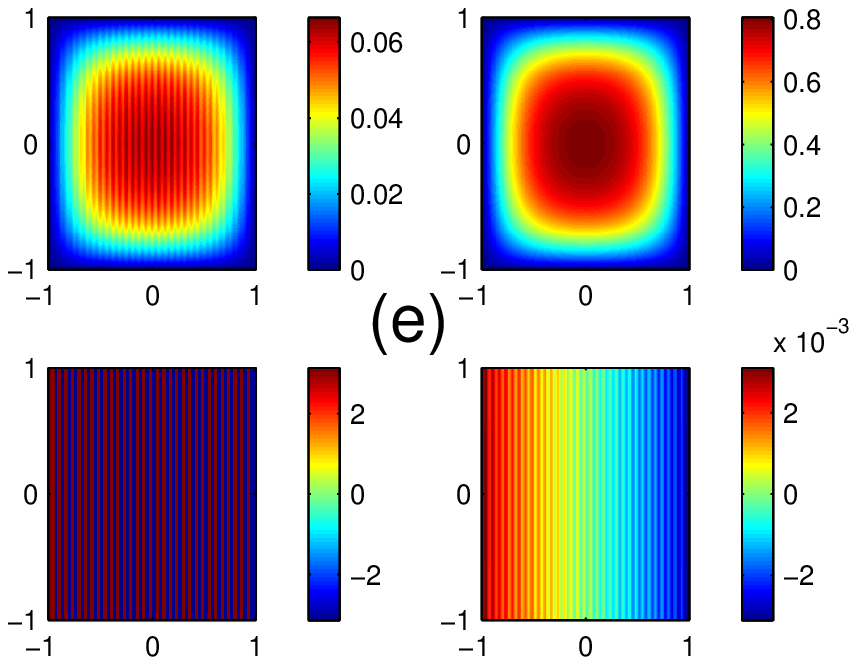}}

\caption{Modulated ground state $\tilde{\Phi}_g=(\tilde{\phi}_\uparrow^g,\tilde{\phi}_\downarrow^g)^T$
in Example 3.1
for a SO-coupled BEC in 2D with $\Omega=50$, $\delta=0$,
$\beta_{11}=10$, $\beta_{12}=\beta_{21}=\beta_{22}=9$ for:
(a) $k_0=0$, (b) $k_0=1$,
(d) $k_0=10$ and (e) $k_0=50$.
In each subplot, top panel shows densities
and bottom panel shows phases of  $\tilde{\phi}_\uparrow^g$ (left column)
and $\tilde{\phi}_\downarrow^g$ (right column). \label{fig:1:sec3}}
\end{figure}

{\it Example 3.1}. To verify the asymptotic property of the ground states
in Theorem \ref{thm:k0change1:sec3},  we take $d=2$,  $\delta=0$,
$\beta_{\uparrow\uparrow}:\beta_{\uparrow\downarrow}:
\beta_{\downarrow\downarrow}=(1:0.9:0.9) \beta$ with $\beta_{\uparrow\uparrow}=10$ in
\eqref{eq:cgpen:sec3}. The potential $V_j(\bx)$ ($j=\uparrow,\downarrow$)
are taken as the box potential given in \eqref{eq:boxpot:sec2} with $U=[-1,1]\times[-1,1]$.
We compute the ground state via the above BESP method with mesh size $h=\frac{1}{128}$
and time step $\tau=0.01$ ($\tau=0.001$ for large $\Omega$).
For the chosen parameters, it is easy to find that when $\Omega=0$, the ground
state $\Phi_g$ satisfies $\phi_\uparrow^g=0$.
Figure \ref{fig:1:sec3} shows the profile of $\tilde{\Phi}_g=(e^{-ik_0x}\phi_\uparrow^g,e^{ik_0x}\phi_\downarrow^g)^T$
where $\Phi_g=(\phi_\uparrow^g,\phi_\downarrow^g)^T$ is a ground state of \eqref{eq:minimize:sec3} with $\Omega=50$ for different $k_0$,
which clearly demonstrates that as $k_0\to+\infty$,  effect of $\Omega$
disappears. This is consistent with Theorem \ref{thm:k0change1:sec3}.

\subsection{Dynamics}
For a SO-coupled BEC described by the CGPEs \eqref{eq:cgpen:sec3}, we consider the dynamics characterized by the center-of-mass $\bx_c(t)$ in \eqref{eq:centerofm:sec2},
momentum ${\bf P}(t)$ in \eqref{eq:momentum:sec2} and spin density $\delta_N$ in \eqref{eq:density:sec2}.

\subsubsection{Dynamical properties}
For the center-of-mass motion, we have the lemma.

\begin{lemma}[dynamics of center-of-mass \cite{Baocai2015}]\label{lem:com:sec3} Let $V_\uparrow(\bx)=V_\downarrow(\bx)$ be the
$d$-dimensional ($d=1,2,3$) harmonic potentials given in \eqref{eq:hopot:sec2},
then the  motion of the center-of-mass $\bx_c(t)$ for the CGPEs \eqref{eq:cgpen:sec3} is governed by
        \be\label{eq:xc:sec3}
\ddot\bx_c(t)=-\Lambda \bx_c(t)-2k_0\Omega\,\mathrm{Im}\left(\int_{\mathbb R^d}\overline{\psi_\uparrow(\bx,t)}\psi_\downarrow(\bx,t)\,d\bx\right)\,{\bf e}_x, \qquad t>0,
        \ee
where $\Lambda$ is a $d\times d$ diagonal matrix with $\Lambda=\gamma_x^2$ in 1D ($d=1$), $\Lambda=\mathrm{diag}(\gamma_x^2,\gamma_y^2)$ in 2D ($d=2$)
and $\Lambda=\mathrm{diag}(\gamma_x^2,\gamma_y^2,\gamma_z^2)$ in 3D ($d=3$),
${\bf e}_x$ is the unit vector for $x$-axis.
The  initial conditions for \eqref{eq:xc:sec3} are given as
        \begin{equation*}
        \bx_c(0)=\int_{\mathbb R^d}\bx\sum_{j=\uparrow,\downarrow}|\psi_j(\bx,0)|^2\,d\bx,\qquad
        \dot{\bx}_c(0)={\bf P}(0)-k_0 \delta_N(0)\,{\bf e}_x.
        \end{equation*}
 In particular, \eqref{eq:xc:sec3} implies that the center-of-mass $\bx_c(t)$ is
 periodic  in $y$-component with frequency $\gamma_y$ when $d=2,3$, and in $z$-component
 with frequency $\gamma_z$ when $d=3$.
 If $k_0\Omega=0$, $\bx_c(t)$ is also periodic in $x$-component with frequency $\gamma_x$.
\end{lemma}
The above lemma leads to the following approximations of $\bx_c(t)$.

\begin{theorem}[approximation of center-of-mass  \cite{Baocai2015}]\label{thm:comapp:sec3} Let $V_\uparrow(\bx)=V_\downarrow(\bx)$ be the harmonic potential as \eqref{eq:hopot:sec2} in $d$ dimensions ($d=1,2,3$) and $k_0\Omega\neq0$. For the $x$-component $x_c(t)$ of the
center-of-mass $\bx_c(t)$ of the  CGPEs \eqref{eq:cgpen:sec3} with any initial data $\Psi(\bx,0):=\Psi_0(\bx)$ satisfying
$\|\Psi_0\|=1$, we have
\be\label{eq:xcc:sec3}
x_c(t)=x_0\cos(\gamma_xt)+\frac{P_0^x}{\gamma_x}\sin(\gamma_xt)-
k_0\int_0^t\cos(\gamma_x(t-s))\delta_N(s)\,ds,
\qquad t\ge0,
\ee
where $x_0=\int_{\mathbb R^d}x\sum_{j=\uparrow,\downarrow}|\psi_j(\bx,0)|^2\,d\bx$ and $P_0^x=\int_{\mathbb R^d}\sum_{j=\uparrow,\downarrow}\mathrm{Im}(\overline{\psi_j(\bx,0)}\p_x\psi_j(\bx,0))\,d\bx$.
In addition, if $\delta\approx 0$, $|k_0|$ is small, $\beta_{\uparrow\uparrow}\approx\beta$, $\beta_{\uparrow\downarrow}=\beta_{\downarrow\uparrow}\approx\beta$ and
$\beta_{\downarrow\downarrow}\approx \beta$ with $\beta$ being a fixed constant, we can approximate the solution $x_c(t)$ as follows:

(i) If $|\Omega|=\gamma_x$,  we can get
\begin{equation}\label{eq:xc1:1:sec3}
x_c(t)\approx\left(x_0-\frac{k_0}{2}\delta_N(0)t\right)\cos(\gamma_xt)+\frac{1}{\gamma_x}
\left(P_0^x-\frac{k_0}{2}\delta_N(0)-
{\rm sgn}(\Omega)\frac{\gamma_xk_0C_0}{2}t\right)\sin(\gamma_xt),
\end{equation}
where  $C_0=2\mathrm{Im}\int_{\mathbb R^d}\overline{\psi_\uparrow(\bx,0)}\psi_\downarrow(\bx,0)\,d\bx$.

(ii) If $|\Omega|\neq \gamma_x$,  we can get
\begin{equation}\label{eq:xc1:2:sec3}
\begin{split}
x_c(t)\approx& \left(x_0+\frac{k_0C_0}{\gamma_x^2-\Omega^2}\right)\cos(\gamma_xt)
+\frac{1}{\gamma_x}\left(P_0^x-\frac{\gamma_x^2k_0\delta_N(0)}{\gamma_x^2-\Omega^2}\right)\sin(\gamma_xt)\\
&-\frac{k_0C_0}{\gamma_x^2-\Omega^2}\cos(\Omega t)+\frac{k_0 \delta_N(0)\Omega}{\gamma_x^2-\Omega^2} \sin(\Omega t).
\end{split}
\end{equation}
Based on the above approximation,  if $|\Omega|=\gamma_x$ or $\frac{\Omega}{\gamma_x}$ is an
irrational number, $x_c(t)$ is not periodic; and if
$\frac{\Omega}{\gamma_x}$ is a rational number, $x_c(t)$ is a periodic function, but its frequency is different
from the trapping frequency $\gamma_x$.
\end{theorem}
As in the experiments, the initial data of CGPEs \eqref{eq:cgpen:sec3} are usually prepared in a special form,
i.e. shift of the ground state $\Phi_g=(\phi_\uparrow^g,\phi_\downarrow^g)^T$
 of \eqref{eq:minimize:sec3} for the CGPEs \eqref{eq:cgpen:sec3}, i.e., the initial condition
 for \eqref{eq:cgpen:sec3} is chosen as
 \be\label{eq:inishift:sec3}
 \psi_\uparrow(\bx,0)=\phi_\uparrow^g(\bx-\bx_0),\quad \psi_\downarrow(\bx,0)=\phi_2^g(\bx-\bx_0), \qquad
 \bx\in {\Bbb R}^d,
 \ee
where $\bx_0=x_0$ in 1D, $\bx_0=(x_0,y_0)^T$ in 2D and $\bx_0=(x_0,y_0,z_0)^T$ in 3D.
Then we have the approximate dynamical law for the center-of-mass in  $x$-direction $x_c(t)$.

\begin{theorem}[approximation of center-of-mass  \cite{Baocai2015}] \label{thm:com2:sec3}Suppose $V_\uparrow(\bx)=V_\downarrow(\bx)$ for $\bx\in{\mathbb R}^d$ are harmonic potentials given in \eqref{eq:hopot:sec2},
and the initial data for the CGPEs \eqref{eq:cgpen:sec3} is taken as \eqref{eq:inishift:sec3}, then we have

(i) when $\frac{|k_0|^2}{|\Omega|}\gg1$,  the dynamics of the center-of-mass $x_c(t)$ can be approximated by the ODE
\be\label{eq:odecomgsk0:sec3}
\ddot{x}_c(t)=-\gamma_x^2x_c(t),\quad x_c(0)=x_0,\quad \dot{x}_c(0)=0,
\ee
i.e., $x_c(t)=x_0\cos(\gamma_xt)$, which is the same as the case without SO coupling $k_0$;

(ii) when $\frac{|k_0|^2}{|\Omega|}\ll1$, $\beta_{\uparrow\uparrow}\approx \beta$, $\beta_{\uparrow\downarrow}=\beta_{\downarrow\uparrow}\approx \beta$ and $\beta_{\downarrow\downarrow}\approx\beta$ with $\beta$ a fixed constant,
the dynamics of the center-of-mass $x_c(t)$ can
be approximated by the following ODE
\be\label{eq:odecomgs:sec3}
\dot x_c(t)=P^x(t)-\frac{k_0[2k_0P^x(t)-\delta]}{\sqrt{[2k_0P^x(t)-
\delta]^2+\Omega^2}},\quad
\dot P^x(t)=-\gamma_x^2x_c(t),\qquad t\ge0,
\ee
with $x_c(0)=x_0$ and $P^x(0)=k_0\delta_N(0)$. In particular, the solution to \eqref{eq:odecomgs:sec3} is periodic,
and, in general, its frequency is different from the trapping frequency $\gamma_x$.
\end{theorem}
\subsubsection{Numerical methods and results}
Different from the pseudo spin-1/2 case in section \ref{dynanum:sec2}, we propose a time splitting Fourier spectral (TSFP)
scheme for solving the CGPEs
\eqref{eq:cgpen:sec3}. Similarly, we truncate the equations onto a bounded computational domain $U$, e.g.
an interval $[a,b]$ in 1D,
a rectangle $[a,b]\times[c,d]$ in 2D and
a box $[a,b]\times[c,d]\times[e,f]$ in 3D, equipped
with periodic boundary conditions. Then from $t_n$ to $t_{n+1}$,
the CGPEs \eqref{eq:cgpen:sec3} can be solved in two steps.
One first solve for $\bx\in U$
\be\label{eq:split1:sec3}
\begin{split}
&i\p_t\psi_\uparrow=\left(-\frac12\nabla^2+ik_0\p_x
+\frac{\delta}{2}\right)\psi_\uparrow+\frac{\Omega}{2}\psi_\downarrow,\\
&i\p_t\psi_\downarrow=-\left(\frac12\nabla^2+ik_0\p_x+\frac{\delta}{2}
\right)\psi_\downarrow+\frac{\Omega}{2}\psi_\uparrow,
\end{split}
\ee
for  time step $\tau$, followed by solving
\be\label{eq:split2:sec3}
i\p_t\psi_j=\left(V_j(\bx)+\beta_{j\uparrow}|\psi_\uparrow|^2+\beta_{j\downarrow}|\psi_\downarrow|^2\right)\psi_j,\qquad
j=\uparrow,\downarrow,
\qquad \bx\in U,
\ee
for another time step $\tau$. Eq. \eqref{eq:split1:sec3} with periodic boundary conditions can be discretized
by the Fourier spectral method in space and then integrated in time {\sl exactly}.
Eq. \eqref{eq:split2:sec3} leaves the densities $|\psi_\uparrow|$ and $|\psi_\downarrow|$ unchanged and it can
be integrated in time {\sl exactly}. Then
a full discretization scheme can be constructed via a
combination of the splitting steps \eqref{eq:split1:sec3} and \eqref{eq:split2:sec3}
with a second-order or higher-order time-splitting method.

For the convenience of the readers, here we present the method in 1D for the simplicity of notations.
Extensions to 2D and 3D are straightforward. In 1D, let $h=\Delta x=(b-a)/L$
($L$ an even positive integer), $x_j=a+jh$ ($j=0,\ldots,L$),
$\Psi_j^n=(\psi_{\uparrow,j}^n,\psi_{\downarrow,j}^n)^T$ be the numerical approximation
of $\Psi(x_j,t_n)=(\psi_\uparrow(x_j,t_n),\psi_{\downarrow}(x_j,t_n))^T$,
and for each fixed $l=\uparrow,\downarrow$, denote $\psi_{l}^n$  to be the vector consisting of
$\psi_{l,j}^n$ for $j=0,1,\ldots,L-1$.
From time $t=t_n$ to $t=t_{n+1}$,  a second-order time-splitting Fourier pseudospectral (TSFP)
method for the CGPEs \eqref{eq:cgpen:sec3} in 1D reads
\begin{equation} \label{TSFPsec3}
\begin{split}
&\Psi^{(1)}_j
 =\sum\limits_{k=-L/2}^{L/2-1}e^{i\mu_k(x_j-a)}
 \,Q_k^T\,e^{-\frac{i\tau}{4} U_k}\;Q_k(\widetilde{\Psi^n})_k,\\
&\Psi_j^{(2)}=e^{-i\tau P_j^{(1)}}\;\Psi_j^{(1)},  \qquad \qquad j=0,1,\ldots, L-1, \\
&\Psi^{n+1}_j
 =\sum\limits_{k=-L/2}^{L/2-1}e^{i\mu_k(x_j-a)}
 \,Q_k^T\,e^{-\frac{i\tau}{4} U_k}\;Q_k(\widetilde{\Psi^{(2)}})_k,
\end{split}
 \end{equation}
where for each fixed $k=-\frac{L}{2},-\frac{L}{2}+1,\ldots,\frac{L}{2}-1$, $\mu_k=\frac{2 k\pi}{b-a}$,
$(\widetilde{\Psi}^n)_k=((\widetilde{\psi_{\uparrow}^n})_k,(\widetilde{\psi}_{\downarrow}^n)_k)^T$
with $(\widetilde{\psi_l^n})_k=\frac{1}{L}\sum_{j=0}^{L-1}(\psi_l^n)_je^{i\frac{2\pi jk}{L}}$  being the discrete Fourier transform
coefficients of $\psi_l^n$ ($l=\uparrow,\downarrow$), $U_k=\text{diag}\left(\mu_k^2+2\lambda_k, \mu_k^2-2\lambda_k\right)$
is a diagonal matrix, and
\[
 Q_k=\begin{pmatrix}\frac{\sqrt{\lambda_k-\chi_k}}{\sqrt{2\lambda_k}}&\frac{\frac{\Omega}{2}}
 {\sqrt{2\lambda_k(\lambda_k-\chi_k)}}\\
  \frac{-\sqrt{\lambda_k+\chi_k}}{\sqrt{2\lambda_k}}
&\frac{\frac{\Omega}{2}}{\sqrt{2\lambda_k(\lambda_k+\chi_k)}}\end{pmatrix}\quad \hbox{with}\quad
\chi_k=k_0\mu_k-\frac{\delta}{2},\quad \lambda_k=\frac{1}{2}\sqrt{4\chi_k^2+\Omega^2},
\]
and
$P_j^{(1)}=\text{diag}\big(V_\uparrow(x_j)+\sum\limits_{l=\uparrow,\downarrow}\beta_{\uparrow l}|\psi_{l,j}^{(1)}|^2,
V_\downarrow(x_j)+\sum\limits_{l=\uparrow,\downarrow}\beta_{\downarrow l}|\psi_{l,j}^{(1)}|^2\big)$ for $j=0,1\ldots,L-1$.

\begin{figure}[htb]
\centerline{
\includegraphics[height=4cm,width=7cm]{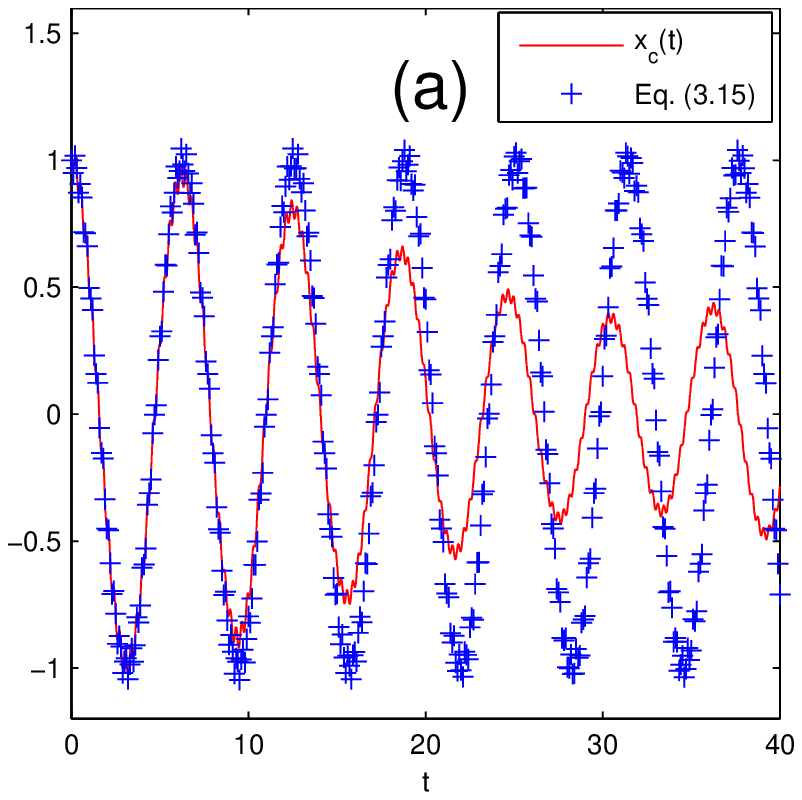} \quad
\includegraphics[height=4cm,width=7cm]{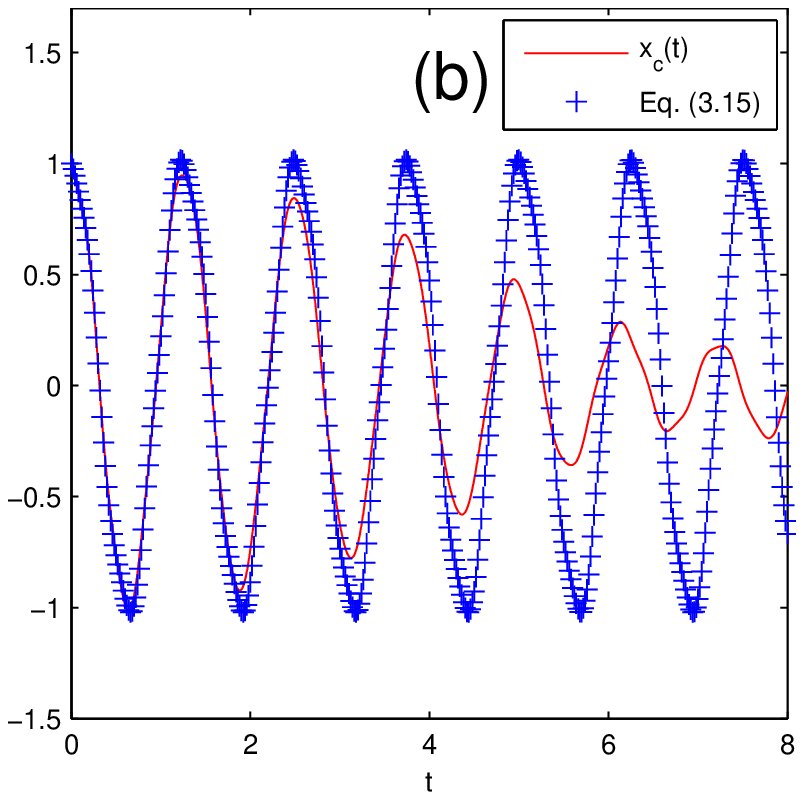}}
\centerline{
\includegraphics[height=4cm,width=7cm]{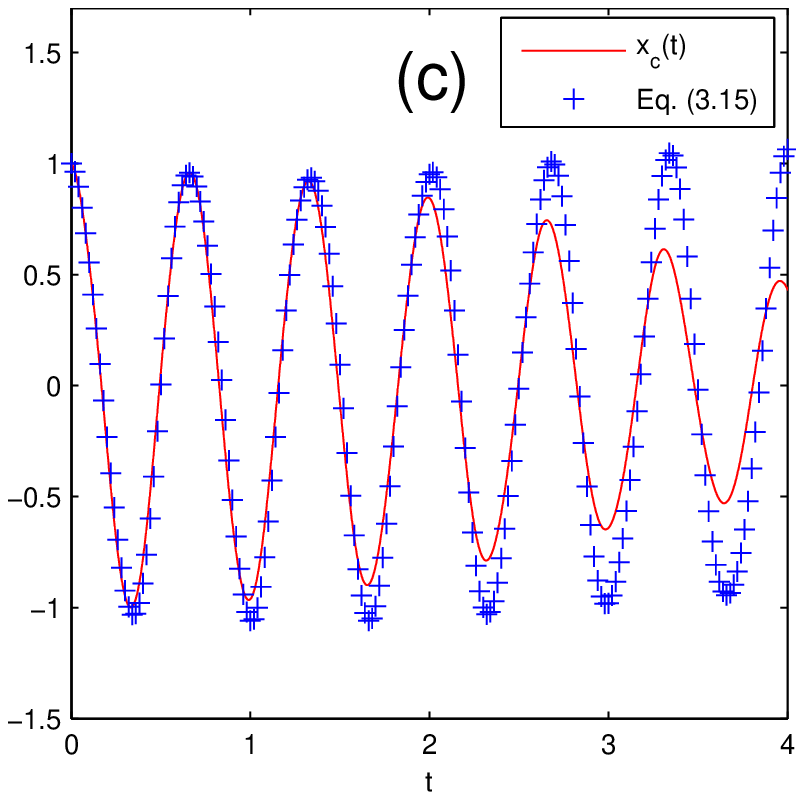} \quad
\includegraphics[height=4cm,width=7cm]{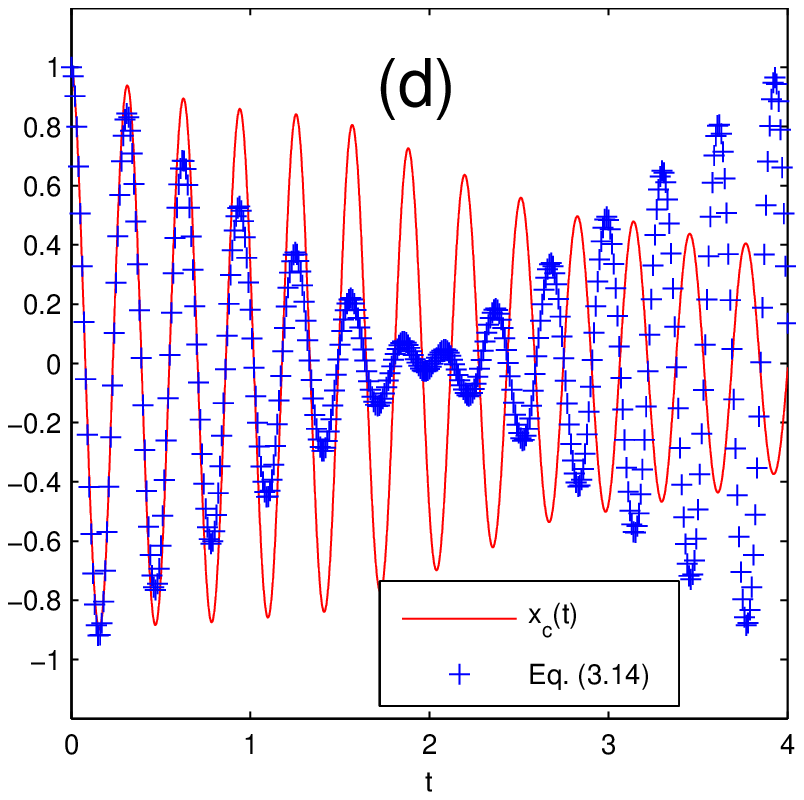}}

\caption{Time evolution of  $x_c(t)$ for the CGPEs \eqref{eq:cgpen:sec3}
as in Example 3.2 obtained numerically from its numerical solution (i.e. labeled by '$x_c(t)$' with solid lines) and
asymptotically as Eqs. \eqref{eq:xc1:1:sec3} and \eqref{eq:xc1:2:sec3} in Theorem \ref{thm:comapp:sec3} (i.e. labeled by 'Eq.'  with `+ + +')
with $\Omega=20$ and  $k_0=1$ for different $\gamma_x$: (a) $\gamma_x=1$, (b) $\gamma_x=5$, (c) $\gamma_x=3\pi$,
 and (d) $\gamma_x=20$.  \label{fig:0:sec3}}
\end{figure}
\begin{figure}[htb]
\centerline{
\includegraphics[height=4cm,width=7cm]{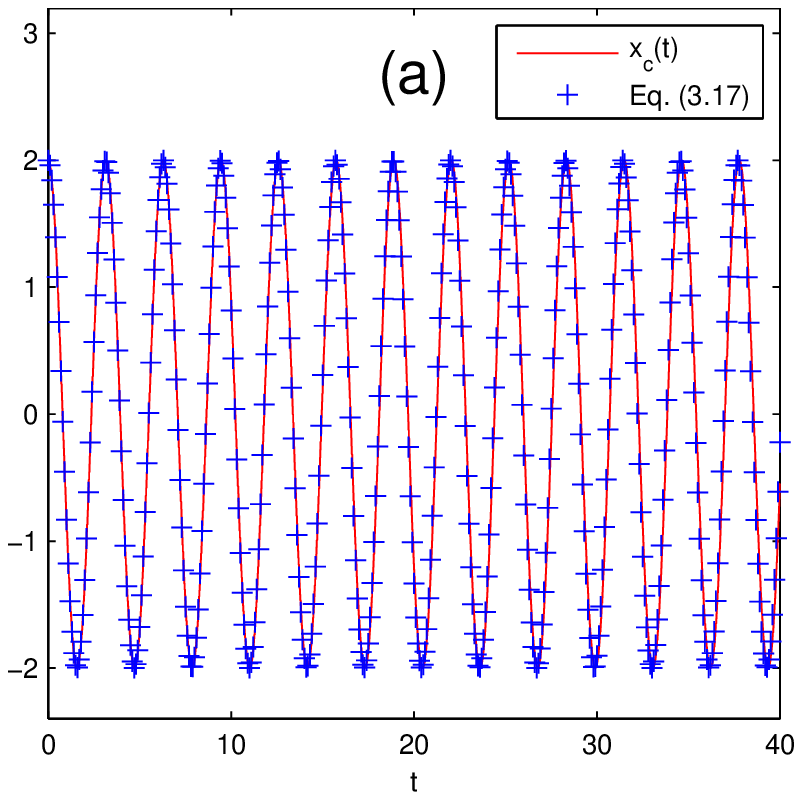} \quad
\includegraphics[height=4cm,width=7cm]{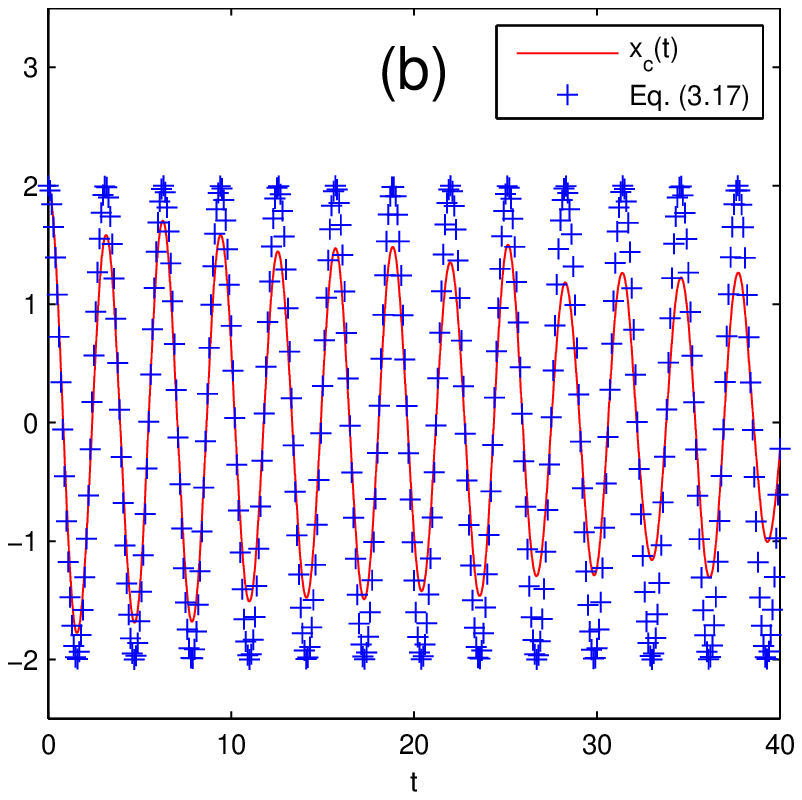}}
\centerline{
\includegraphics[height=4cm,width=7cm]{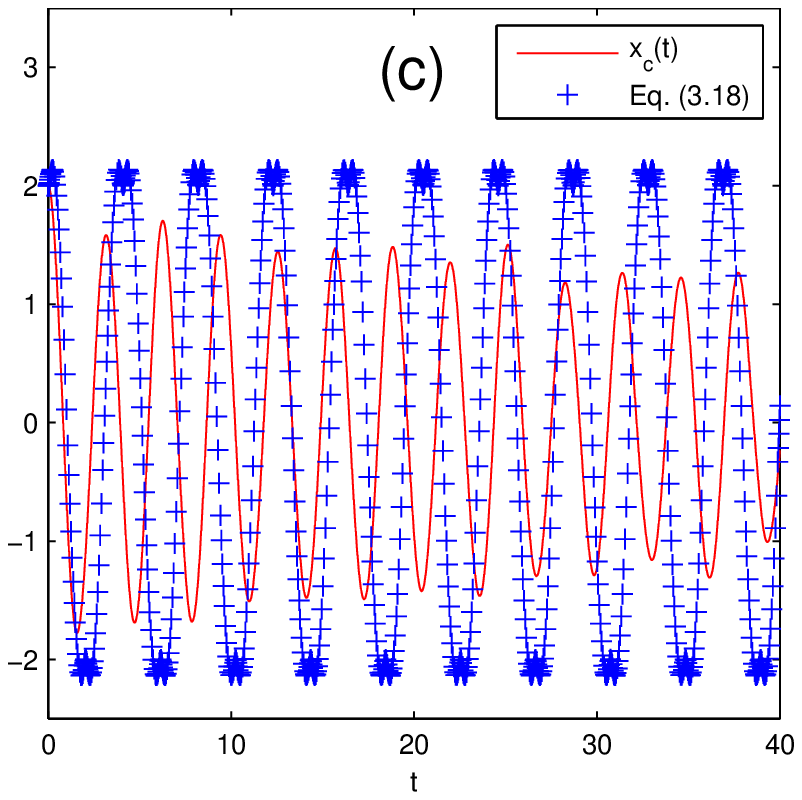}
\quad \includegraphics[height=4cm,width=7cm]{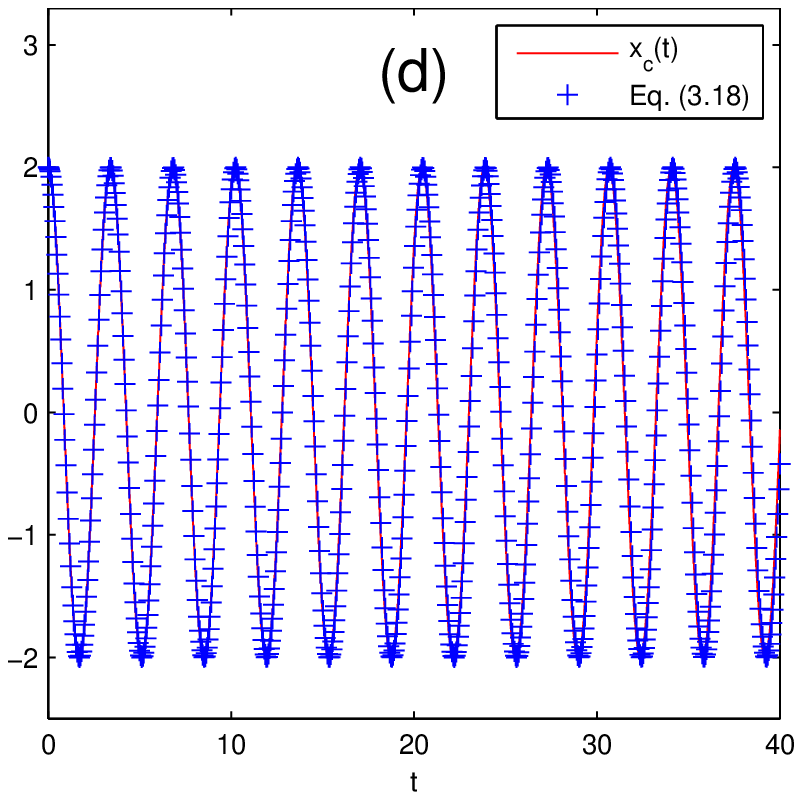}
}
\caption{\label{fig:3:sec3}
Time evolution of  $x_c(t)$ for the CGPEs \eqref{eq:cgpen:sec3} as in Example 3.2 obtained numerically from its numerical solution (i.e. labeled as $x_c(t)$ with solid lines) and
asymptotically as Eqs. \eqref{eq:odecomgsk0:sec3} and \eqref{eq:odecomgs:sec3}
in Theorem  \ref{thm:com2:sec3} (i.e. labeled as 'Eq.' with `+ + +')
for different sets of parameters:
(a) $(\Omega,k_0)=(50,20)$, (b) and (c) $(\Omega,k_0)=(2,2)$,
 and (d) $(\Omega,k_0)=(50,2)$.   }
\end{figure}

{\it Example 3.2}.  To verify the
the asymptotic (or approximate) results for $x_c(t)$ in Theorem \ref{thm:comapp:sec3}, we take $d=2$, $\delta=0$ and $\beta_{\uparrow\uparrow}=\beta_{\uparrow\downarrow}=
\beta_{\downarrow\downarrow}=1$ in
\eqref{eq:cgpen:sec3},  and choose the initial data  as
\be
\psi_\uparrow(\bx,0)=\pi^{-1/2}e^{-\frac{|\bx-\bx_0|^2}{2}},\quad \psi_\downarrow(\bx,0)=0,\quad \bx\in{\Bbb R}^2,
\ee
where $\bx_0=(1,1)^T$. The problem is solved numerically
on a bounded domain $[-16,16]^2$ by the TSFP method \eqref{TSFPsec3}
with mesh size $h=1/128$ and time step $\tau=10^{-4}$.
Figure \ref{fig:0:sec3} depicts time evolution of $x_c(t)$ obtained numerically and asymptotically as in Theorem \ref{thm:comapp:sec3}
with $\Omega=20$ and $k_0=1$ for different $\gamma_x$.
From this figure, we see that: for short time $t$, the approximation given in
Theorem \ref{thm:comapp:sec3} is very accurate; and when $t\gg1$, it becomes inaccurate. Similarly, to verify the asymptotic (or approximate) results for $x_c(t)$ in Theorem \ref{thm:com2:sec3}, we take $d=2$, $\delta=0$, $\gamma_x=\gamma_y=2$ and
$\beta_{\uparrow\uparrow}=\beta_{\uparrow\downarrow}=
\beta_{\downarrow\downarrow}=10$ in
  the CGPEs \eqref{eq:cgpen:sec3}, and choose the initial data in  \eqref{eq:hopot:sec2} as
\eqref{eq:inishift:sec3} with $\bx_0=(2,2)^T$ and the ground state computed numerically.
Figure \ref{fig:3:sec3} depicts time evolution of $x_c(t)$ obtained numerically and asymptotically as in Theorem
\ref{thm:com2:sec3} with different $\Omega$ and $k_0$.

\subsection{Bogoliubov excitation}
Similar to section \ref{sec:bdg:2}, to determine the Bogoliubov excitation spectrum, we consider small perturbations
around the ground state of the CGPEs \eqref{eq:cgpen:sec3} with $\Omega\neq0$.  Assume $\Phi_g(\bx)=(\phi_\uparrow^g(\bx),\phi_\downarrow^g(\bx))^T$ is a ground state of the CGPEs \eqref{eq:cgpen:sec3} with chemical potential $\mu_g$,
we write perturbed wave function $\Psi(\bx,t)$ as
\be\label{eq:bogo:sec3}
\Psi(\bx,t)=e^{-i\mu_g t}\left[\Phi_g(\bx)+u(\bx)e^{-i\omega t}+\overline{v(\bx)}e^{i\overline{\omega} t}\right],
\ee
where $\omega$ is the frequency of perturbation and $u(\bx)=(u_\uparrow,u_\downarrow)^T$ and $v(\bx)=(v_\uparrow,v_\downarrow)^T$ are the two vector amplitude functions.
Plugging \eqref{eq:bogo:sec3} into the CGPEs \eqref{eq:cgpen:sec3} and keep  the linear terms (w.r.t $u$ and $v$), separating the $e^{-i(\mu_g-\omega)t}$ and $e^{-i(\mu_g+\overline{\omega})t}$ parts, we could find
\begin{equation}\label{eq:Bdg:sec3}
\begin{pmatrix}
\mathcal{L}_1+ik_0\partial_x&\beta_{\uparrow\uparrow}(\phi_{\uparrow}^g)^2
&\beta_{\uparrow\downarrow}\overline{\phi_{\downarrow}^g}\phi_\uparrow^g
+\frac{\Omega}{2}&\beta_{\uparrow\downarrow}\phi_\downarrow^g\phi_\uparrow^g\\
-\beta_{\uparrow\uparrow}(\overline{\phi_\uparrow^g})^2&-\mathcal{L}_1
+ik_0\partial_x&-\beta_{\uparrow\downarrow}\overline{\phi_\downarrow^g}
\overline{\phi_\uparrow^g}&
-\beta_{\uparrow\downarrow}\phi_\downarrow^g\overline{\phi_\uparrow^g}
-\frac{\Omega}{2}\\
\beta_{\uparrow\downarrow}\phi_\downarrow^g\overline{\phi_\uparrow^g}
+\frac{\Omega}{2}&\beta_{\uparrow\downarrow}\phi_\uparrow^g\phi_\downarrow^g&
\mathcal{L}_2-ik_0\partial_x&\beta_{\downarrow\downarrow}(\phi_\downarrow^g)^2\\
-\beta_{\uparrow\downarrow}\overline{\phi_\uparrow^g}
\overline{\phi_\downarrow^g}&-\beta_{\uparrow\downarrow}
\overline{\phi_\downarrow^g}\phi_\uparrow^g-\frac{\Omega}{2}&
-\beta_{\downarrow\downarrow}(\overline{\phi_\downarrow^g})^2
&-\mathcal{L}_2-ik_0\partial_x
\end{pmatrix}\begin{pmatrix}u_{\uparrow}\\ v_{\uparrow} \\ u_{\downarrow} \\ v_{\downarrow}\end{pmatrix}=\omega\begin{pmatrix}u_{\uparrow}\\ v_{\uparrow} \\ u_{\downarrow} \\ v_{\downarrow}\end{pmatrix},
\end{equation}
where
\begin{align}
\mathcal{L}_1=-\frac{1}{2}\nabla^2+V_\uparrow+\frac{\delta}{2}+2
\beta_{\uparrow\uparrow}|\phi_\uparrow^g|^2+\beta_{\uparrow\downarrow}
|\phi_\downarrow^g|^2-\mu_g,\\
\mathcal{L}_2=-\frac{1}{2}\nabla^2+V_\downarrow-\frac{\delta}{2}
+\beta_{\uparrow\downarrow}|\phi_\uparrow^g|^2+2\beta_{\downarrow\downarrow}
|\phi_\downarrow^g|^2-\mu_g.
\end{align}
The Bogoliubov excitations are then determined by the Bogoliubov-de Gennes (BdG) equations \eqref{eq:Bdg:sec3}.
\subsection{Semiclassical scaling and limits}\label{sec:sem-so}
For strong interaction $\beta_{jl}\gg1$ ($j,l=\uparrow,\downarrow$) and harmonic trapping potentials \eqref{eq:hopot:sec2}, we could rescale \eqref{eq:cgpen:sec3} by choosing
$\bx\to\bx \vep^{-1/2}$ , $\psi_{j}\to\psi^\vep_j\vep^{d/4}$, $\vep=1/\beta^{2/(d+2)}$, $
\beta_{\max}=\max\{|\beta_{\uparrow\uparrow}|,|\beta_{\uparrow\downarrow}|,|\beta_{\downarrow\downarrow}|\}$, which gives the following CGPEs
\be\label{eq:cgpe1:semc:sec3}
\begin{split} &i\vep\partial_t \psi_{\uparrow}^\vep=\biggl[-\frac{\vep^2}{2}\nabla^2
+V_\uparrow(\bx)+ik_0\vep^{3/2}\p_x+\frac{\vep\delta}{2}
+\sum_{j=\uparrow,\downarrow}\beta_{\uparrow j}^\vep|\psi^\vep_{j}|^2\biggr]\psi^\vep_{\uparrow}+\frac{\vep\Omega}{2}
\psi^\vep_{\downarrow}, \\
&i\vep\partial _t \psi^\vep_{\downarrow}=\biggl[-\frac{\vep^2}{2}\nabla^2
+V_\downarrow(\bx)-ik_0 \vep^{3/2}\p_x-\frac{\vep\delta}{2}+\sum_{j=\uparrow,\downarrow}\beta_{\downarrow j}^\vep|\psi^\vep_{j}|^2\biggr]\psi^\vep_{\downarrow}
+\frac{\vep\Omega}{2}
\psi^\vep_{\uparrow},\end{split} \ee
where $\beta_{jl}^\vep=\frac{\beta_{jl}}{\beta_{\max}}$ with $\beta_{jl}^\vep\to\beta_{jl}^0$ as $\vep\to0^+$ and the potential functions are given in \eqref{eq:hopot:sec2}. It is
of great interest to study the behavior of \eqref{eq:cgpe1:semc:sec3} when the small parameter $\vep$ tends to 0, i.e. the semiclassical limit.

In the linear case, i.e. $\beta_{jl}^\vep=0$ for $j,l=\uparrow,\downarrow$, \eqref{eq:cgpe1:semc:sec3} collapses to
\be\label{eq:cgpe1:semc:linear:sec3}
i\vep\p_t\Psi^\vep=\begin{bmatrix}\frac{-\vep^2}{2}\Delta+ik_0\vep^{3/2}\p_x
+\frac{\vep\delta}{2}+V_\uparrow&\frac{\vep\Omega}{2}\\
\frac{\vep\Omega}{2}&\frac{-\vep^2}{2}\Delta-ik_0\vep^{3/2}\p_x-
\frac{\vep\delta}{2}+V_\downarrow
\end{bmatrix}\Psi^\vep, \ee
where $\Psi^\vep=(\psi_\uparrow^\vep,\psi_\downarrow^\vep)^T$.
We now describe the limit as $\vep\to0^+$ using the Wigner transform instead of WKB approach in section 2,
\be
W^\vep(\Psi^\vep)(\bx,\xi)=(2\pi)^{-d}\int_{\mathbb R^d}\Psi^\vep(\bx-\vep v/2)\otimes\Psi^\vep(\bx+\vep v/2)e^{iv\cdot\xi}\,dv,
\ee
where $W^\vep$ is a $2\times2$ matrix-valued function.
The symbol corresponds to \eqref{eq:cgpe1:semc:linear:sec3} can be written as
\be
P^\vep(\bx,\xi)=\frac{i}{2}|\xi|^2+i\begin{bmatrix}k_0\vep^{1/2}\xi_1+
V_\uparrow(\bx)+\frac{\vep\delta}{2}&\frac{\vep\Omega}{2}\\
\frac{\vep\Omega}{2}&-k_0\vep^{1/2}\xi_1+V_\downarrow(\bx)-
\frac{\vep\delta}{2}\end{bmatrix},
\ee
where $\xi=(\xi_1,\xi_2,\ldots, \xi_d)^T$.
Let us consider the principal part $P$ of $P^\vep=P+O(\vep)$, i.e., we omit small $O(\vep)$ terms,
and we know that $-iP(\bx,\xi)$ has two eigenvalues $
\lambda_\uparrow(\bx,\xi)$ and $\lambda_\downarrow(\bx,\xi)$.
 Let $\Pi_{j}$ ($j=\uparrow,\downarrow$) be the projection matrix from $\mathbb C^2$ to the eigenvector space associated with $\lambda_{j}$. If $\lambda_{\uparrow,\downarrow}$ are
well separated, then $W^\vep(\Psi^\vep)$ converges to the
Wigner measure $W^0$ which can be decomposed as
\be\label{eq:decwg:sec3}
W^0=u_\uparrow(\bx,\xi,t)\Pi_\uparrow+u_\downarrow(\bx,\xi,t)\Pi_\downarrow,
\ee
where $u_{j}(\bx,\xi,t)$ ($j=\uparrow,\downarrow$) satisfies the Liouville equation
\be \label{Liou:sec3}
\p_tu_j(\bx,\xi,t)+\nabla_{\xi}\lambda_j(\bx,\xi,t)\cdot\nabla_{\bx}u_j(\bx,\xi,t)-
\nabla_{\bx}\lambda_j(\bx,\xi,t)\cdot\nabla_{\xi}u_j(\bx,\xi,t)=0.
\ee
It is known that such semi-classical limit fails at regions when $\lambda_\uparrow$ and $\lambda_\downarrow$ are close.

  Specifically, when $k_0=O(1)$, $\delta=O(1)$ and $\Omega=O(1)$, the limit of the
  Wigner transform $W^\vep(\Psi^\vep)$
  only has diagonal elements, and we have
\be
\label{weakwg:sec3}
 P=
\frac{i}{2}|\xi|^2+i\begin{bmatrix}
V_\uparrow(\bx)&0\\
0&V_\downarrow(\bx)\end{bmatrix}, \quad \lambda_\uparrow=\frac{1}{2}|\xi|^2+V_\uparrow(\bx), \quad \lambda_\downarrow=
\frac{1}{2}|\xi|^2+V_\downarrow(\bx).\ee
In the limit of this case, $W^0$ in \eqref{eq:decwg:sec3},  $\Pi_\uparrow$ and $\Pi_\downarrow$ are diagonal matrices,
which means the two components of $\Psi^{\vep}$ in \eqref{eq:cgpe1:semc:linear:sec3} are decoupled as $\vep\to0^+$.
In addition, the Liouville equation \eqref{Liou:sec3} is valid with $\lambda_\uparrow$ and $\lambda_\downarrow$ defined in
\eqref{weakwg:sec3}.

Similarly, when $k_0=O(1/\vep^{1/2})$, $\delta=O(1/\vep)$ and $\Omega=O(1/\vep)$, e.g.
$k_0=\frac{k_\infty}{\vep^{1/2}}$, $\Omega=\frac{\Omega_\infty}{\vep}$ and
 $\delta=\frac{\delta_\infty}{\vep}$ with $k_\infty$, $\Omega_\infty$ and
$\delta_\infty$ nonzero constants,  the limit of the Wigner transform $W^\vep(\Psi^\vep)$
  has nonzero diagonal and off-diagonal elements, and we have
\be\label{weakwg3:sec3}
 P=\frac{i}{2}|\xi|^2+i\begin{bmatrix}k_\infty\xi_1+
V_\uparrow(\bx)+\frac{\delta_\infty}{2}&\frac{\Omega_\infty}{2}\\
\frac{\Omega_\infty}{2}&-k_\infty\xi_1+V_\downarrow(\bx)-\frac{\delta_\infty}{2}\end{bmatrix},
\ee
and
\be\label{weakwg5:sec3}
\lambda_{\uparrow,\downarrow}=\frac{|\xi|^2}{2}+\frac{V_1(\bx)+V_2(\bx)}{2}\pm\frac{\sqrt{[V_\uparrow(\bx)-
V_\downarrow(\bx)+2k_\infty\xi_1+\delta_\infty]^2+\Omega_\infty^2}}{2}.
\ee
In the limit of this case, $W^0$ in \eqref{eq:decwg:sec3}, $\Pi_1$ and $\Pi_2$ are full matrices,
which means that the two components of $\Psi^{\vep}$ in \eqref{eq:cgpe1:semc:linear:sec3}
are coupled as $\vep\to0^+$. Again, the Liouville equation \eqref{Liou:sec3} is
valid with $\lambda_\uparrow$ and $\lambda_\downarrow$ defined in
\eqref{weakwg5:sec3}.

Of course, for the nonlinear case, i.e. $\beta_{jl}^0\ne0$ for $j,l=\uparrow,\downarrow$,
only the case when $\Omega=0$ and $k_0=0$ has been addressed \cite{Lee}.
For $\Omega\neq0$ and $k_0\neq0$, it is
still not clear about the semi-classical limit of the CGPEs
\eqref{eq:cgpe1:semc:sec3}.

\section{Spin-1 BEC}
\setcounter{equation}{0}
\setcounter{figure}{0}
When an optical trap is used to confine  the particles instead of a magnetic trap,  a spinor condensate can be achieved experimentally. In current
experiments, alkali atoms are widely used and have an electron spin of $1/2$ with a nuclear spin of $3/2$ for $^{23}$Na, $^{87}$Rb and $^{41}$Kb. Therefore,
the hyperfine spin  for these atoms as a consequence of the interaction between the electron spin and the nuclear spin can be 1 or 2. In this section,
we focus on the spin-1 BEC.
\subsection{The mathematical model}\label{sec:mf:spin-1}
At  temperature $T$ much lower than the critical
temperature $T_c$,  a spin-1 condensate subject to an external  uniform magnetic field $\mathbb{B}$, can be described by the three-component wave functions $\Psi(\bx,t)
= (\psi_1(\bx,t), \psi_0(\bx,t), \psi_{-1}(\bx,t)^T)$  ($\bx\in\mathbb{R}^3$) ($\psi_l$ for the $m_F=l$ state, $l=-1,0,1$) governed by the following CGPEs in 3D
\cite{Ho1,Ohmi,Ueda,StamperUeda}   as
\begin{equation}\label{eq:cgpe:sec4}
 \begin{split} &i\hbar\p_t\psi_{1}=\left[\widetilde{H}
    -p_0+q_0 + c_0 \rho
    +c_1(\rho_1+\rho_0-\rho_{-1})\right]\psi_1+c_1\,\bar{\psi}_{-1}\,\psi_0^2,\\
  &i\hbar\p_t\psi_{0} =\left[\widetilde{H}
     + c_0 \rho
    +c_1(\rho_1+\rho_{-1})\right]\psi_0 +2c_1\,\psi_{-1}\,\bar{\psi}_{0}\,\psi_1,\\
&i\hbar\p_t\psi_{-1}=\left[\widetilde{H}
     +p_0+q_0+ c_0 \rho
    +c_1(\rho_{-1}+\rho_0-\rho_1)\right]\psi_{-1}+c_1\,\psi_0^2\,\bar{\psi}_1,
\end{split}
\end{equation}
where the single Hamiltonian $\widetilde{H}=-\frac{\hbar^2}{2m}\nabla^2+\tilde{V}(\bx)$ with the trapping potential $\tilde{V}(\bx)$ usually chosen as the harmonic potential in \eqref{eq:hot:sec2}, $p_0=-\frac{\mu_B\mathbb{B}}{2}$ and $q_0=\frac{\mu_B^2\mathbb{B}^2}{4E_{\mathrm{hfs}}}$ are the linear and quadratic Zeeman energy shifts, respectively.
Here $\mu_B=e\hbar/m_e$ is the Bohr magneton, $e>0$ is the elementary charge and $m_e$ is the electron mass, $E_{\mathrm{hfs}}$ is the hyperfine energy splitting \cite{Ueda}.
$\rho=\sum_{l=-1,0,1}\rho_l$ and $\rho_l=|\psi_l|^2$ ($l=-1,0,1$) is the density of $l$-th component; $c_0=\frac{g_0+2g_2}{3}$  characterizes the spin-independent
interaction (positive for repulsive interaction and negative for attractive interaction)
while  $c_1=\frac{g_2-g_0}{3}$ characterizes the spin-exchange interaction
(negative for ferromagnetic interaction and positive for antiferromagnetic interaction) with
$g_0=\frac{4\pi\hbar^2}{m}a_0$ ($g_2=\frac{4\pi\hbar^2}{m}a_2$), and $a_0$ ($a_2$) being the s-wave scattering length for scattering channel of total hyperfine spin 0 (spin 2).

Similar to the CGPEs \eqref{eq:cgpe:sec2} for pseudo spin-1/2 condensate, introducing the  scaling:
$t\to t/\omega_s$ with $\omega_s=\min\{\omega_x,\omega_y,\omega_z\}$, $\bx\to \bx/ x_s $ with
$x_s=\sqrt{\frac{\hbar}{m\omega_s}}$, $\psi_l\to\psi_l x_s^{3/2}/\sqrt{N}$ ($l=-1,0,1$) with
$N$ being the total number of particles in the system,  after a proper dimension reduction process in 1D and 2D, the dimensionless CGPEs are obtained in
$d$ dimensions ($d=1,2,3$) for $\Psi=(\psi_1,\psi_0,\psi_{-1})^T$ as
\begin{equation} \label{eq:cgpen:sec4}
\begin{split}
  &i\p_t\psi_{1}=\left[H
    + q-p + \beta_0 \rho
    +\beta_1(\rho_1+\rho_0-\rho_{-1})\right]\psi_1+\beta_1\,\bar{\psi}_{-1}\,\psi_0^2,\\
  &i\p_t\psi_{0} =\left[H
    + \beta_0 \rho
    +\beta_1(\rho_1+\rho_{-1})\right]\psi_0 +2\beta_1\,\psi_{-1}\,\bar{\psi}_{0}\,\psi_1,\\
  &i\p_t\psi_{-1}=\left[H
    + q+p + \beta_0 \rho
    +\beta_1(\rho_{-1}+\rho_0-\rho_1)\right]\psi_{-1}+\beta_1\,\psi_0^2\,\bar{\psi}_1,
\end{split}
\end{equation}
where $H=-\fl{1}{2}\nabla^2+V(\bx)$ and the dimensionless harmonic trapping potential
$V(\bx)$ is given in \eqref{eq:hotn:sec2}. The linear and quadratic Zeeman terms are scaled according to $p = \frac{p_0}{\hbar\omega_s}$
and $q = \frac{q_0}{\hbar\omega_s}$.
The dimensionless mean-field and spin-exchange interaction terms are now given by
$\beta_0 = \frac{Nc_0}{x_s^3\hbar\omega_s}=\frac{4\pi N(a_0+2a_2)}{3x_s}$ and
$\beta_1 = \frac{Nc_2}{x_s^3\hbar\omega_s}=\frac{4\pi N(a_2-a_0)}{3x_s}$ in 3D;
$\beta_0 =\frac{4\pi N(a_0+2a_2)}{3x_s}\frac{\sqrt{\gamma_z}}{\sqrt{2\pi}}$ and
$\beta_1 = \frac{4\pi N(a_2-a_0)}{3x_s}\frac{\sqrt{\gamma_z}}{\sqrt{2\pi}}$ in 2D; $\beta_0 =\frac{4\pi N(a_0+2a_2)}{3x_s}\frac{\sqrt{\gamma_z\gamma_y}}{2\pi}$ and
$\beta_1 = \frac{4\pi N(a_2-a_0)}{3x_s}\frac{\sqrt{\gamma_z\gamma_y}}{2\pi}$ in 1D.

Introduce the spin-1 matrices $\mathbf{f}=(\mathrm{f}_x,\mathrm{f}_y,\mathrm{f}_z)^T$ as
\be\label{eq:spin1m:sec4}
\mathrm{f}_x=\frac{1}{\sqrt{2}}\begin{pmatrix}0&1&0\\
1&0&1\\
0&1&0\end{pmatrix},\quad \mathrm{f}_y=\frac{i}{\sqrt{2}}\begin{pmatrix}0&-1&0\\
1&0&-1\\
0&1&0\end{pmatrix},\quad\mathrm{f}_z=\frac{1}{\sqrt{2}}\begin{pmatrix}1&0&0\\
0&0&0\\
0&0&-1\end{pmatrix}
\ee
and the spin vector $\mathbf{F}(\Psi)=(F_x(\Psi),F_y(\Psi),F_z(\Psi))^T=(F_x,F_y,F_z)^T=(\Psi^*\mathrm{f}_x\Psi,\Psi^*\mathrm{f}_y\Psi,\Psi^*\mathrm{f}_z\Psi)^T$
($\Psi^*=\overline{\Psi}^T$ is the conjugate transpose) of the condensate can be expressed as
\be\label{eq:spin1v:sec4}
\begin{split}
&F_x=\frac{1}{\sqrt{2}}\left[\overline{\psi}_1\psi_0+\overline{\psi}_0
(\psi_1+\psi_{-1})+\overline{\psi}_{-1}\psi_0\right],\\
&F_y=\frac{i}{\sqrt{2}}\left[-\overline{\psi}_1\psi_0+\overline{\psi}_0
(\psi_1-\psi_{-1})+\overline{\psi}_{-1}
\psi_0\right],\\
&F_z=|\psi_1|^2-|\psi_{-1}|^2,
\end{split}
\ee
and the CGPEs \eqref{eq:cgpen:sec4} can be written in the compact form as
\be\label{eq:cgpenc:sec4}
i\partial_t\Psi=[H+\beta_0 \rho-p \mathrm{f}_z+q \mathrm{f}_z^2+\beta_1 \mathbf{F}\cdot\mathbf{f}]\Psi,
\ee
where $\mathbf{F}\cdot\mathbf{f}=F_x\mathrm{f}_x+F_y\mathrm{f}_y+F_z\mathrm{f}_z$ and $\rho=|\Psi|^2=\sum_{l=-1}^1|\psi_l|^2$.

The CGPEs \eqref{eq:cgpen:sec4} (or \eqref{eq:cgpenc:sec4}) conserve the following three important quantities, i.e.
the {\sl mass} (or {\sl normalization})
\be\label{eq:norm:sec4}
N(\Psi(\cdot,t)):=\|\Psi(\cdot,t)\|^2=\int_{\Bbb R^d}\sum_{l=-1,0,1}|\psi_l(\bx,t)|^2\,d\bx=N(\Psi(\cdot,0))=1,
\ee
the {\sl magnetization} (with $M\in[-1,1]$)
\be\label{eq:mag:sec4}
M(\Psi(\cdot,t)):=\int_{\Bbb R^d}\sum_{l=-1,0,1}l|\psi_l(\bx,t)|^2\,d\bx
=M(\Psi(\cdot,0))=M,
\ee
and the {\sl energy per particle}
\bea\label{eq:energy:sec4}
E(\Psi(\cdot,t))&=&\int_{\mathbb{R}^d
}\left\{\sum_{l=-1}^{1} \left(\frac{1}{2}|\nabla
\psi_l|^2+(V(\bx)-pl+ql^2)|\psi_l|^2\right)
+\frac{\beta_0}{2}|\Psi|^4+\frac{\beta_1}{2}|\mathbf{F}|^2\right\}\; d\bx \nonumber\\
&\equiv&E(\Psi(\cdot,0)), \qquad t\ge0.
\eea
In practice, introducing $\psi_l\to e^{-ilpt}\psi_l$ in the CGPEs \eqref{eq:cgpen:sec4} (or \eqref{eq:cgpenc:sec4}), the system is unchanged and it is thus
reasonable to assume the linear Zeeman term $p=0$ in the subsequent discussion. On the other hand, it is easy to observe from \eqref{eq:energy:sec4} that
the linear Zeeman term does not contribute to the energy due to the magnetization conservation \eqref{eq:mag:sec4}.
\subsection{Ground states}
The ground state $\Phi_g(\bx)$ of the spin-1 BEC described by the CGPEs \eqref{eq:cgpen:sec4}
is obtained from the minimization of
the energy functional subject to the conservation of total mass
and magnetization:
\begin{quote}
  Find $\left(\Phi_g \in S_M\right)$ such that
\end{quote}
  \begin{equation}\label{eq:minimize:sec4}
    E_g := E\left(\Phi_g\right) = \min_{\Phi \in S_M}
    E\left(\Phi\right),
  \end{equation}
\noindent where the nonconvex set $S_M$ is defined as
\be \label{eq:cons:sec4}
S_M=\left\{\Phi=(\phi_1,\phi_0,\phi_{-1})^T\ |\ \|\Phi\|=1, \
\int_{{\Bbb R}^d} \left[|\phi_1(\bx)|^2
-|\phi_{-1}(\bx)|^2\right]\,d\bx=M, \ E(\Phi)<\infty\right\}. \ee This is
a nonconvex minimization problem and the  Euler-Lagrange equations associated to the minimization problem
\eqref{eq:minimize:sec4} read:
\begin{align}
(\mu+\lambda)\;\phi_{1}(\bx)=&\left[-\frac{1}{2}\nabla^2
+V(\bx)-p+q+(\beta_0+\beta_1)\left(|\phi_1|^2+|\phi_0|^2\right)
+(\beta_0-\beta_1)|\phi_{-1}|^2\right]\phi_1\nn\\
&+\beta_1\,\bar{\phi}_{-1}\,\phi_0^2,\label{eq:el:1} \\
\mu\;\phi_{0}(\bx)=&\left[-\frac{1}{2}\nabla^2
+V(\bx)+ (\beta_0+\beta_1)\left(|\phi_1|^2+|\phi_{-1}|^2\right)
+\beta_0|\phi_{0}|^2\right]\phi_0 \nn \\
&+2\beta_1\,\phi_{-1}\,\bar{\phi}_{0}\,\phi_1,\label{eq:el:2} \\
(\mu-\lambda)\;\phi_{-1}(\bx)=&\left[-\frac{1}{2}\nabla^2 +V(\bx)+p+q+
(\beta_0+\beta_1)\left(|\phi_{-1}|^2+|\phi_0|^2\right)
+(\beta_0-\beta_1)|\phi_{1}|^2\right]\phi_{-1} \nn\\
&+\beta_1\,\phi_0^2\bar{\phi}_1.\label{eq:el:3}
\end{align}
Here $\mu$ and $\lambda$ are the Lagrange multipliers (or chemical
potentials) corresponding to the normalization constraint \eqref{eq:norm:sec5} and the magnetization constraint \eqref{eq:mag:sec5},
respectively.
\subsubsection{Mathematical theories}
We collect the existence and uniqueness results on the ground state \eqref{eq:minimize:sec4} below.
\begin{theorem}[existence and uniqueness \cite{LinChern}]\label{thm:gs:sec4} Suppose $\lim\limits_{|\bx|\to\infty}V(\bx)=+\infty$, there exists a ground state $\Phi_g=(\phi_1^g,\phi_0^g,\phi_{-1}^g)^T\in S_M$ of \eqref{eq:minimize:sec4} for the spin-1 BEC
governed by the CGPEs \eqref{eq:cgpen:sec4}, if one of the following conditions hold,
\begin{enumerate}\renewcommand{\labelenumi}{(\roman{enumi})}
\item $d=1$;
\item $d=2$, $M=\pm1$ and $\beta_0+\beta_1>-C_b$ or $M\in(-1,1)$ and $\beta_0+\beta_1> -C_b$ with $\beta_1\leq0$, or
$M\in(-1,1)$, $\beta_0+\beta_1> \frac{-2C_b}{1+|M|}$ and $\beta_0\ge-\frac{C_b^2+\beta_1C_b}{\beta_1(1-M^2)+C_b}$ with $\beta_1>0$;
\item $d=3$, $M=\pm1$ and $\beta_0+\beta_1\ge0$ or $M\in(-1,1)$ and $\beta_0\ge0$, $\beta_0+\beta_1\ge0$.
\end{enumerate}
In particular, $(e^{i\theta_{1}}\phi^g_1,e^{i\theta_0}\phi_0^g, e^{i\theta_{-1}}\phi_{-1}^g)^T\in S_M$  with real constants $\theta_1+\theta_{-1}-2\theta_0=2k\pi$ ($k\in\mathbb{Z}$) if
$\beta_1\neq0$ or arbitrary constants $\theta_l$ ($l=-1,0,1$) if $\beta_1=0$, is
also a ground state of \eqref{eq:minimize:sec4}. Moreover, the ground state can be chosen as $(|\phi^g_1|,-\mathrm{sgn}(\beta_1)|\phi_0^g|, |\phi_{-1}^g|)^T$ ($\beta_1\neq0$)
and  $(|\phi^g_1|,|\phi_0^g|, |\phi_{-1}^g|)^T$ ($\beta_1=0$). These special ground states  are unique if
\begin{enumerate}\renewcommand{\labelenumi}{(\roman{enumi})$^\prime$}
\item when $M=\pm1$, $\beta_0+\beta_1\ge0$;
\item when $M\in(-1,1)$ and $q=0$, $\beta_1>0$ and $\beta_0\ge0$ or $\beta_1<0$ and $\beta_0+\beta_1\ge0$.
\end{enumerate}
When $M\in(-1,1)$, $q=0$ and $\beta_1=0$, the ground states with nonnegative components are of the form $(\sqrt{1+M-\alpha_0/2}\phi,\sqrt{\alpha_0}\phi,\sqrt{1-M-\alpha_0/2}\phi)$ with $\phi:=\phi(\bx)\ge0$ and $0\leq \alpha_0\leq 1-|M|$.
On the other hand, there exists no ground state of \eqref{eq:minimize:sec4},  i.e. $\inf_{\Phi\in S_M}E(\Phi)=-\infty$, if one of the following conditions hold,
\begin{enumerate}\renewcommand{\labelenumi}{(\roman{enumi})$^{\prime\prime}$}
\item $d=2$, if $M=\pm1$ and $\beta_0+\beta_1\leq-C_b$ or $M\in(-1,1)$ and $\beta_0+\beta_1\leq-C_b$ with $\beta_1\leq0$, or
$M\in(-1,1)$, $\beta_1>0$, $\beta_0+\beta_1\leq \frac{-2C_b}{1+|M|}$ or $\beta_0<-C_b-M^2\beta_1$;
\item $d=3$,  $\beta_0+\beta_1<0$ or $M\in(-1,1)$ and $\beta_0<0$.
\end{enumerate}
\end{theorem}
When $M=\pm1$, the ground state of \eqref{eq:minimize:sec4} collapses to a single component case which has been widely studied \cite{Baocai2013}.
Most of the results in Theorem \ref{thm:gs:sec4} can be drawn from the following observations when the quadratic Zeeman term is absent in the CGPEs \eqref{eq:cgpen:sec4}, i.e. $q=0$.
Firstly, when $q=0$ and $M\in(-1,1)$,  for the ferromagnetic system $\beta_1<0$, we have the single mode approximation (SMA), i.e. each component of the ground state $\Phi_g$ is identical up to a constant factor \cite{LinChern,BaoChernZ}.
\begin{theorem}[single mode approximation (SMA) \cite{LinChern}]\label{thm:sma:sec4}  Suppose $\lim\limits_{|\bx|\to\infty}V(\bx)=+\infty$, $q=0$, $M\in(-1,1)$, $\beta_1<0$ and the existence conditions in Theorem \ref{thm:gs:sec4} hold, the ground state $\Phi_g=(\phi_{1}^g,\phi_0^g,\phi_{-1}^g)^T\in S_M$
satisfies $\phi_l^g=e^{i\theta_l}\alpha_l\phi_g$ ($\theta_{1}+\theta_{-1}-2\theta_{0}=(2k+1)\pi$, $k\in\mathbb{Z}$), where $\phi_g$ is the unique positive minimizer of the energy functional
\be
E_{\rm SMA}(\phi)=\int_{\mathbb{R}^d}\left[\frac{1}{2}|\nabla\phi|^2+V(\bx)|\phi|^2+\frac{\beta_0+\beta_1}{2}|\phi|^4\right]\,d\bx,
\ee
under the constraint $\|\phi\|=1$, and $\alpha_1=\frac{1+M}{2}$, $\alpha_{-1}=\frac{1-M}{2}$, $\alpha_0=\sqrt{\frac{1-M^2}{2}}$.
\end{theorem}
Secondly, when $q=0$ and $M\in(-1,1)$,  for the anti-ferromagnetic system $\beta_1>0$, we have the vanishing phenomenon, i.e.   the ground state $\Phi_g=(\phi_1^g,\phi_0^g,\phi_{-1}^g)^T$
satisfies $\phi_0^g=0$.
\begin{theorem}[two-component case \cite{LinChern}]\label{thm:van:sec4} Suppose $\lim\limits_{|\bx|\to\infty}V(\bx)=+\infty$, $q<0$, $M\in(-1,1)$, $\beta_1>0$ and the existence conditions in Theorem \ref{thm:gs:sec4} hold, the ground state $\Phi_g=(\phi_{1}^g,\phi_0^g,\phi_{-1}^g)^T\in S_M$
satisfies $\phi_0^g=0$, and $\tilde{\Phi}_g=(\phi_{1}^g,\phi_{-1}^g)^T$ is a  minimizer of the pseudo spin-1/2 system given in section \ref{sec:2} described by \eqref{eq:minim2:sec2} with $\delta=0$, $\nu=\frac{1+M}{2}$ and $\beta_{\uparrow\uparrow}=\beta_{\downarrow\downarrow}=\beta_{0}+\beta_{1}$, $\beta_{\uparrow\downarrow}=\beta_0-\beta_1$.
\end{theorem}
When quadratic Zeeman effects are considered, the above Theorems \ref{thm:sma:sec4}\&\ref{thm:van:sec4} are generally no longer valid, more rich phases will appear in the spin-1 system.
\subsubsection{Numerical methods and results}
To  compute the ground state \eqref{eq:minimize:sec4}, we generalize the GFDN method in section \ref{numeric-gs:sec2}.
We start with the following CNGF for $\Phi=(\phi_1,\phi_0,\phi_{-1})^T$ \cite{BaoWang2}
\begin{align}
\p_t\phi_{1}(\bx,t)=&\left[\fl{1}{2}\nabla^2 -V(\bx)-q-
(\beta_0+\beta_1)\left(|\phi_1|^2+|\phi_0|^2\right)
-(\beta_0-\beta_1)|\phi_{-1}|^2\right]\phi_1\nn\\
&-\,\beta_1\,\bar{\phi}_{-1}\,\phi_0^2
+\left[\mu_\Phi(t)+\lambda_\Phi(t)\right]\phi_1 , \label{eq:cngf1:sec4}\\
\p_t\phi_{0}(\bx,t)=&\left[\fl{1}{2}\nabla^2
-V(\bx)- (\beta_0+\beta_1)\left(|\phi_1|^2+|\phi_{-1}|^2\right)
-\beta_0|\phi_{0}|^2\right]\phi_0 \nn \\
&-\,2\beta_1\,\phi_{-1}\,\bar{\phi}_{0}\,\phi_1
+\mu_\Phi(t)\;\phi_0 , \label{eq:cngf2:sec4}\\
\p_t\phi_{-1}(\bx,t)=&\left[\fl{1}{2}\nabla^2 -V(\bx)-q-
(\beta_0+\beta_1)\left(|\phi_{-1}|^2+|\phi_0|^2\right)
-(\beta_0-\beta_1)|\phi_{1}|^2\right]\phi_{-1} \nn\\
&-\,\beta_1\,\phi_0^2\,\bar{\phi}_{1}
+\left[\mu_\Phi(t)-\lambda_\Phi(t)\right]\phi_{-1},\label{eq:cngf3:sec4}
\end{align}
where $\mu_{\Phi}(t)$ and $\lambda_{\Phi}(t)$ are Lagrange multipliers such that the normalization condition \eqref{eq:norm:sec4} and the magnetization constraint \eqref{eq:mag:sec4} are preserved during the evolution. For  given initial data $\Phi(\bx,t=0)=\Phi^{(0)}$ satisfying $\|\Phi(\bx,t=0)\|=1$ and $M(\Phi(\cdot,t=0))=M$,  the above CNGF \eqref{eq:cngf1:sec4}-\eqref{eq:cngf3:sec4} will preserve the constraints \eqref{eq:norm:sec4} and \eqref{eq:mag:sec4}, while the energy is diminishing.

We  approximate the CNGF \eqref{eq:cngf1:sec4}-\eqref{eq:cngf3:sec4} following the   spin-1/2 system case in section \ref{numeric-gs:sec2}, resulting in the  GFDN below, for $t\in[t_{n-1},t_{n})$ ($n\ge1$), one solves
\begin{align}\label{eq:GFDN_1:sec4}
  \p_t\phi_{1}=&\left[\fl{1}{2}\nabla^2 -V(\bx)-q-
(\beta_0+\beta_1)(|\phi_1|^2+|\phi_0|^2)
-(\beta_0-\beta_1)|\phi_{-1}|^2\right]\phi_1-\,\beta_1\,\bar{\phi}_{-1}\,\phi_0^2, \\
  \label{eq:GFDN_2:sec4}
   \p_t\phi_{0}=&\left[\fl{1}{2}\nabla^2
-V(\bx)- (\beta_0+\beta_1)\left(|\phi_1|^2+|\phi_{-1}|^2\right)
-\beta_0|\phi_{0}|^2\right]\phi_0-\,2\beta_1\,\phi_{-1}\,\bar{\phi}_{0}\,\phi_1,\\
  \label{eq:GFDN_3:sec4}
\p_t\phi_{-1}=&\left[\fl{1}{2}\nabla^2 -V(\bx)-q-
(\beta_0+\beta_1)(|\phi_{-1}|^2+|\phi_0|^2)
-(\beta_0-\beta_1)|\phi_{1}|^2\right]\phi_{-1}-\,\beta_1\,\phi_0^2\,\bar{\phi}_{1},
\end{align}
followed by a projection step as
\be\label{eq:projection:sec4}
\phi_l(\bx,t_{n}):=\phi_l(\bx,t_{n}^+)= \sigma_l^{n}\;
  \phi_l(\bx,t_{n}^-), \qquad \bx\in{\Bbb
R}^d,\qquad n\ge1,\qquad l=-1,0,1,
\ee
where $\phi_l(\bx,t_n^\pm)=\lim_{t\to t_n^\pm} \phi_l(\bx,t)$
($l=-1,0,1$) and the projection constants $\sigma_l^{n}$ ($l=-1,0,1$) are chosen such that \be\label{eq:cong1:sec4}
\|\Phi(\cdot,t_{n})\|^2=\sum_{l=-1}^1\|\phi_l(\cdot,t_{n})\|^2
=1,\qquad   \|\phi_1(\cdot,t_{n})\|^2-
\|\phi_{-1}(\cdot,t_{n})\|^2=M. \ee
Similar to the pseudo spin-$1/2$ BEC case, there are three projection constants to be determined,
i.e. $\sigma_l^{n}$ ($l=-1,0,1$) in \eqref{eq:projection:sec4}, and there are only two equations, i.e. \eqref{eq:cong1:sec4}, to fix them, we need to find another condition
so that the three projection constants are uniquely determined.
Again, in fact, the above GFDN can be viewed as a split-step discretization of the CNGF \eqref{eq:cngf1:sec4}-\eqref{eq:cngf3:sec4} and the projection step is equivalent to solve
$\partial_t\phi_l(\bx,t)=[\mu_\Phi(t)+l\lambda_\Phi(t)]\phi_l(\bx,t)$ ($l=-1,0,1$). From this observation,  we can find another condition for the projection constants
in \eqref{eq:projection:sec4} as \cite{BaoLim,Lim}
\be\label{eq:cong2:sec4}
\sigma_1^{n}\sigma_{-1}^{n}=(\sigma_0^{n})^2.
\ee
 Solving \eqref{eq:cong1:sec4}-\eqref{eq:cong2:sec4},
 we get explicitly the
projection constants
\be\label{eq:constant1:sec4}
  \sigma_0^{n}=\frac{\sqrt{1-M^2}}
  {\left[\|\phi_0(\cdot,t_{n}^-)\|^2 +
  \sqrt{4(1-M^2)\|\phi_1(\cdot,t_{n}^-)\|^2
  \|\phi_{-1}(\cdot,t_{n}^-)\|^2
  + M^2\|\phi_0(\cdot,t_{n}^-)\|^4}\right]^{1/2}},
  \ee
  and
  \be
  \label{eq:constant2:sec4}
  \sigma_1^{n}=\frac{\sqrt{1+M-(\sigma_0^n)^2
  \|\phi_0(\cdot,t_{n}^-)\|^2}}
  {\sqrt{2}\ \|\phi_1(\cdot,t_{n}^-)\|}, \qquad
  \sigma_{-1}^{n}=\frac{\sqrt{1-M-(\sigma_0^n)^2
  \|\phi_0(\cdot,t_{n}^-)\|^2}}
  {\sqrt{2}\ \|\phi_{-1}(\cdot,t_{n}^-)\|}.
 \ee

To fully discretize the GFDN \eqref{eq:GFDN_1:sec4}-\eqref{eq:projection:sec4} together with the
projection constants \eqref{eq:constant1:sec4}-\eqref{eq:constant2:sec4}, we use backward Euler scheme in temporal discretization and one can choose
finite difference method, spectral method and finite element method for spatial discretization \cite{BaoLim,BaoWang2}. For the simplicity, we omit the detail here. 

\bigskip

{\it  Example 4.1}. To show the ground states of the spin-1 BEC, we take $d=1$, $p=q=0$, $V(x)=x^2/2+25\sin^2\left(\frac{\pi x}{4}\right)$ in
\eqref{eq:cgpen:sec4}. Two different types of interaction strengths
are chosen as

\begin{itemize}
\item Case I. For $^{87}$Rb with dimensionless quantities in
\eqref{eq:cgpen:sec4} used as
$\beta_0=0.0885N$,
 and $\beta_1=-0.00041N$ with $N$ the total number of atoms in the condensate
and the dimensionless length unit
$a_s=2.4116\times 10^{-6}$ [m] and time unit
$t_s=0.007958$ [s].

\item Case II. For $^{23}$Na with dimensionless quantities in
\eqref{eq:cgpen:sec4} used as $\beta_0=0.0241N$,
 and $\beta_1=0.00075N$, with $N$ the total number of atoms in the condensate
and the dimensionless length unit $a_s=4.6896\times 10^{-6}$ [m]
and time unit $t_s=0.007958$ [s].
\end{itemize}

\begin{figure}[tb]
\centerline{\includegraphics[height=5cm,width=7cm]{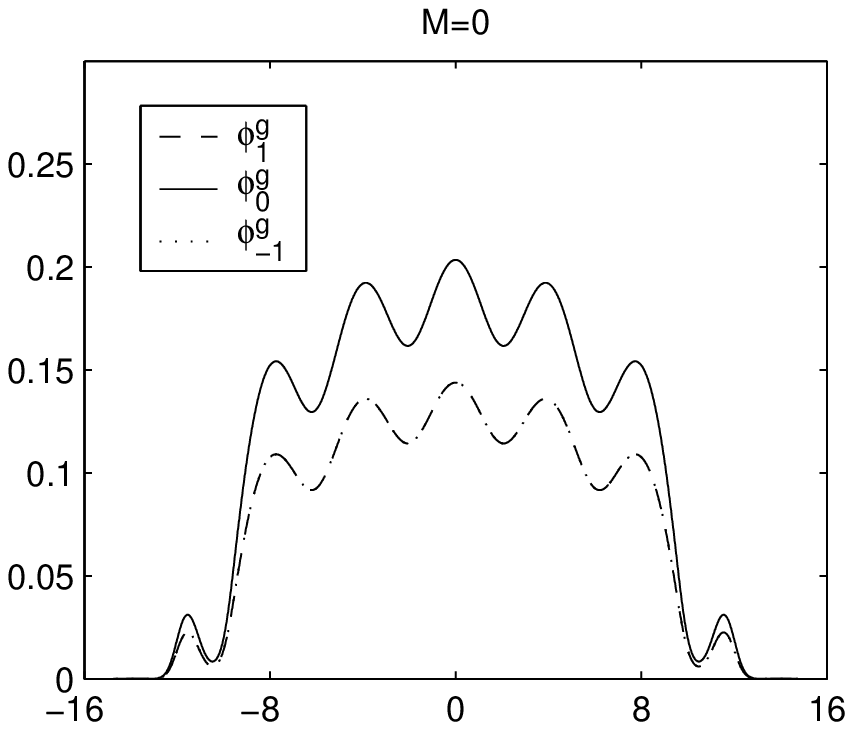}
\quad
\includegraphics[height=5cm,width=7cm]{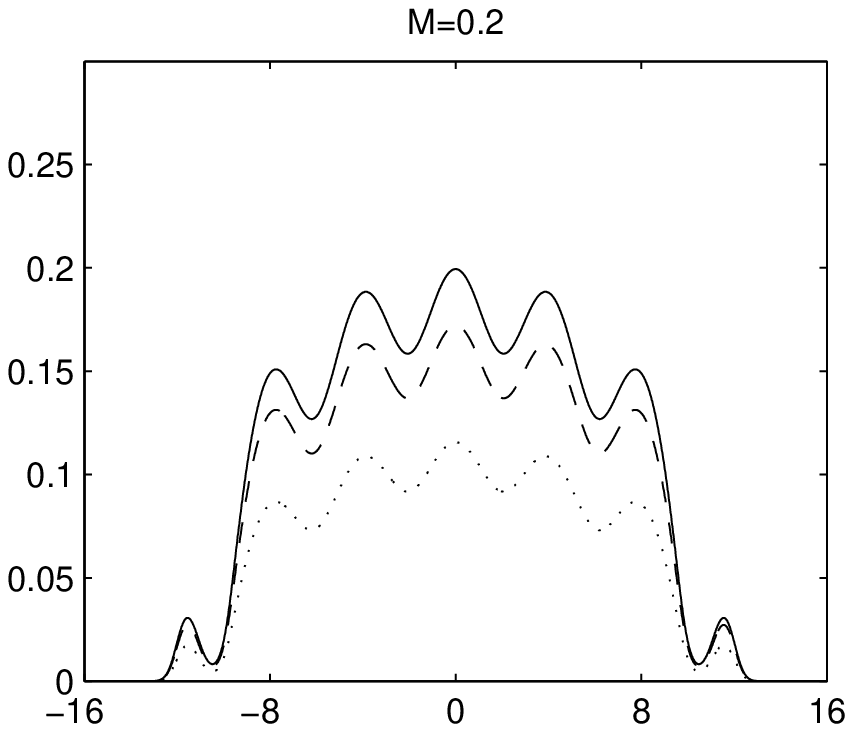}}
\centerline{\includegraphics[height=5cm,width=7cm]{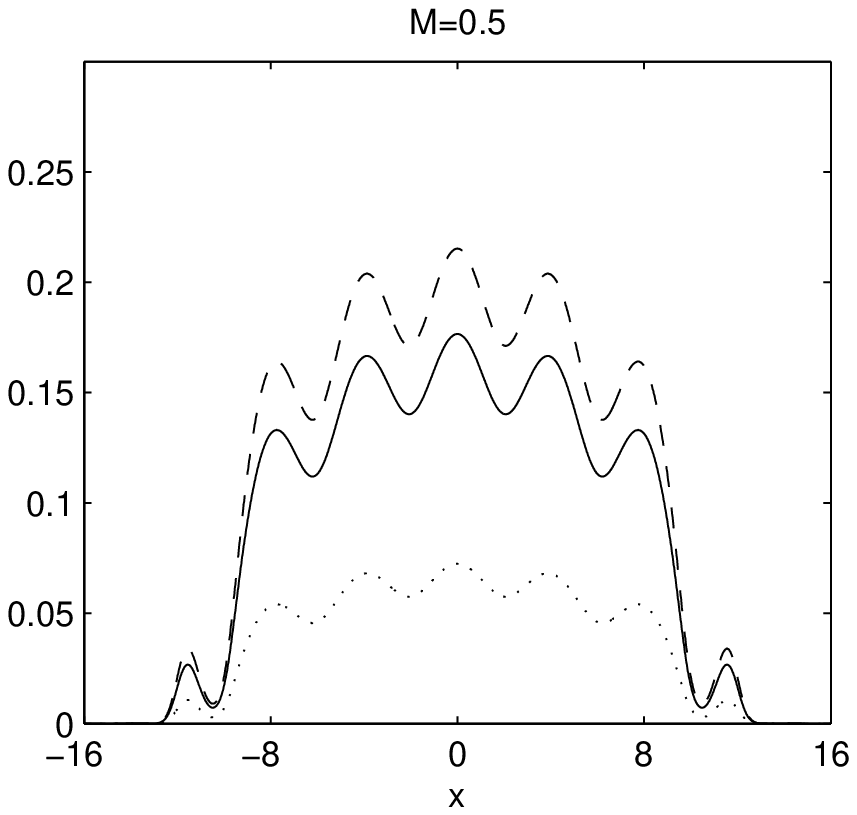}
\quad
\includegraphics[height=5cm,width=7cm]{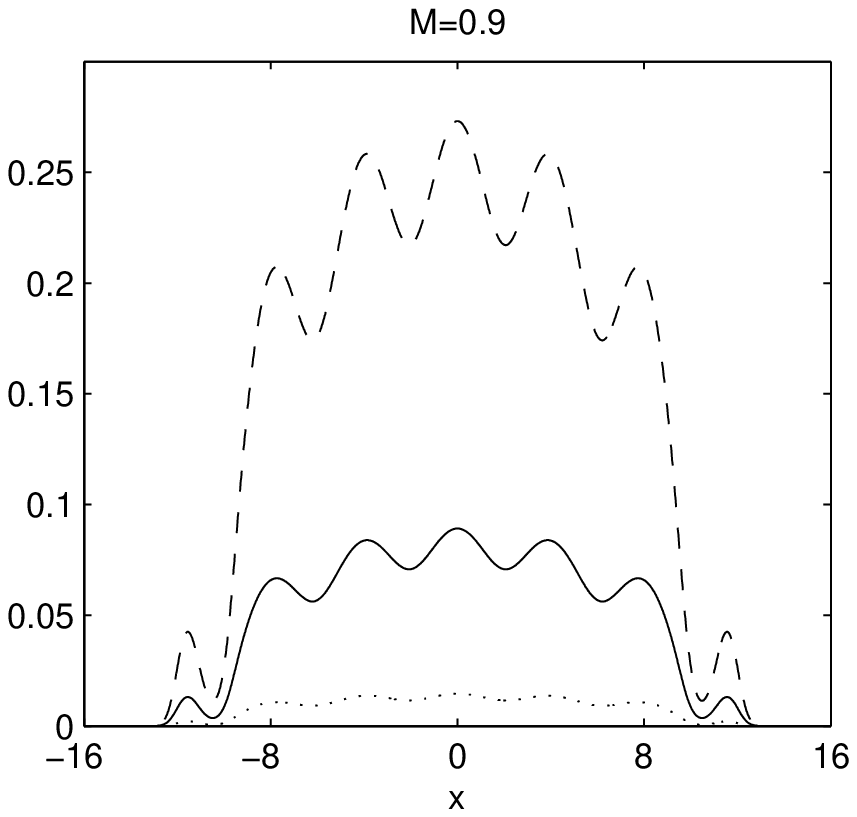}}
\caption{Wave functions of the ground state, i.e., $\phi_1^g(x)$
(dashed line), $\phi_0^g(x)$ (solid line), and $\phi_{-1}^g(x)$ (dotted
line), of $^{87}$Rb in Example 4.1 case {\rm I} with a fixed number of particles
$N=10^4$  for different magnetizations $M=0,0.2,0.5,0.9$ in an
optical lattice potential.} \label{fig:1o:sec4}
\end{figure}

\begin{figure}[tb]
\centerline{\includegraphics[height=5cm,width=7cm]{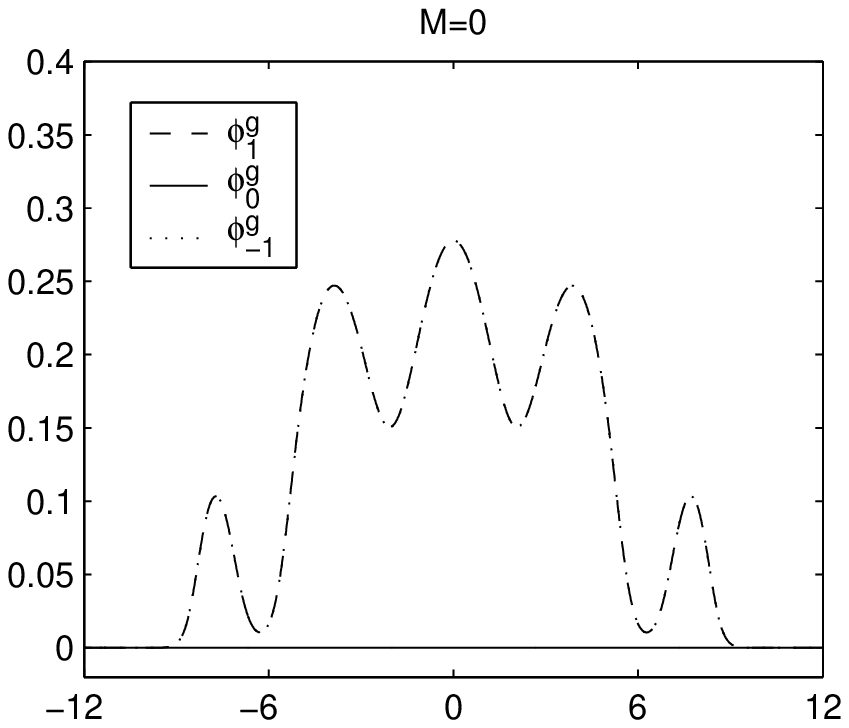}
\quad
\includegraphics[height=5cm,width=7cm]{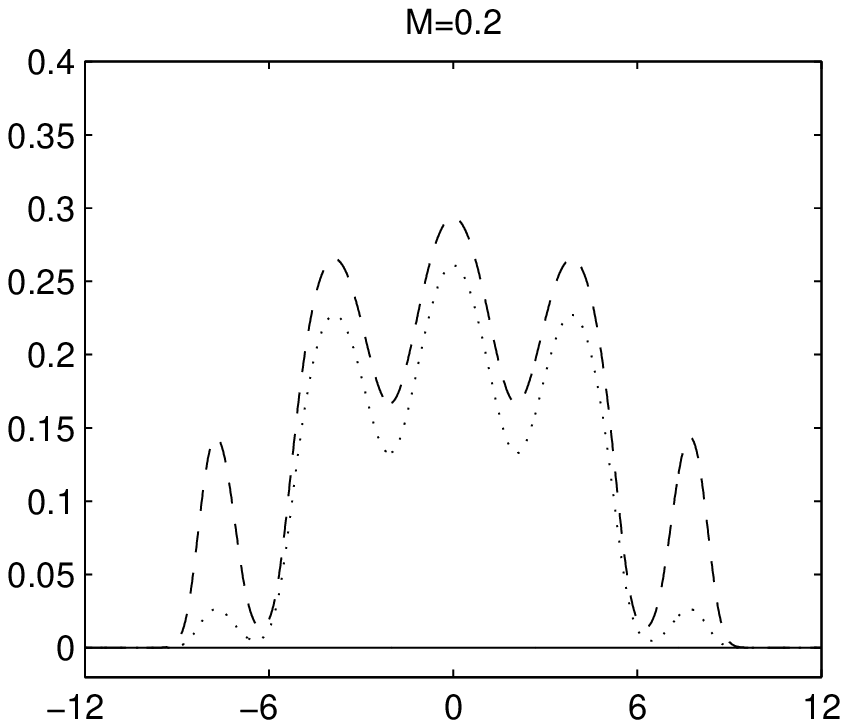}}
\centerline{\includegraphics[height=5cm,width=7cm]{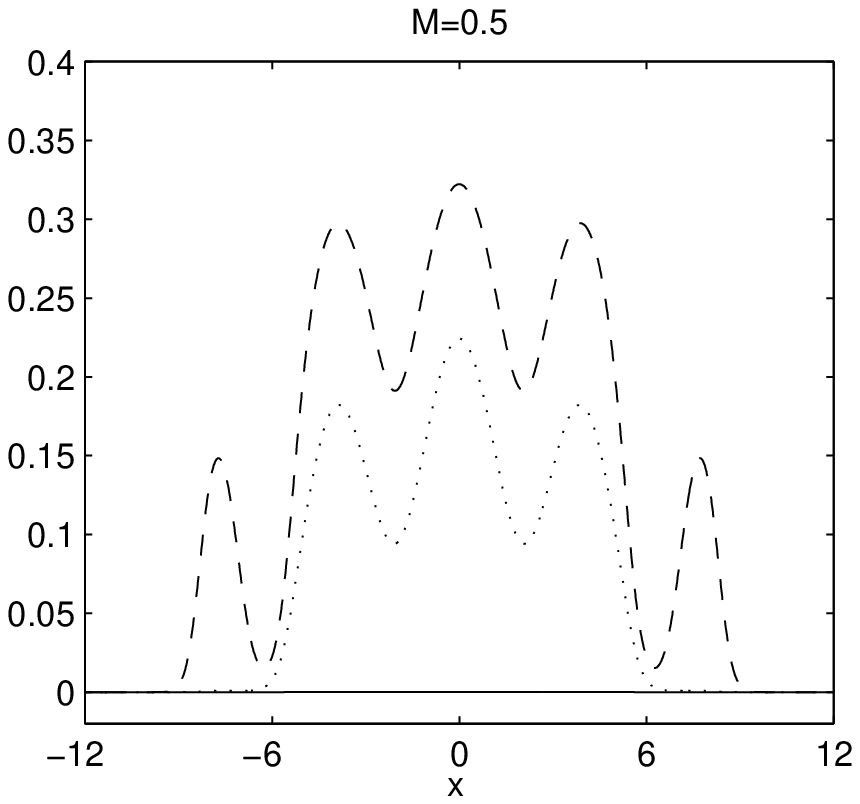}
\quad
\includegraphics[height=5cm,width=7cm]{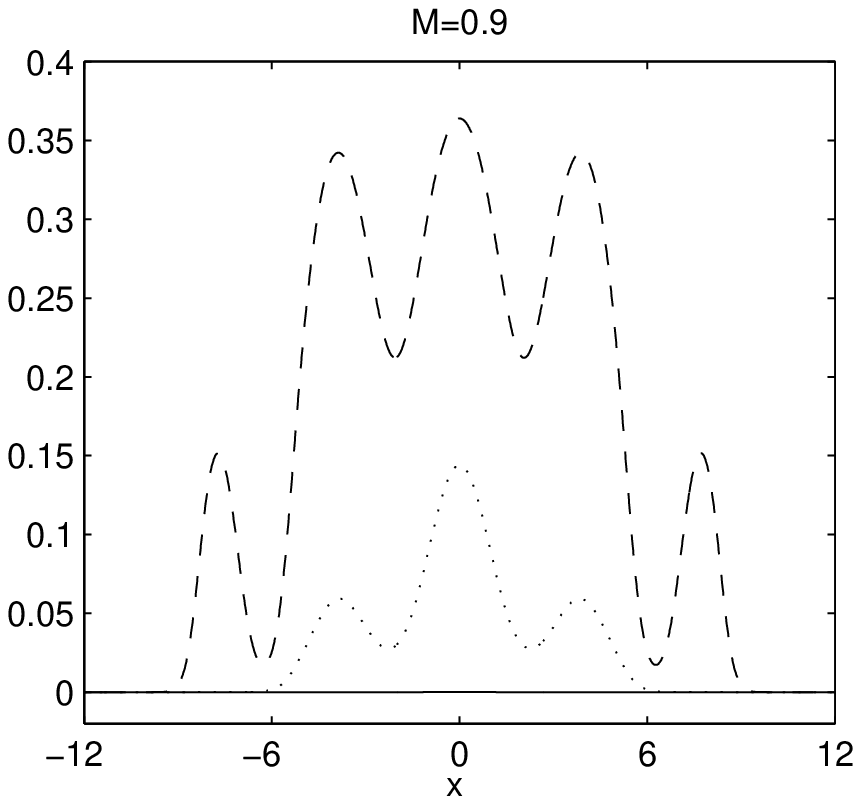}}
\caption{Wave functions of the ground state, i.e., $\phi_1^g(x)$
(dashed line), $\phi_0^g(x)$ (solid line), and $\phi_{-1}^g(x)$ (dotted
line), of $^{23}$Na in Example 4.1 case {\rm II} with $N=10^4$  for different
magnetizations $M=0,0.2,0.5,0.9$ in an optical lattice potential.}
\label{fig:2o:sec4}
\end{figure}

The ground states are computed numerically
by the backward Euler sine pseudospectral method
presented in \cite{BaoLim}.
Figure \ref{fig:1o:sec4} shows the ground state solutions of $^{87}$Rb
in case I with $N=10^4$ for different magnetizations $M$. Figure \ref{fig:2o:sec4}  shows similar results for $^{23}$Na in case II.

For the cases when $q=0$ in Theorems \ref{thm:sma:sec4}\&\ref{thm:van:sec4}, the minimization problem \eqref{eq:minimize:sec4} can be reduced to a single component and  a two component system, respectively, where the numerical methods could be simplified \cite{BaoChernZ}.

\

We remark here that there is another type of ground state
of the spin-1 BEC, especially with an Ioffe-Pritchard magnetic field
$B(\bx)$,  in the literatures \cite{BaoChernZ,Ernst1998}, which is defined as
the minimizer of
the energy functional subject to the conservation of total mass:
\begin{quote}
  Find $\left(\tilde{\Phi}_g \in S\right)$ such that
\end{quote}
  \begin{equation}\label{eq:minimize32:sec4}
    \tilde{E}_g := E\left(\tilde{\Phi}_g\right) = \min_{\Phi \in S}
    E\left(\Phi\right),
  \end{equation}
\noindent where the nonconvex set $S$ is defined as
\be \label{eq:cons32:sec4}
S=\left\{\Phi=(\phi_1,\phi_0,\phi_{-1})^T\ |\ \|\Phi\|=1,  \ E(\Phi)<\infty\right\}. \ee
For the analysis and numerical simulation of
this type of the ground state of spin-1 BEC, we refer
to \cite{BaoChernZ,Ernst1998,Chen2011} and references therein.
In addition, when there is no Ioffe-Pritchard magnetic field
in the spin-1 BEC, the ground state $\tilde{\Phi}_g$ defined in
\eqref{eq:minimize32:sec4} can be computed from the ground state
$\Phi_g$ in \eqref{eq:minimize:sec4} as
\be
\tilde{E}_g := E\left(\tilde{\Phi}_g\right)=\min_{\Phi \in S}
    E\left(\Phi\right)=\min_{-1\le M\le 1} E\left(\Phi_g\right)
    =\min_{-1\le M\le 1}\;\min_{\Phi \in S_M}
    E\left(\Phi\right).
\ee

\subsection{Dynamics}
For a spin-1 system governed by the CGPEs \eqref{eq:cgpen:sec4} (or \eqref{eq:cgpenc:sec4}),
define the {\sl mass} (or density) of each spin component  as \be
\label{eq:massec:sec4} N_l(t):=\int_{{\Bbb R}^d} |\psi_l(\bx,t)|^2\;d\bx,
\qquad t\ge0, \quad l=-1,0,1, \ee
and the condensate width  as
\be
\label{eq:dtap9:sec4}\sigma_\alpha(t) =
\sqrt{\dt_\alpha(t)}=\sqrt{\dt_{\alpha,1}(t)+\dt_{\alpha,0}(t)+\dt_{\alpha,-1}(t)},\qquad
\alpha = x, y, z, \ee where \be \label{eq:dtj:sec4}\dt_{\alpha,j}(t)=\langle
\alpha^2\rangle_l(t) = \int_{{\Bbb R}^d}\alpha^2|\psi_j(\bx,t)|^2d\bx,
\qquad t\geq0, \quad j = 1, 0, -1. \ee
\subsubsection{Dynamical properties}
The most interesting property would be the change of density $N_l(t)$ w.r.t time $t$ \cite{BaoZhang}.
\begin{lemma}\label{lem:1:sec4}
Suppose $\Psi(\bx,t)$ is the solution of the CGPEs
\eqref{eq:cgpen:sec4} for spin-1 BEC, then we have
\be
\dot{N}_1(t)=\dot{N}_{-1}(t) = \widetilde{F}(t), \qquad
\dot{N}_0(t)=-2\widetilde{F}(t),
\ee
with $N_l(0)=\int_{\mathbb{R}^d}|\psi_l(\bx,0)|^2\,d\bx$ ($l=-1,0,1$) and
  \be  \widetilde{F}(t) = 2\beta_1\;{\rm Im}
\int_{{\Bbb R}^d}\bar{\psi}_{-1}\psi_0^2\bar{\psi}_{1}\; d\bx, \qquad t\ge0.\ee
\end{lemma}
For the condensate width, we have the following results.
\begin{lemma}
Suppose $\Psi(\bx,t)$ is the solution of the CGPEs
\eqref{eq:cgpen:sec4}, then we have
\begin{align}\label{eq:dtap91:sec4}
&\ddot{\delta}_{\alpha}(t)=\int_{{\Bbb R}^d}\left[\sum_{j=-1}^1
\left(2|\p_\alpha\psi_j|^2-2\alpha|\psi_j|^2 \p_\alpha V(\bx)\right)+\beta_0
|\Psi|^2+\beta_1|\mathbf{F}(\Psi)|^2\right]\,d\bx,
\quad t\geq 0,\\
\label{eq:dtap92:sec4} &\delta_\alpha(0)=\int_{{\Bbb R}^d}\alpha^2
\left(|\psi_1(\bx,0)|^2+|\psi_0(\bx,0)|^2
+|\psi_{-1}(\bx,0)|^2\right)d\bx, \quad \alpha = x, y, z,\\
\label{eq:dtap93:sec4} &\dot{\delta}_\alpha(0) = 2\sum_{j=-1}^1\int_{{\Bbb
R}^d}\alpha{\rm Im}
\left(\bar{\psi}_j^{(0)}\partial_\alpha\psi_j^0\right)d\bx.
 \end{align}
\end{lemma}
\begin{lemma} Suppose $\Psi(\bx,t)$ is the solution of the CGPEs
\eqref{eq:cgpen:sec4}, $q=0$ and $V(\bx)$ is the harmonic potential in \eqref{eq:hopot:sec2}, then we have

(i) In 1D without nonlinear terms, i.e. $d=1$, $\beta_0=\beta_1=0$
in \eqref{eq:cgpen:sec4}, for any initial data
$\Psi(\bx,0)=\Psi^{(0)}(x)$, we have,
\be \label{eq:cdw61:sec5}
\delta_x(t)=\frac{E(\Psi^{(0)})}{\gamma_x^2}\left[1-\cos(2\gamma_x
t)\right] +\delta_x^{(0)}\cos(2\gamma_x
t)+\frac{\delta_x^\prime(0)}{2\gamma_x}\sin(2\gamma_x t). \ee

(ii) In 2D with a radially symmetric trap, i.e. $d=2$,
$\gamma_x=\gamma_y:=\gamma_r$ and $\beta_1=0$ in \eqref{eq:cgpen:sec5}, for
any initial data $\Psi(\bx,0)=\Psi^{(0)}(x,y)$, we
have, for any $t\ge0$, \be \label{eq:cdw61t:sec4}
\delta_r(t)=\frac{E(\Psi^{(0)})}{\gamma_r^2}\left[1-\cos(2\gamma_r t)\right]
+\delta_r^{(0)}\cos(2\gamma_r t)+\frac{\delta_r^{(1)}(0)}{2\gamma_r}\sin(2\gamma_r
t), \ee where $\delta_r(t)=\delta_x(t)+\delta_y(t)$,
$\delta_r^{(0)}:=\delta_x(0)+\delta_y(0)$ and
$\delta_r^{(1)}:=\delta^\prime_x(0)+\delta_y^\prime(0)$.
\end{lemma}

Let
$\Phi^s:=\Phi^s(\bx)$
be a stationary state of the CGPEs \eqref{eq:cgpen:sec4}, i.e. $\Phi^s$
solves the Euler-Lagrange system \eqref{eq:el:1}-\eqref{eq:el:3}.
If the initial data $\Psi(\bx,0)$  for  the CGPEs \eqref{eq:cgpen:sec4}
 is chosen as a stationary state with its
center-of-mass shifted from the trap center, we can construct an
exact solution of the CGPEs \eqref{eq:cgpen:sec4} with a
harmonic oscillator potential \eqref{eq:hopot:sec2}.
\begin{lemma}
If the initial data $\Psi(\bx,0)$ for the CGPEs \eqref{eq:cgpen:sec4}  is chosen as \be
\label{eq:init5:sec4} \Psi(\bx,0)=\Phi^s(\bx-\bx_0)e^{i({\mathbf{a}}^{(0)}\cdot
\bx+b^{(0)})}, \qquad \bx \in {\Bbb R}^d, \ee where $\bx_0$ is a
given point in $\mathbb{R}^d$,
$\mathbf{a}^{(0)}=(a_1^{(0)},\ldots,a_d^{(0)})^T$ is a given vector in
$\mathbb{R}^d$ and $b^{(0)}$ is a given real number, then the exact
solution of \eqref{eq:cgpen:sec4} with the initial data
\eqref{eq:init5:sec4} satisfies:
\be \label{eq:exacts1:sec4}
\psi_l(\bx,t)=\phi_l^s(\bx-\bx(t))\;e^{-i\mu_l^s t}\;
e^{i(\mathbf{a}(t)\cdot \bx +b(t))}, \quad \bx\in{\Bbb R}^d, \quad t\ge
0,\qquad l=-1,0,1, \ee where for any time $t\ge0$, $\bx(t)$
satisfies the following second-order ODE system: \begin{align}
&\ddot{\bx}(t)+\Lambda \bx(t)=0,\\
\label{govern_eq3} &\bx(0) = \bx_0,  \quad\dot{\bx}(0) = \mathbf{a}^{(0)}.
 \end{align} In addition,
$\mathbf{a}(t)=(a_1(t), \ldots,a_d(t))^T$ and $b(t)$ satisfy \begin{align}
\label{eq:exact3:sec4} &\dot{\mathbf{a}}(t) =-\Lambda\,\bx(t), \quad b^\prime(t)=-\frac{1}{2}|\mathbf{a}(t)|^2-\frac{1}{2}\bx(t)^T\,
\Lambda\,\bx(t),
\quad t>0,
 \end{align}
 with initial data $\mathbf{a}(0)=\mathbf{a}^{(0)}$, $b(0)=b^{(0)}$
 and $\Lambda=\mathrm{diag}(\gamma_x^2)$ in 1D, $\Lambda=\mathrm{diag}(\gamma_x^2,\gamma_y^2)$ in 2D and
 $\Lambda=\mathrm{diag}(\gamma_x^2,\gamma_y^2,\gamma_z^2)$ in 3D.
\end{lemma}
\subsubsection{Numerical methods and results}
To compute the dynamics, we numerically solve the Cauchy problem for the CGPEs \eqref{eq:cgpen:sec4}.
The time splitting technique can be used here, and the basic idea is to divide the evolution of the CGPEs \eqref{eq:cgpen:sec4}
into several subproblems which are much easier to deal with.  In \cite{BaoZhang,Wang2007}, the nonlinear subproblem are not integrated exactly
and are solved by numerical quadratures. Recently, Symes et al. \cite{SMB2016} proposed a time-splitting scheme where the nonlinear subproblem are solved exactly and we will
introduce the procedure below.

After truncating the CGPEs \eqref{eq:cgpen:sec4} (or \eqref{eq:cgpenc:sec4}) onto a bounded domain with homogeneous Dirichlet boundary conditions or periodic boundary conditions,
we solve \eqref{eq:cgpen:sec4} (or \eqref{eq:cgpenc:sec4}) from time $t_n=n\tau$ to $t_{n+1}=t_n+\tau$ through the following subproblems. One first
solves \be \label{eq:ODE1:sec4} i\frac{\p\psi_l}{\p
t}=(-\frac{1}{2}\nabla^2+pl+ql^2)\psi_l, \qquad l=-1,0,1, \ee
for the time step of length $\tau$, followed by solving \be
\label{eq:ODE3:sec4} i\partial_t\Psi=[V(\bx)+\beta_0|\Psi|^2+\beta_1(F_x^n\mathrm{f}_x+F_y^n\mathrm{f}_y+F_z^n\mathrm{f}_z)]\Psi
\ee for the same time step.
Again, \eqref{eq:ODE1:sec4} can be integrated exactly in phase space. For \eqref{eq:ODE3:sec4}, noticing $\mathrm{f}_\alpha$ ($\alpha=x,y,z$) are Hermitian matrices and satisfy the commutator relations $[\mathrm{f}_x,\mathrm{f}_y]=\mathrm{f}_x\mathrm{f}_y-\mathrm{f}_y\mathrm{f}_x=i\mathrm{f}_z$, $[\mathrm{f}_y,\mathrm{f}_z]=i\mathrm{f}_x$ and
$[\mathrm{f}_z,\mathrm{f}_x]=i\mathrm{f}_y$, we find
$\partial_t|\Psi(t,\bx)|^2=0$ ($t\in(t_n,t_{n+1})$) and for $\alpha=x,y,z$,
\begin{align*}
\partial_t(\Psi^\ast f_\alpha\Psi)=&\text{Im}(\Psi^\ast f_\alpha[V(\bx)+\beta_0|\Psi|^2+\beta_1(F_x(\Psi)\mathrm{f}_x+F_y(\Psi)\mathrm{f}_y+F_z(\Psi)\mathrm{f}_z)]\Psi)\\
&-\text{Im}(\Psi^\ast[V(\bx)+\beta_0|\Psi|^2+\beta_1(F_x(\Psi)\mathrm{f}_x+F_y(\Psi)\mathrm{f}_y+F_z(\Psi)\mathrm{f}_z)]^*f_\alpha\Psi)\\
=&2\text{Im} \left(F_x(\Psi)\Psi^\ast[\mathrm{f}_\alpha,\mathrm{f}_x]\Psi+F_y(\Psi)\Psi^\ast[\mathrm{f}_\alpha,\mathrm{f}_y]\Psi
+F_z(\Psi)\Psi^\ast[\mathrm{f}_\alpha,\mathrm{f}_z]\Psi\right)=0,
\end{align*}
which implies that the spin vector components $F_\alpha(\Psi(\bx,t))=F_\alpha(\Psi(\bx,t_n))$ ($t\in(t_n,t_{n+1})$). Now, it is clear that \eqref{eq:ODE3:sec4} becomes a linear
ODE with solution
\be\label{eq:odes1:sec4}
\Psi(\bx,t)=e^{-i(t-t_n)[V(\bx)+\beta_0|\Psi(\bx,t_n)|^2+\beta_1(F_x^n\mathrm{f}_x+F_y^n\mathrm{f}_y+F_z^n\mathrm{f}_z)]}\Psi(\bx,t_n),\quad t\in[t_n,t_{n+1}],
\ee
where the spin vector $\mathbf{F}^n=(F_x^n,F_y^n,F_z^n)^T$ is evaluated using $\Psi(\bx,t_n)$, i.e. $F_\alpha^n=F_\alpha(\Psi(\bx,t_n))$. Denote the matrix $\mathbf{S}^n=F_x^n\mathrm{f}_x+F_y^n\mathrm{f}_y+F_z^n\mathrm{f}_z$ with detailed form as
\begin{equation}\label{eq:Sma:sec4}
\mathbf{S}^n=\begin{pmatrix}
 F_z^n& \frac{1}{\sqrt{2}}F_-^n&0\\
\frac{1}{\sqrt{2}}F_+^n&0&\frac{1}{\sqrt{2}}F_-^n\\
0&\frac{1}{\sqrt{2}}F_+^n& -F_z^n
\end{pmatrix},\quad F_+^n=\overline{F}_-^n=F_x^n+iF_y^n,
\end{equation}
and then $\widetilde{\Psi}(\bx,t)=e^{i(t-t_n)(V(\bx)+\beta_0|\Psi(\bx,t_n)|^2)}\Psi(\bx, t)$ ($t\in[t_n,t_{n+1}]$) satisfies
\be
\widetilde{\Psi}(\bx,t)=e^{-i(t-t_n)\beta_1\mathbf{S}^n}\widetilde{\Psi}(\bx,t_n),\quad t\in[t_n,t_{n+1}].
\ee
Now $\mathbf{S}^n$ has eigenvalues $0$ and $\pm|\mathbf{F}^n|$, and the eigenvector corresponding to 0 is
\be
\mathbf{e}^n=\frac{1}{|\mathbf{F}^n|}\left(-\frac{1}{\sqrt{2}}F_-^n,F_z^n,\frac{1}{\sqrt{2}}F_+^n\right)^T.
\ee
By computation, it is easy to verify $(\mathbf{e}^n)^*\widetilde{\Psi}(\bx,t)=\left((\mathbf{e}^n)^*\Psi^n\right)=0$ ($t\in[t_n,t_{n+1}]$).
Therefore, we have $(\mathbf{S}^n)^2\widetilde{\Psi}=|\mathbf{F}^n|^2\widetilde{\Psi}$  and the exponential of $\mathbf{S}^n$ can be computed as
\be
\widetilde{\Psi}(\bx,t)=\cos(\beta_1(t-t_n)|\mathbf{F}^n|)\Psi(\bx,t_n)-i\frac{\sin(\beta_1(t-t_n)|\mathbf{F}^n|)}{|\mathbf{F}^n|}\mathbf{S}^n
\Psi(\bx,t_n),
\ee
where we used the fact that $\widetilde{\Psi}(\bx,t_n)=\Psi(\bx,t_n)$.
The ODE \eqref{eq:ODE3:sec4} can be solved exactly as
\be\label{eq:solode:sec4}
\Psi(\bx,t_{n+1})=e^{-i\tau(V(\bx)+\beta_0|\Psi(\bx,t_n)|^2)}
\left(\cos(\beta_1\tau|\mathbf{F}^n|)\Psi(\bx,t_n)-i\frac{\sin(\beta_1\tau|\mathbf{F}^n|)}{|\mathbf{F}^n|}\mathbf{S}^n\Psi(\bx,t_n)\right).
\ee
Let $\Psi^n=(\psi_1^n,\psi_0^n,\psi_{-1}^n)^T$ be the numerical approximation of $\Psi(\bx,t)$ at $t_n=n\tau$.
Combining the subproblems \eqref{eq:ODE1:sec4} and \eqref{eq:ODE3:sec4} via Strang Splitting, we obtain a second order semi-discretization in time as:
from $t_n$ to $t_{n+1}$,
\begin{align}
\label{eq:tssps1:sec4}&\psi_l^{(1)}
=e^{i\tau(\frac{1}{2}\nabla^2-pl-ql^2)/2}\psi_l^{n}, \qquad l=-1,0,1, \, \\
\label{eq:tssps2:sec4}&\Psi^{(2)}=e^{-i\tau(V(\bx)+\beta_0|\Psi^{(1)}|^2)}
\left(\cos(\beta_1\tau|\mathbf{F}^{(1)}|)\Psi^{(1)}-i\frac{\sin(\beta_1\tau|\mathbf{F}^{(1)}|)}{|\mathbf{F}^{(1)}|}\mathbf{S}^{(1)}\Psi^{(1)}\right),\\
\label{eq:tssps3:sec4}&\psi^{n+1}_l
=e^{i\tau(\frac{1}{2}\nabla^2-pl-ql^2)/2}\psi_l^{(2)}, \qquad l=-1,0,1,
\end{align}
where the spin vector $\mathbf{F}^{(1)}=\mathbf{F}(\Psi^{(1)})=(F_x^{(1)},F_y^{(1)},F_z^{(1)})^T$ and $\mathbf{S}^{(1)}$ is given in \eqref{eq:Sma:sec4} with elements
computed from $\mathbf{F}^{(1)}$.
The above splitting procedure \eqref{eq:tssps1:sec4}-\eqref{eq:tssps3:sec4} can be implemented with Fourier/sine/cosine spectral method for periodic/homogeneou Dirichlet/ homogeneous Neumann boundary conditions for spatial discretizations, and we refer to \cite{Baocai2013} for detail.

\bigskip

\begin{figure}[htb]
\centerline{ 
\includegraphics[height=5cm,width=7cm]{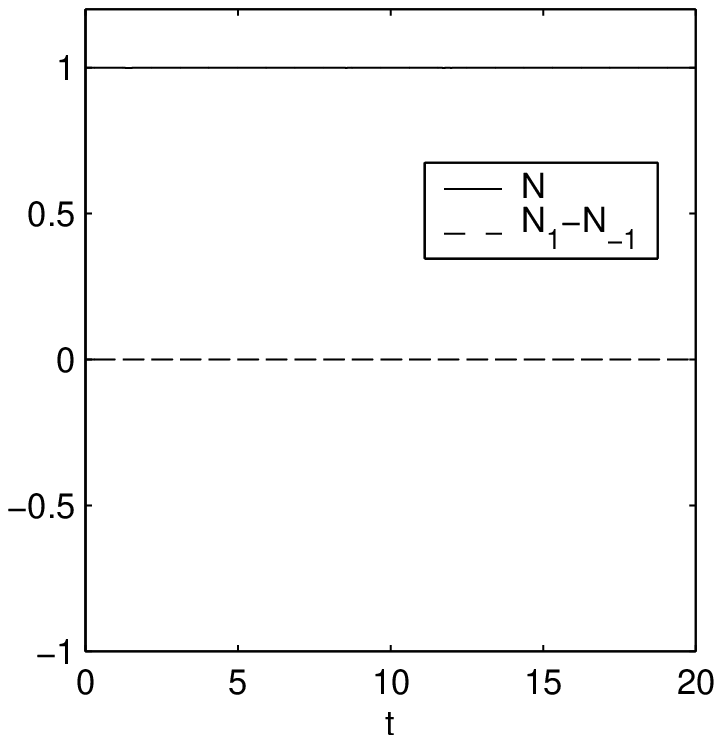} \quad 
\includegraphics[height=5cm,width=7cm]{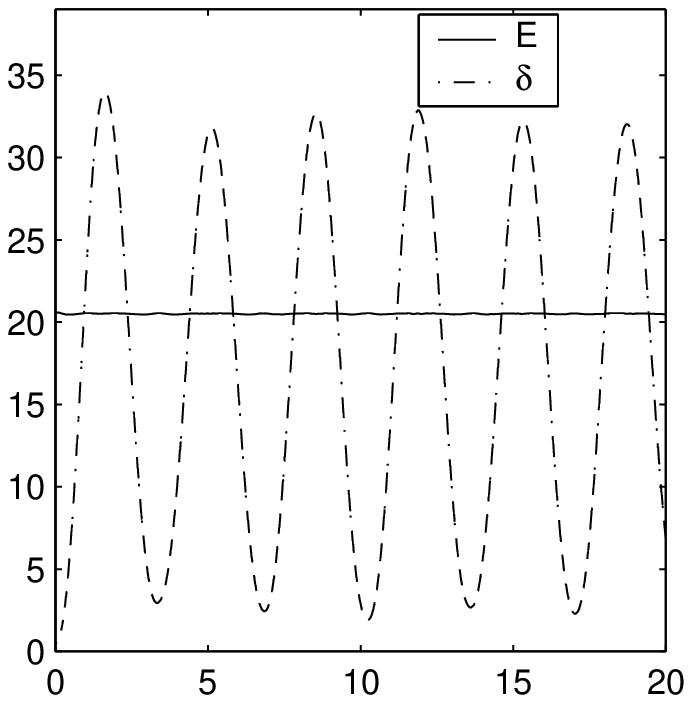} }\caption{
Time evolutions of the total normalization $N(t)$
and the magnetization $M(t)=N_1-N_{-1}$ (left); and the energy
$E$ and condensate width
$\delta:=\delta_x$ (right) in Example 4.2. } \label{fig:dy:sec4}
\end{figure}

{\it Example 4.2}. To show the dynamics of spin-1 BEC,
we take $d=1$, $p=q=0$, $\beta_0=100$, $\beta_1=2$ and
$V(x)=\frac{1}{2}x^2$ in the CGPEs \eqref{eq:cgpen:sec4}. The initial data is taken as
\begin{equation*}
\psi_1(x,0)=\psi_{-1}(x,0)=\frac{\sqrt{0.05}}{\pi^{\frac{1}{4}}}
e^{\frac{-x^2}{2}}, \quad \psi_{0}(x,0)=\frac{\sqrt{0.9}}{\pi^{\frac{1}{4}}}
e^{\frac{-x^2}{2}},\qquad x\in {\mathbb R}.
\end{equation*}
The problem is solved on a bounded domain $[-10,10]$ with $h=5/128$ and $\tau=10^{-3}$ by the TSSP method \cite{Baocai2013}.
 Fig. \ref{fig:dy:sec4}  shows that the total
normalization $N=N_1+N_0+N_{-1}$, the magnetization $M(t)=N_1-N_{-1}$ and the energy $E:=E(t)$ are conserved very well.

\section{Spin-2 BEC}
\setcounter{equation}{0}
\setcounter{figure}{0}
In recent experiments, spin-2 BEC has been realized in $F=2$ $^{87}$Rb and $^{23}$Na systems. The spin-2 condensates exhibited rich phenomenon, and  many properties have been investigated including  magnetic phases, charged vortices, and the phase separation, etc. Below, we briefly discuss the mathematical model, mathematical theories and numerical methods.
\subsection{The mathematical model}
At temperature $T$ much smaller than the critical temperature $T_c$ \cite{Ueda}, a spin-2 BEC  can be well described by the spin-2 vector wave function
$\Psi:=\Psi(\bx,t)=(\psi_2,\psi_1,\psi_0,\psi_{-1},\psi_{-2})^T$ ($\psi_l$ for $m_F=l$ state, $l=-2,-1,0,1,2$) governed by the CGPEs in the compact form as \cite{Ho1,Ohmi,Ueda,StamperUeda}:
\be\label{eq:cgpe:sec5}
i\hbar\partial_t\Psi=[\tilde{H}+c_0 \rho-\tilde{p} \mathrm{f}_z+\tilde{q} \mathrm{f}_z^2+c_1 \mathbf{F}\cdot\mathbf{f}]\Psi+c_2A_{00}\mathbf{A}\overline{\Psi},
\ee
where as described in section \ref{sec:mf:spin-1}, $\tilde{H}=-\frac{\hbar^2}{2m}\nabla^2+\tilde{V}(\bx)$ is the single particle Hamiltonian, $\tilde{p}$ and $\tilde{q}$ are the linear and quadratic Zeeman energy shifts, respectively.
$\rho=|\Psi|^2=\sum_{l=-2}^2|\psi_l|^2$ is the total density. $c_0=\frac{4g_2+3g_4}{7}$, $c_1=\frac{g_4-g_2}{7}$ and $c_2=\frac{7g_0-10g_2+3g_4}{7}$  characterizes the spin-independent
interaction, spin-exchange interaction and spin-singlet interaction, respectively,
with $g_k=\frac{4\pi\hbar^2}{m}a_k$ ($k=0,2,4$) and $a_k$ being the s-wave scattering length for scattering channel of total hyperfine spin $k$.
The spin-2 matrices $\mathbf{f}=(\mathrm{f}_x,\mathrm{f}_y,\mathrm{f}_z)^T$ are given as
\be
\mathrm{f}_x=\begin{pmatrix}0&1&0&0&0\\
1&0&\sqrt{\frac{3}{2}}&0&0\\
0&\sqrt{\frac{3}{2}}&0&\sqrt{\frac{3}{2}}&0\\
0&0&\sqrt{\frac{3}{2}}&0&1\\
0&0&0&1&0\end{pmatrix},\quad \mathrm{f}_y=i\begin{pmatrix}0&-1&0&0&0\\
1&0&-\sqrt{\frac{3}{2}}&0&0\\
0&\sqrt{\frac{3}{2}}&0&-\sqrt{\frac{3}{2}}&0\\
0&0&\sqrt{\frac{3}{2}}&0&-1\\
0&0&0&1&0\end{pmatrix}
\ee
and
\be
\mathrm{f}_z=\text{diag}(2,1,0,-1,-2),
\ee
and the spin vector $\mathbf{F}=\mathbf{F}(\Psi)=(F_x,F_y,F_z)^T$ with $F_\alpha=\Psi^*\mathrm{f}_\alpha\Psi$ ($\alpha=x,y,z$) is given as
\be\label{eq:F:sec5}\begin{split}
&F_x=\overline{\psi}_2\psi_1+\overline{\psi}_1\psi_2+\overline{\psi}_{-2}\psi_{-1}+\overline{\psi}_{-1}\psi_{-2}+\frac{\sqrt{6}}{2}(\overline{\psi}_1\psi_0+
\overline{\psi}_0\psi_1+\overline{\psi}_0\psi_{-1}+\overline{\psi}_{-1}\psi_0),\\
&F_y=i\left[\overline{\psi}_1\psi_2-\overline{\psi}_2\psi_1+\overline{\psi}_{-2}\psi_{-1}-\overline{\psi}_{-1}\psi_{-2}+\frac{\sqrt{6}}{2}(
\overline{\psi}_0\psi_1-\overline{\psi}_1\psi_0+\overline{\psi}_{-1}\psi_0-\overline{\psi}_0\psi_{-1})\right],\\
&F_z=2|\psi_2|^2+|\psi_1|^2-|\psi_{-1}|^2-2|\psi_{-2}|^2,
\end{split}
\ee
with $\mathbf{F}\cdot\mathbf{f}=F_x\mathrm{f}_x+F_y\mathrm{f}_y+F_z\mathrm{f}_z$. The matric $\mathbf{A}$ is
\be
\mathbf{A}=\frac{1}{\sqrt{5}}\begin{pmatrix}0&0&0&0&1\\
0&0&0&-1&0\\
0&0&1&0&0\\
0&-1&0&0&0\\
1&0&0&0&0\end{pmatrix}
\ee
and $A_{00}:=A_{00}(\Psi)=\Psi^T \mathbf{A}\Psi$ with
\be\label{eq:A:sec5}
A_{00}=\frac{1}{\sqrt{5}}(2\psi_2\psi_{-2}-2\psi_{1}\psi_{-1}+\psi_0^2).
\ee
Here, we adopt  notations similar to the spin-1 case, while there should be no confusion about such abuse of notations.

 Similar to section \ref{sec:mf:spin-1}, introducing the  scaling:
$t\to t/\omega_s$ with $\omega_s=\min\{\omega_x,\omega_y,\omega_z\}$, $\bx\to \bx/ x_s $ with
$x_s=\sqrt{\frac{\hbar}{m\omega_s}}$, $\psi_l\to\psi_lx_s^{3/2}/\sqrt{N}$ ($l=-2,-1,0,1,2$) with
$N$ being the total number of particles in the system,  after a proper dimension reduction process in 1D and 2D, the dimensionless CGPEs for spin-2 BEC are obtained in
$d$ dimensions ($d=1,2,3$) for $\Psi=(\psi_2,\psi_1,\psi_0,\psi_{-1},\psi_{-2})^T$ as
\be\label{eq:cgpen:sec5}
i\partial_t\Psi=[H+\beta_0 \rho-p \mathrm{f}_z+q \mathrm{f}_z^2+\beta_1 \mathbf{F}\cdot\mathbf{f}]\Psi+\beta_2A_{00}\mathbf{A}\overline{\Psi},
\ee
where the Hamiltonian $H$, linear and quadratic Zeeman parameters $p$ and $q$, $\rho=|\Psi|^2$ are the same as the spin-1 case in section \ref{sec:mf:spin-1}, and interaction parameters
$\beta_0 =\frac{4\pi N(4a_2+3a_4)}{7x_s}$, $\beta_1 =\frac{4\pi N(a_4-a_2)}{7x_s}$ and
$\beta_2 =\frac{4\pi N(7a_0-10a_2+3a_4)}{7x_s}$ in 3D;
$\beta_0 =\frac{4\pi N(4a_2+3a_4)}{7x_s}\frac{\sqrt{\gamma_z}}{\sqrt{2\pi}}$, $\beta_1 =\frac{4\pi N(a_4-a_2)}{7x_s}\frac{\sqrt{\gamma_z}}{\sqrt{2\pi}}$ and
$\beta_2 =\frac{4\pi N(7a_0-10a_2+3a_4)}{7x_s}\frac{\sqrt{\gamma_z}}{\sqrt{2\pi}}$;
$\beta_2 =\frac{4\pi N(7a_0-10a_2+3a_4)}{7x_s}\frac{\sqrt{\gamma_z\gamma_y}}{2\pi}$, $\beta_0 =\frac{4\pi N(4a_2+3a_4)}{7x_s}\frac{\sqrt{\gamma_z\gamma_y}}{2\pi}$ and$\beta_1 =\frac{4\pi N(a_4-a_2)}{7x_s}\frac{\sqrt{\gamma_z\gamma_y}}{2\pi}$
in 1D.

The CGPEs \eqref{eq:cgpen:sec5} conserve the following three important quantities, i.e.
the {\sl mass} (or {\sl normalization})
\be\label{eq:norm:sec5}
N(\Psi(\cdot,t)):=\|\Psi(\cdot,t)\|^2=\int_{\Bbb R^d}\sum_{l=-2}^2|\psi_l(\bx,t)|^2\,d\bx=N(\Psi(\cdot,0))=1,
\ee
the {\sl magnetization} (with $M\in[-2,2]$)
\be\label{eq:mag:sec5}
M(\Psi(\cdot,t)):=\int_{\Bbb R^d}\sum_{l=-2}^2l|\psi_l(\bx,t)|^2\,d\bx
=M(\Psi(\cdot,0))=M,
\ee
and the {\sl energy per particle}
\begin{align}
E(\Psi(\cdot,t))&=\int_{\mathbb{R}^d
}\bigg\{\sum_{l=-2}^{2} \left(\frac{1}{2}|\nabla
\psi_l|^2+(V(\bx)-pl+ql^2)|\psi_l|^2\right)
+\frac{\beta_0}{2}|\Psi|^4+\frac{\beta_1}{2}|\mathbf{F}|^2
+\frac{\beta_2}{2}|A_{00}|^2\bigg\}\; d\bx \nonumber\\
&\equiv E(\Psi(\cdot,0)).\label{eq:energy:sec5}
\end{align}
In practice, introducing $\psi_l\to e^{-ilpt}\psi_l$ in the CGPEs \eqref{eq:cgpen:sec5}, the system is unchanged and it is thus
reasonable to assume the linear Zeeman term $p=0$ in the subsequent discussion. On the other hand, it is easy to observe from \eqref{eq:energy:sec5} that
the linear Zeeman term does not contribute to the energy due to the magnetization conservation \eqref{eq:mag:sec5}.

\subsection{Ground states}
The ground state $\Phi_g(\bx)$ of the spin-2 BEC described by the CGPEs \eqref{eq:cgpen:sec5}
can be obtained from the minimization of
the energy functional \eqref{eq:energy:sec5} subject to the conservation of total mass
and magnetization:
\begin{quote}
  Find $\left(\Phi_g \in S_M\right)$ such that
\end{quote}
  \begin{equation}\label{eq:minimize:sec5}
    E_g := E\left(\Phi_g\right) = \min_{\Phi \in S_M}
    E\left(\Phi\right),
  \end{equation}
\noindent where the nonconvex set $S_M$ is defined as
\be \label{eq:cons:sec5}
S_M=\left\{\Phi=(\phi_2,\phi_1,\phi_0,\phi_{-1},\phi_{-2})^T\ |\ \|\Phi\|=1, \
\int_{{\Bbb R}^d} \sum_{l=-2}^2l|\phi_1(\bx)|^2\,d\bx=M, \ E(\Phi)<\infty\right\}. \ee This is
a nonconvex minimization problem and the  Euler-Lagrange equations associated to the minimization problem
\eqref{eq:minimize:sec4} reads:
\begin{align}
(\mu\pm2\lambda)\;\phi_{\pm2}(\bx)&=\left[H\mp 2p+4q+\beta_0\rho\pm2\beta_1F_z\right]\phi_{\pm2}+\,\beta_1\,F_{\mp}\phi_{\pm1}
+\frac{\beta_2}{\sqrt{5}}A_{00}\overline{\phi}_{\mp2}, \label{eq:el:1:sec5} \\
(\mu\pm\lambda)\;\phi_{\pm1}(\bx)&=\left[H\mp p+q+\beta_0\rho\pm\beta_1F_z\right]\phi_{\pm1}+\,\beta_1\left(\frac{\sqrt{6}}{2}
F_{\mp}\phi_{0}+F_{\pm}\phi_{\pm2}\right)
-\frac{\beta_2}{\sqrt{5}}A_{00}\overline{\phi}_{\mp1},\nonumber \\
\mu\;\phi_{0}(\bx)&=\left[H+\beta_0\rho\right]\phi_{0}+\frac{\sqrt{6}}{2}\beta_1\left(F_{+}
\phi_{1}+F_{-}\phi_{-1}\right)+\frac{\beta_2}{\sqrt{5}}A_{00}
\overline{\phi}_{0}, \label{eq:el:3:sec5}
\end{align}
where $H=-\fl{1}{2}\nabla^2
+V(\bx)$,  $F_+=\overline{F}_-=F_x+iF_y$.
Here $\mu$ and $\lambda$ are the Lagrange multipliers (or chemical
potentials) corresponding to the normalization constraint \eqref{eq:norm:sec5} and the magnetization constraint \eqref{eq:mag:sec5},
respectively.
\subsubsection{Mathematical theories}
We collect the existence and uniqueness results on the ground state \eqref{eq:minimize:sec5} below.

\begin{theorem}[existence and uniqueness \cite{Cai}]\label{thm:gs:sec5} Suppose $\lim\limits_{|\bx|\to\infty}V(\bx)=+\infty$, there exists a ground state $\Phi_g=(\phi_2^g,\phi_1^g,\phi_0^g,\phi_{-1}^g,\phi_{-2}^g)^T\in S_M$ of \eqref{eq:minimize:sec5} of the spin-2 BEC
governed by the CGPEs \eqref{eq:cgpen:sec5}, if one of the following conditions hold
\begin{enumerate}\renewcommand{\labelenumi}{(\roman{enumi})}
\item $d=1$;
\item $d=2$, if $M=\pm2$ and $\beta_0+4\beta_1>-C_b$; or $M\in(-2,2)$, $\beta_0+4\beta_1> -C_b$ with $\frac{\beta_2}{20}>\beta_1$ and $\beta_1<0$; or
$M\in(-2,2)$, $\beta_0+4\beta_1> -4C_b/(2+|M|)$, $\beta_0\ge-\frac{C_b^2+4C_b\beta_{1}+\frac{4-M^2}{100}\beta_2(20\beta_1-\beta_2)}{C_b+\frac{4-M^2}{20}(20\beta_1-\beta_2)}$ with $\beta_2<0$ and $\frac{\beta_2}{20}\leq \beta_1$; or $M\in(-2,2)$, $\beta_{0}+4\beta_{1}>-4C_b/(2+|M|)$, $\beta_{0}\ge-\frac{C_b^2+4\beta_1C_b}{\beta_1(4-M^2)+C_b}$ with $\beta_{1}\ge0$ and $\beta_2\ge0$;
\item $d=3$, $M=\pm2$ and $\beta_0+4\beta_1\ge0$; or $M\in(-2,2)$  $\beta_0+4\beta_1\ge0$ with $\frac{\beta_2}{20}>\beta_1$ and $\beta_1<0$; or
$M\in(-2,2)$, $\beta_0+\frac{\beta_2}{5}\ge0$ with $\beta_2<0$ and $\frac{\beta_2}{20}\leq \beta_1$; or
$M\in(-2,2)$, $\beta_{0}\ge0$, $\beta_{1}\ge0$ and $\beta_2\ge0$.
\end{enumerate}
In particular, $(e^{i(\theta_1+2\theta_{2})}\phi^g_2,e^{i(\theta_1+\theta_2)}\phi_1^g, e^{i\theta_1}\phi_{0}^g,e^{i(\theta_1-\theta_2)}\phi_{-1}^g, e^{i(\theta_1-2\theta_{2})}\phi_{-2}^g)^T\in S_M$  with real constants $\theta_1$ and $\theta_2$  is
also a ground state of \eqref{eq:minimize:sec5}.  We could obtain uniqueness of the ground state under the constant phase factors satisfying the aforementioned conditions in the following cases.
\begin{enumerate}\renewcommand{\labelenumi}{(\roman{enumi})$^\prime$}
\item  $M=\pm2$, $\beta_0+4\beta_1\ge0$, the ground state can be chosen as $(|\phi_2^g|,0,0,0,0)^T$ ($M=2$) or $(0,0,0,0,|\phi_{-2}^g|)^T$ ($M=-2$), and such special form is unique.
\item  $M\in(-2,2)$ and $q=0$, for the ferromagnetic interactions, i.e., $\beta_{1}<0$ and $\beta_1\leq\frac{\beta_2}{20}$. Under the above conditions for the existence,  the ground state can be chosen as $\vec{\alpha}\phi_g\in S_M$ with $\vec{\alpha}=(\alpha_2,\alpha_1,\alpha_0,\alpha_{-1},\alpha_{-2})^T$, $\alpha_2=\frac{(2+M)^2}{16}$, $\alpha_1=\frac{(2+M)\sqrt{4-M^2}}{8}$, $\alpha_0=\frac{\sqrt{6}(4-M^2)}{16}$, $\alpha_{-1}=\frac{(2-M)\sqrt{4-M^2}}{8}$ and $\alpha_{-2}=\frac{(2-M)^2}{16}$, where $\phi_g$ is a positive function satisfying $\|\phi_g\|=1$.
    If $\beta_0+4\beta_1\ge0$, $\phi_g$ is unique.
 \item   $M\in(-2,2)$ and $q=0$, for the nematic interactions, i.e. $\beta_2<0$ and $\beta_1>\frac{\beta_2}{20}$. Under the above conditions for the existence,
 when $M\neq0$, the ground state (real valued) can be chosen as $(|\phi_2^g|,0,0,0,|\phi_{-2}^g|)^T\in S_M$ and such ground state is unique if $\beta_0+\frac{\beta_1}{5}\ge0$. When $M=0$,
 the  ground state (real valued) can be chosen as $(\alpha_2,\alpha_1,\alpha_0,-\alpha_{1},\alpha_{2})^T\phi_g\in S_M$ with $2\alpha_2^2+2\alpha_1^2+\alpha_0^2=1$ ($\alpha_l\in\mathbb{R}$, $l=0,1,2$) or $(\alpha_2,\alpha_1,0,\alpha_{1},-\alpha_{2})^T\phi_g\in S_M$ if $\alpha_0=0$. $\phi_g$ is a positive function satisfying $\|\phi_g\|=1$ and
 is unique if $\beta_0+\frac{\beta_1}{5}\ge0$.
 \item $M\in(-2,2)$ and $q=0$, for the cyclic interactions, i.e. $\beta_2\ge0$ and $\beta_1\ge0$. Under the above conditions for the existence, if $M=0$, the ground state (real valued) can be chosen as $(\alpha_2,\alpha_1,\alpha_0,\alpha_{-1},\alpha_{-2})^T\phi_g\in S_M$ ($\alpha_l\in\mathbb{R}$, $l=-2,\ldots,2$) such that with $\sum_l|\alpha_l|^2=1$, $\sum_ll|\alpha|_l^2=0$, $\alpha_0^2-2\alpha_{1}\alpha_{-1}+2\alpha_2\alpha_{-2}=0$ and $\alpha_1\alpha_2+\alpha_{-2}\alpha_{-1}+\frac{\sqrt{6}}{2}(\alpha_0\alpha_1+\alpha_0\alpha_{-1})=0$. $\phi_g$ is a positive function satisfying $\|\phi_g\|=1$ and
 is unique if $\beta_0\ge0$.
\end{enumerate}
On the other hand, there exists no ground state of \eqref{eq:minimize:sec5} if one of the following conditions holds, i.e. $\inf_{\Phi\in S_M}E(\Phi)=-\infty$
\begin{enumerate}\renewcommand{\labelenumi}{(\roman{enumi})$^{\prime\prime}$}
\item $d=2$,  $M=\pm2$ and $\beta_0+4\beta_1\leq-C_b$; or $M\in(-2,2)$, $\beta_0+4\beta_1\leq -C_b$ with $\frac{\beta_2}{20}\leq\beta_1$ and $\beta_1<0$; or
$M\in(-2,2)$, $\beta_0+4\beta_1\leq-4C_b/(2+|M|)$ or $\beta_0<-M^2\beta_1-\frac{4-M^2}{20}\beta_2-C_b$, with $\beta_2<0$ and $\frac{\beta_2}{20}\leq \beta_1$; or $M\in(-2,2)$, $\beta_{0}+4\beta_{1}\leq-4C_b/(2+|M|)$ or $\beta_{0}<-M^2\beta_1-C_b$ with $\beta_{1}\ge0$ and $\beta_2\ge0$.
\item $d=3$, $M=\pm2$ and $\beta_0+4\beta_1<0$; or $M\in(-2,2)$  $\beta_0+4\beta_1<0$ with $\frac{\beta_2}{20}>\beta_1$ and $\beta_1<0$; or
$M\in(-2,2)$, $\beta_0+\frac{\beta_2}{5}<0$ with $\beta_2<0$ and $\frac{\beta_2}{20}\leq \beta_1$; or
$M\in(-2,2)$, $\beta_{0}<0$, $\beta_{1}\ge0$ and $\beta_2\ge0$.
\end{enumerate}
\end{theorem}
As in section 4 for the spin-1 case, most of the results in Theorem \ref{thm:gs:sec5} can be derived from the following observations when the quadratic Zeeman term is absent in the CGPEs \eqref{eq:cgpen:sec5}, i.e. $q=0$.
Firstly, when $q=0$ and $M\in(-2,2)$,  for the ferromagnetic interactions $\beta_{1}<0$ and $\beta_1\leq\frac{\beta_2}{20}$, we have the single mode approximation (SMA), i.e. each component of the ground state $\Phi_g$ is identical up to a constant factor.
\begin{theorem}[single mode approximation \cite{Cai}]\label{thm:sma:sec5}  Suppose $\lim\limits_{|\bx|\to\infty}V(\bx)=+\infty$, $q=0$, $M\in(-2,2)$, $\beta_{1}<0$, $\beta_1\leq\frac{\beta_2}{20}$ and the existence conditions in Theorem \ref{thm:gs:sec5} hold, the ground state $\Phi_g=(\phi_{2}^g,\phi_{1}^g, \phi_0^g,\phi_{-1}^g,\phi_{-2}^g)^T\in S_M$
satisfies $\phi_l^g=e^{i\theta_1+il\theta_2}\alpha_l\phi_g$ with $\alpha_2=\frac{(2+M)^2}{16}$, $\alpha_1=\frac{(2+M)\sqrt{4-M^2}}{8}$, $\alpha_0=\frac{\sqrt{6}(4-M^2)}{16}$, $\alpha_{-1}=\frac{(2-M)\sqrt{4-M^2}}{8}$ and $\alpha_{-2}=\frac{(2-M)^2}{16}$ and $\theta_{1},\theta_2\in\mathbb{R}$, and $\phi_g$ is the unique positive minimizer of the energy functional
\be
\tilde{E}_{\rm SMA}(\phi)=\int_{\mathbb{R}^d}\left[\frac{1}{2}|\nabla\phi|^2+V(\bx)|\phi|^2+\frac{\beta_0+4\beta_1}{2}|\phi|^4\right]\,d\bx,
\ee
under the constraint $\|\phi\|=1$.
\end{theorem}
Secondly, when $q=0$ and $M\in(-2,2)$,  for the anti-ferromagnetic interactions $\beta_2<0$ and $\beta_1\ge\frac{\beta_2}{20}$, we have  similar simplification for  the ground state $\Phi_g$.
\begin{theorem}[two-component case \cite{Cai}]\label{thm:van:sec5} Suppose $\lim\limits_{|\bx|\to\infty}V(\bx)=+\infty$, $q\leq0$, $M\in(-2,2)$ and $M^2+q^2\neq0$, $\beta_{1}<0$, $\beta_1\ge\frac{\beta_2}{20}$ and the existence conditions in Theorem \ref{thm:gs:sec5} hold, the ground state $\Phi_g=(\phi_{2}^g,\phi_1^g,\phi_{0}^g,\phi_{-1}^g,\phi_{-2}^g)^T\in S_M$
satisfies $\phi_0^g=\phi_1^g=\phi_{-1}^g=0$, and $\tilde{\Phi}_g=(\phi_{2}^g,\phi_{-2}^g)^T$ is a  minimizer of the pseudo spin-1/2 system given in section \ref{sec:2} described by \eqref{eq:minim2:sec2} with $\delta=0$, $\nu=\frac{2+M}{4}$ and $\beta_{\uparrow\uparrow}=\beta_{\downarrow\downarrow}=\beta_{0}+4\beta_{1}$, $\beta_{\uparrow\downarrow}=\beta_0-4\beta_1+\frac{2}{5}\beta_2$.

If $M=q=0$, the ground state satisfies
 $\phi_l^g=e^{i\theta_1+il\theta_2}\alpha_l\phi_g$ ($\theta_{1},\theta_2\in\mathbb{R}$) where $\alpha_1=-\alpha_{-1}$, $\alpha_2=\alpha_{-2}$
or $\alpha_1=\alpha_{-1}$, $\alpha_2=-\alpha_{-2}$ if $\alpha_0=0$,
 with $2\alpha_2^2+2\alpha_1^2+\alpha_0^2=1$. $\phi_g$ is the unique positive minimizer of the energy functional
 \be
 E_{\rm c}(\phi)=\int_{\mathbb{R}^d}\left[\frac{1}{2}|\nabla\phi|^2+
 V(\bx)|\phi|^2+\frac{\beta_0}{2}|\phi|^4\right]\,d\bx,
 \ee
 under the constraint $\|\phi\|=1$.
\end{theorem}
For the cyclic interactions $\beta_2\ge0$ and $\beta_1\ge0$, the classification of the ground states become more complicated and we leave the discussion somewhere else.
\subsubsection{Numerical methods and results}
To  compute the ground state \eqref{eq:minimize:sec5}, we generalize the GFDN method in section \ref{numeric-gs:sec2} for  $M\in(-2,2)$.
We start with the following CNGF for $\Phi=(\phi_2,\phi_1,\phi_0,\phi_{-1},\phi_{-2})^T$ \cite{Wang2014}
\be\label{eq:cngf:sec5}
\partial_t\Phi=-[H+\beta_0 \rho-p \mathrm{f}_z+q \mathrm{f}_z^2+\beta_1 \mathbf{F}\cdot\mathbf{f}]\Phi-\beta_2A_{00}\mathbf{A}\Phi
+\mu_{\Phi}(t)\Phi+\mu_{\Phi}(t)\mathrm{f}_z\Phi,
\ee
where  $\mathbf{F}=\mathbf{F}(\Phi)$ is given in \eqref{eq:F:sec5} and $A_{00}=A_{00}(\Phi)$ is defined in \eqref{eq:A:sec5},
$\mu_{\Phi}(t)$ and $\lambda_{\Phi}(t)$ are the Lagrange multipliers  to make the flow preserve the mass constraint \eqref{eq:norm:sec5} and magnetization constraint
\eqref{eq:mag:sec5}, respectively. It is not difficult to see that if initially $\Phi(\bx,0)$ satisfies \eqref{eq:norm:sec5} and \eqref{eq:mag:sec5}, the continuous flow
\eqref{eq:cngf:sec5} is globally wellposed under appropriate assumptions. Moreover, the CNGF \eqref{eq:cngf:sec5} is energy diminishing.

The GFDN method to compute the ground state \eqref{eq:minimize:sec5} then can be regarded as applying a first order splitting algorithm to discretize the above CNGF.
We present a semi-discretization in time as follows. Let $\Phi^n=(\phi_2^n,\phi_1^n,\phi_0^n,\phi_{-1}^n,\phi_{-2}^n)^T$, from $t_n$ to $t_{n+1}$, we first solve
\be\label{eq:dfgn1:sec5}
\frac{\Phi^{(1)}-\Phi^n}{\tau}=-[H+\beta_0 \rho^n-p \mathrm{f}_z+q \mathrm{f}_z^2+\beta_1 \mathbf{F}^n\cdot\mathbf{f}]\Phi^{(1)}-\beta_2A_{00}^n\mathbf{A}\overline{\Phi^{(1)}},
\ee
where $\rho^n=|\Phi^n|^2$, $\mathbf{F}^n=\mathbf{F}(\Phi^n)$ and $A_{00}^n=A_{00}(\Phi^n)$. Then we have the projection step for $\Phi^{(1)}=(\phi_2^{(1)},\phi_1^{(1)},\phi_0^{(1)},\phi_{-1}^{(1)},\phi_{-2}^{(1)})^T$
\be\label{eq:dfgn2:sec5}
\Phi^{n+1}=\text{diag}(\alpha_2,\alpha_1,\alpha_0,\alpha_{-1},\alpha_{-2}) \Phi^{(1)},
\ee
and the projection constants $\alpha_l$ ($-2\le l\le 2$) are chosen such that $\Phi^{n+1}$ satisfies the constraints \eqref{eq:norm:sec5} and \eqref{eq:mag:sec5}.

Similar to the spin-1 BEC, here there are five projection constants to be determined, i.e. $\sigma_l^{n}$ ($-2\le l\le 2$) in \eqref{eq:dfgn2:sec5}, and there are only two equations to fix them, we need to find three more conditions
so that the five projection constants are uniquely determined.
Again, in fact, the projection step \eqref{eq:dfgn2:sec5} can be regarded as
an approximation of the  ODE $\partial_t\Phi=(\mu(t)+\lambda(t)\mathrm{f}_z)\Phi$ whose solution can be written as
$\Phi(t)=e^{\int_{t_n}^t(\mu(s)+\lambda(s)\mathrm{f}_z)\,ds}\Phi(t_n)$.
From this observation, three additional equations are proposed
for determining the projection constants in \eqref{eq:dfgn2:sec5} as \cite{BaoTY}
\be
\alpha_2\alpha_{-2} =\alpha_0^2, \qquad
\alpha_1\alpha_{-1} =\alpha_0^2, \qquad
\alpha_2\alpha_{0} =\alpha_1^2.
\ee
The above equations suggest that the projection constants can be assumed as $\alpha_l=\tilde{c}_0\tilde{c}_1^{l}$ ($l=-2,\ldots,2$) with $\tilde{c}_0,\tilde{c}_1>0$ \cite{BaoTY} satisfying
\be\label{projspin21}
\begin{split}
&\tilde{c}_0^2\left(\tilde{c}_1^4\|\phi_2^{(1)}\|^2
+\tilde{c}_1^2\|\phi_1^{(1)}\|^2+\|\phi_0^{(1)}\|^2+
\tilde{c}_1^{-2}\|\phi_{-1}^{(1)}\|^2+\tilde{c}_1^{-4}
\|\phi_{-2}^{(1)}\|^2\right)=1,\quad \\
&\tilde{c}_0^2\left(2\tilde{c}_1^4\|\phi_2^{(1)}\|^2
+\tilde{c}_1^2\|\phi_1^{(1)}\|^2-\tilde{c}_1^{-2}\|\phi_{-1}^{(1)}
\|^2-2\tilde{c}_1^{-4}\|\phi_{-2}^{(1)}\|^2\right)=M.
\end{split}
\ee
It is proved that the equations \eqref{projspin21} admit a unique positive solution
$(\tilde{c}_0,\tilde{c}_1)$ \cite{BaoTY}. Of course,
it is a little tedious to write down the solution explicitly since it can
be reduced to find the positive root of a fourth-order polynomial and this approach is very hard and/or tedious to extend to the computation of the ground state of spin-$F$ ($F\ge3$) BEC.

Based on the observation that $\Phi(t)=e^{\int_{t_n}^t(\mu(s)+\lambda(s)\mathrm{f}_z)\,ds}\Phi(t_n)$ can
be approximated (by Taylor expansion) as
$\Phi(t)\approx (Id+\int_{t_n}^t(\mu(s)Id+\lambda(s)\mathrm{f}_z)\,ds)\Phi(t_n)$ and
the projection step \eqref{eq:dfgn2:sec5} can be regarded as
an approximation of
$\Phi(t)=e^{\int_{t_n}^t(\mu(s)+\lambda(s)\mathrm{f}_z)\,ds}\Phi(t_n)$,
alternatively, here we propose the following
three additional equations to
determine the projection constants in \eqref{eq:dfgn2:sec5} as \cite{CTY}
\be
\alpha_2+\alpha_{-2} =2\alpha_0, \qquad
\alpha_1+\alpha_{-1} =2\alpha_0, \qquad
\alpha_2+\alpha_{0} =2\alpha_1.
\ee
Again, the above equations suggest that the projection constants can be assumed as $\alpha_l=c_0+lc=c_0(1+lc_1)$ ($l=-2,\ldots,2$) with $c_0,c_1>0$
satisfying
\begin{align}\label{eq:c0:sec5}
&(1+2c_1)^2\|\phi_2^{(1)}\|^2+(1+c_1)^2\|\phi_1^{(1)}\|^2+
\|\phi_0^{(1)}\|^2+(1-c_1)^2\|\phi_{-1}^{(1)}\|^2+(1-2c_1)^2
\|\phi_{-2}^{(1)}\|^2=\frac{1}{c_0^2}, \\
&2(1+2c_1)^2\|\phi_2^{(1)}\|^2+(1+c_1)^2\|\phi_1^{(1)}
\|^2-(1-c_1)^2\|\phi_{-1}^{(1)}\|^2-2(1-2c_1)^2\|\phi_{-2}^{(1)}\|^2
=\frac{M}{c_0^2}.\label{eq:c1:sec5}
\end{align}
The above equations turn out that $c_1$ satisfies a quadratic equation which can be solved very easily  \cite{CTY}. Thus, \eqref{eq:dfgn2:sec5} with $\alpha_l=c_0+lc=c_0(1+lc_1)$ ($l=-2,\ldots,2$) and
\eqref{eq:c0:sec5}-\eqref{eq:c1:sec5} complete the projection step. Then a full discretization can be constructed similarly as the pseudo spin-1/2 case in section 2, and we omit the details here.
\begin{remark} The idea of determining the projection constants through \eqref{eq:c0:sec5}-\eqref{eq:c1:sec5} can be generalized to other spin-$F$ system very easily \cite{CTY}.
\end{remark}

\begin{figure}[htb]
\centerline{\includegraphics[height=4cm,width=7cm]{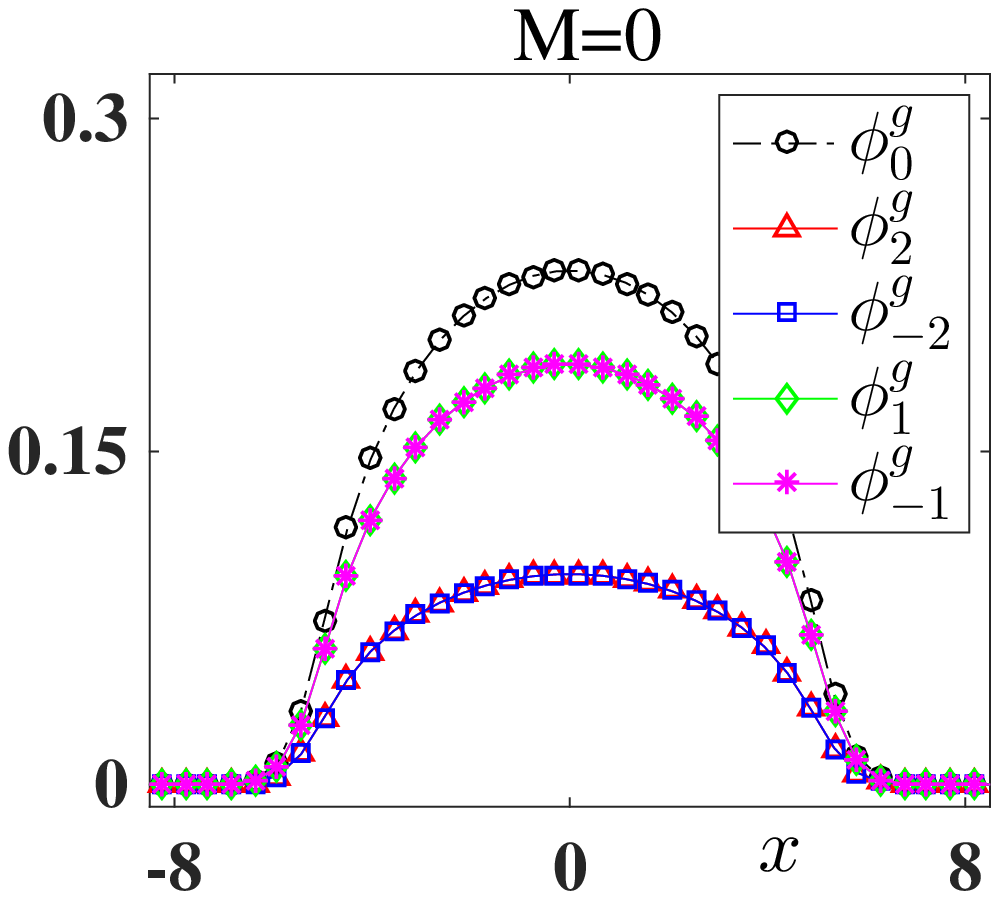}
\includegraphics[height=4cm,width=7cm]{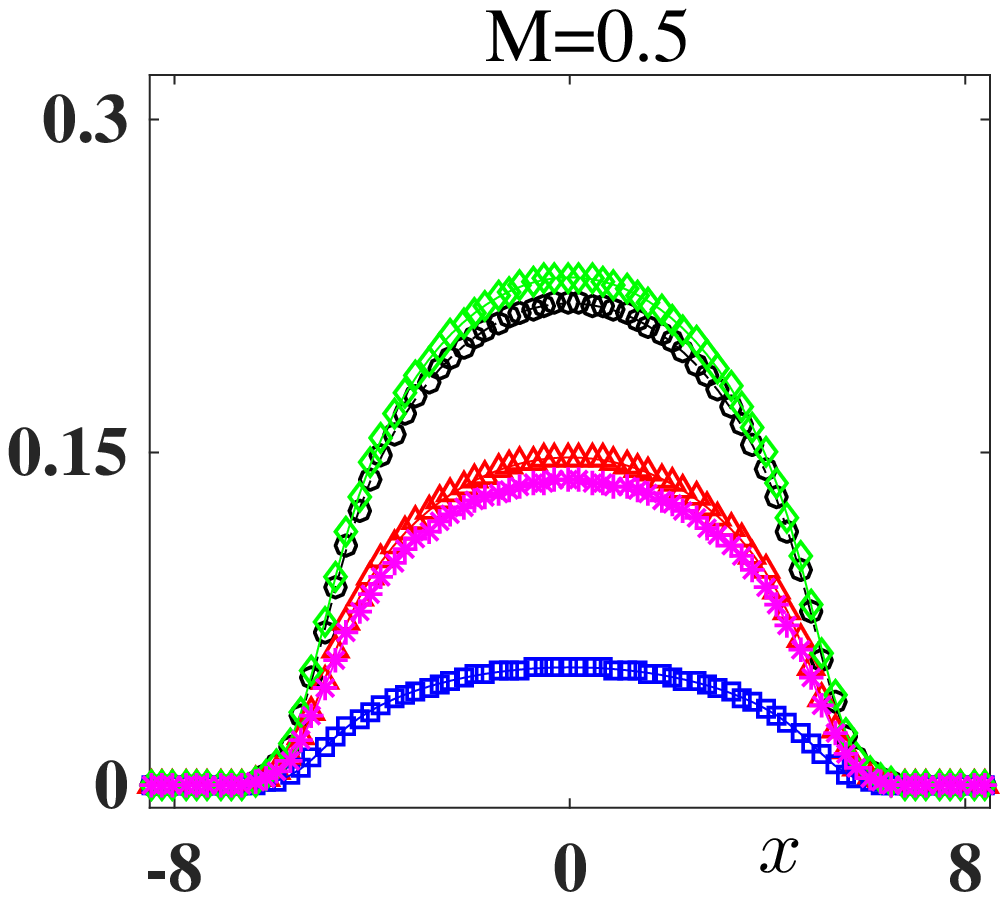}}
\centerline{
\includegraphics[height=4cm,width=7cm]{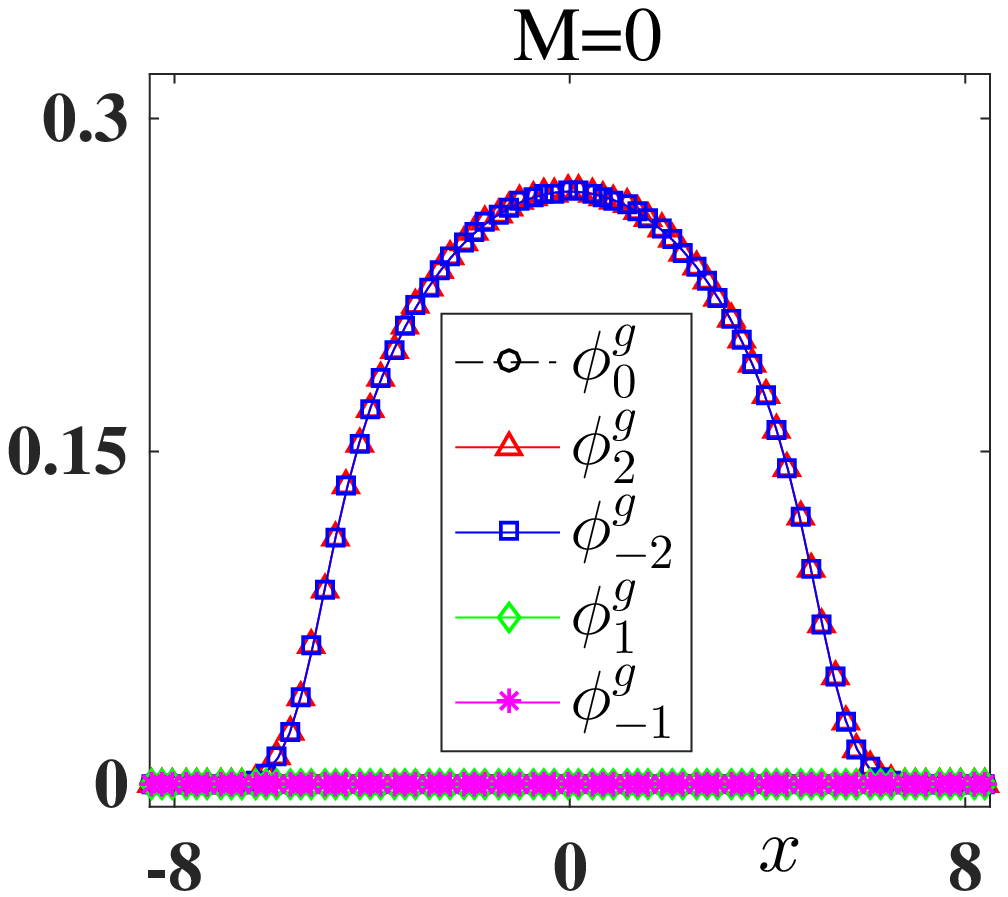}
\includegraphics[height=4cm,width=7cm]{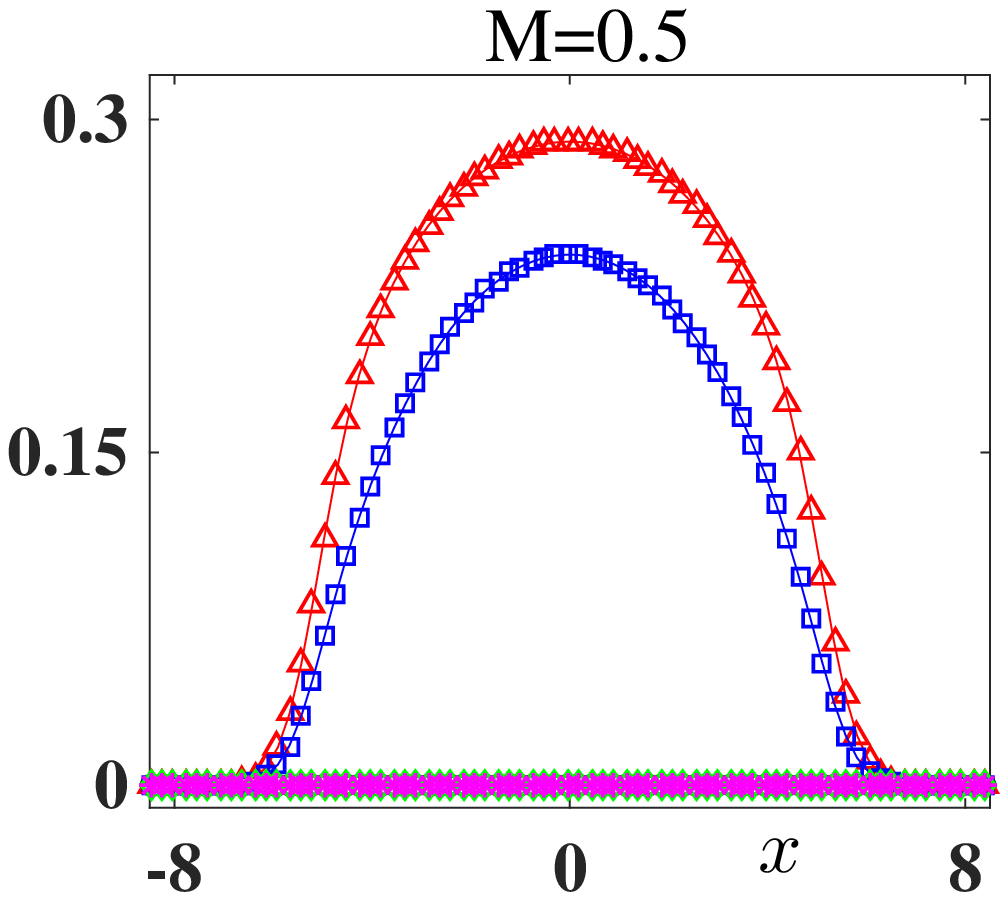}}
\centerline{
\includegraphics[height=4cm,width=7cm]{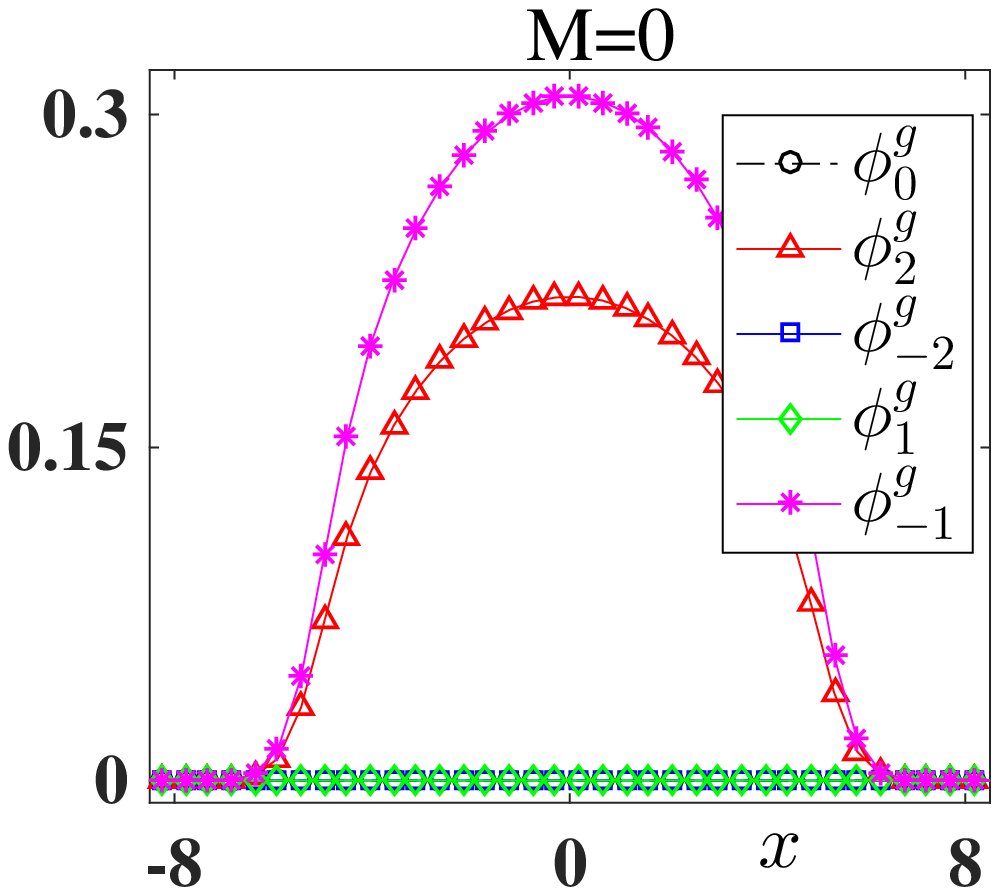}
\includegraphics[height=4cm,width=7cm]{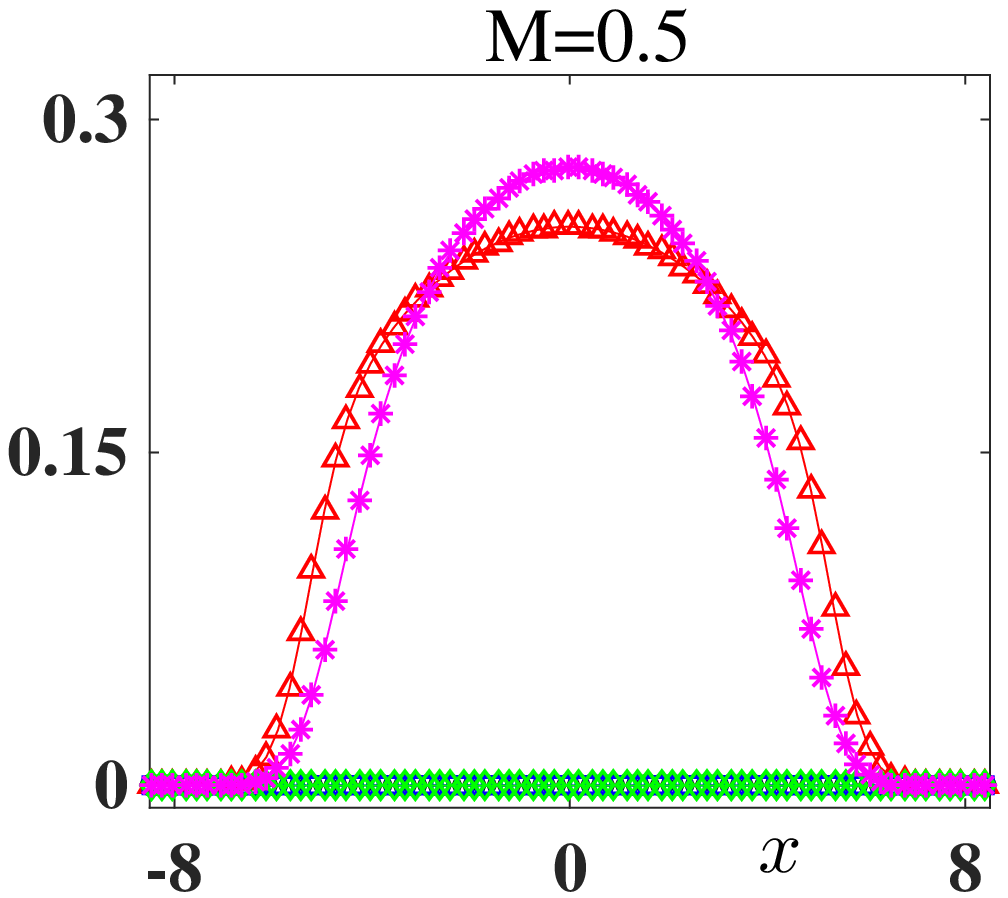}}

\caption{Ground states of spin-2 BEC in Example 5.1 for different magnetization $M=0$ (left column) and $M=0.5$ (right column). The set of parameters are those in
case (i) for the top panel, case (iii) for the bottom panel and case (ii) for the middle panel. \label{fig:gs:sec5}}
\end{figure}

{\it Example 5.1}. To show the ground state of  a spin-2 BEC,
we take $d=1$, $p=q=0$ and $V(\bx)=\frac12x^2$ in
 \eqref{eq:cgpen:sec5}  and consider three types of interactions, i.e. (i) $\beta_0=100$, $\beta_1=-1$ and $\beta_2=2$ (ferromagnetic interaction); (ii) $\beta_0=100$, $\beta_1=1$ and $\beta_2=-2$ (nematic interaction); (iii)  $\beta_0=100$, $\beta_1=10$ and $\beta_2=2$ (cyclic interaction).
Figure \ref{fig:gs:sec5} depicts the numerical ground state profiles under different types of interactions, which shows very rich structures. In particular, we find that the single mode approximation in Theorem \ref{thm:sma:sec5} and the vanishing components approximation in Theorem \ref{thm:van:sec5} hold for the ferromagnetic interactions and the nematic interactions, respectively.

\subsection{Dynamics}
For the CGPEs \eqref{eq:cgpen:sec5},
we consider the {\sl mass} (or density) of each component  as \be
\label{eq:massec:sec5} N_l(t):=\int_{{\Bbb R}^d} |\psi_l(\bx,t)|^2\;d\bx,
\qquad t\ge0, \quad l=-2,-1,0,1,2, \ee
and the condensate width  as
\be
\label{eq:dtap9:sec5}\sigma_\alpha(t) =
\sqrt{\delta_\alpha(t)}=\sqrt{\sum_{j=-2}^2\delta_{\alpha,j}(t)},\qquad
\alpha = x, y, z, \ee
where \be \label{eq:dtj:sec6}\delta_{\alpha,j}(t) = \int_{{\Bbb R}^d}\alpha^2|\psi_j(\bx,t)|^2d\bx,
\qquad t\geq0, \quad j = 2,1, 0, -1,-2. \ee
\subsubsection{Dynamical properties}
We have the spin dynamics of spin-2 BEC as follows.
\begin{lemma}\label{lem:1:sec5}
Suppose $\Psi(\bx,t)$ is the solution of the CGPEs
\eqref{eq:cgpen:sec5}, then we have
\be
\dot{N}_{l}(t)= \widetilde{F}_l(t), \qquad l=-2,-1,0,1,2,
\ee
with $N_l(0)=\int_{\mathbb{R}^d}|\psi_l(\bx,0)|^2\,d\bx$ and
  \begin{align*}  \widetilde{F}_2(t) = &2\beta_1\;\mathrm{Im}
\int_{\mathbb{R}^d}F_-(\Psi)\psi_1\bar{\psi}_2\, d\bx+\frac{2}{\sqrt{5}}\beta_2\;\mathrm{Im}\int_{\mathbb{R}^d}A_{00}(\Psi)\bar{\psi}_{-2}\bar{\psi}_2\;d\bx, \\
\widetilde{F}_1(t) = &2\beta_1\;\mathrm{Im}
\int_{\mathbb{R}^d}\left(\frac{\sqrt{6}}{2}F_-(\Psi)\psi_0\bar{\psi}_1+F_+(\Psi)\psi_2\bar{\psi}_1\right) d\bx
-\frac{2}{\sqrt{5}}\beta_2\;\mathrm{Im}\int_{\mathbb{R}^d}A_{00}(\Psi)\bar{\psi}_{-1}\bar{\psi}_1\;d\bx,
\end{align*}
\begin{align*}
\widetilde{F}_0(t)=&2\beta_1\;\mathrm{Im}
\int_{\mathbb{R}^d}\left(\frac{\sqrt{6}}{2}F_+(\Psi)\psi_1\bar{\psi}_0+\frac{\sqrt{6}}{2}F_-(\Psi)\psi_{-1}\bar{\psi}_0\right)\, d\bx
-\frac{2}{\sqrt{5}}\beta_2\;\mathrm{Im}\int_{\mathbb{R}^d}A_{00}(\Psi)\bar{\psi}_{0}^2\;d\bx,\\
\widetilde{F}_{-1}(t) = &2\beta_1\;\mathrm{Im}
\int_{\mathbb{R}^d}\left(\frac{\sqrt{6}}{2}F_+(\Psi)\psi_0\bar{\psi}_{-1}+F_-(\Psi)\psi_{-2}\bar{\psi}_{-1}\right) d\bx
-\frac{2}{\sqrt{5}}\beta_2\;\mathrm{Im}\int_{\mathbb{R}^d}A_{00}(\Psi)\bar{\psi}_{-1}\bar{\psi}_1\;d\bx,\\
\widetilde{F}_{-2}(t) = &2\beta_1\;\mathrm{Im}\int_{\mathbb{R}^d}F_+(\Psi)\psi_{-1}\bar{\psi}_{-2}\, d\bx+\frac{2}{\sqrt{5}}\beta_2\;\mathrm{Im}\int_{\mathbb{R}^d}A_{00}(\Psi)\bar{\psi}_{-2}\bar{\psi}_2\;d\bx.
\end{align*}
\end{lemma}
For the condensate width, we have the following results.
\begin{lemma}
Suppose $\Psi(\bx,t)$ is the solution of the CGPEs
\eqref{eq:cgpen:sec5}, then we have
\begin{align}\label{eq:dtap91:sec5}
&\ddot\delta_{\alpha}(t)=\int_{{\Bbb R}^d}\left[
\left(2|\p_\alpha\Psi|^2-2\alpha|\Psi|^2 \p_\alpha V(\bx)\right)+\beta_0
|\Psi|^2+\beta_1|\mathbf{F}|^2+\beta_2|A_{00}|^2\right]\,d\bx,
\quad t\geq 0,\\
\label{eq:dtap92:sec5} &\delta_\alpha(0)=\int_{{\Bbb R}^d}\alpha^2
\sum_{j=-2}^{2}|\psi_j(\bx,0)|^2\,d\bx, \quad \alpha = x, y, z,\\
\label{eq:dtap93:sec5} &\dot\delta_\alpha(0) = 2\sum_{j=-2}^2\int_{{\Bbb
R}^d}\alpha{\rm Im}
\left(\bar{\psi}_j^{(0)}\partial_\alpha\psi_j^0\right)d\bx,
 \end{align}
 where $\mathbf{F}(\Psi)$ and $A_{00}(\Psi)$ are defined in \eqref{eq:F:sec5} and \eqref{eq:A:sec5}, respectively.
\end{lemma}
\begin{lemma} Suppose $\Psi(\bx,t)$ is the solution of the CGPEs
\eqref{eq:cgpen:sec5}, $q=0$ and $V(\bx)$ is the harmonic potential in \eqref{eq:hopot:sec2}, then we have

(i) In 1D without nonlinear terms, i.e. $d=1$, $\beta_0=\beta_1=\beta_2=0$
in \eqref{eq:cgpen:sec4}, for any initial data
$\Psi(\bx,0)=\Psi^{(0)}(x)$, we have,
\be \label{eq:cdw61:sec4}
\delta_x(t)=\frac{E(\Psi^{(0)})}{\gamma_x^2}\left[1-\cos(2\gamma_x
t)\right] +\delta_x^{(0)}\cos(2\gamma_x
t)+\frac{\dot\delta_x(0)}{2\gamma_x}\sin(2\gamma_x t). \ee

(ii) In 2D with a radial symmetric trap, i.e. $d=2$,
$\gamma_x=\gamma_y:=\gamma_r$ and $\beta_1=0$ in \eqref{eq:cgpen:sec4}, for
any initial data $\Psi(\bx,0)=\Psi^{(0)}(x,y)$, we
have, for any $t\ge0$, \be \label{eq:cdw61tt:sec4}
\delta_r(t)=\frac{E(\Psi^{(0)})}{\gamma_r^2}\left[1-\cos(2\gamma_r t)\right]
+\delta_r^{(0)}\cos(2\gamma_r t)+\frac{\delta_r^{(1)}(0)}{2\gamma_r}\sin(2\gamma_r
t), \ee where $\delta_r(t)=\delta_x(t)+\delta_y(t)$,
$\delta_r^{(0)}:=\delta_x(0)+\delta_y(0)$ and
$\delta_r^{(1)}:=\dot\delta_x(0)+\dot\delta_y(0)$.
\end{lemma}
Let
$\Phi^s:=\Phi^s(\bx)=(\phi_2^s(\bx),\phi_1^s(\bx),\phi_0^s(\bx),\phi_{-1}^s(\bx),\phi_{-2}^s(\bx))^T$
be a stationary state of the CGPEs \eqref{eq:cgpen:sec5}, i.e. $\Phi^s$
solves the Euler-Lagrange system \eqref{eq:el:1:sec5}-\eqref{eq:el:3:sec5}.
 If the initial data $\Psi(\bx,0)$  for  the CGPEs \eqref{eq:cgpen:sec5}
 is chosen as a stationary state with its
center-of-mass shifted from the trap center, we can construct an
exact solution of the CGPEs \eqref{eq:cgpen:sec5} with a
harmonic potential \eqref{eq:hopot:sec2}.
\begin{lemma}
If the initial data $\Psi(\bx,0)$ for the CGPEs \eqref{eq:cgpen:sec5}  is chosen as \be
\label{eq:init5:sec5} \Psi(\bx,0)=\Phi^s(\bx-\bx_0)e^{i({\mathbf{a}}^{(0)}\cdot
\bx+b^{(0)})}, \qquad \bx \in {\Bbb R}^d, \ee where $\bx_0$ is a
given point in $\mathbb{R}^d$,
$\mathbf{a}^{(0)}=(a_1^{(0)},\ldots,a_d^{(0)})^T$ is a given vector in
$\mathbb{R}^d$ and $b^{(0)}$ is a given real number, then the exact
solution of \eqref{eq:cgpen:sec5} with the initial data
\eqref{eq:init5:sec5} satisfies:
\be \label{eq:exacts1:sec5}
\psi_l(\bx,t)=\phi_l^s(\bx-\bx(t))\;e^{-i\mu_l^s t}\;
e^{i(\mathbf{a}(t)\cdot \bx +b(t))}, \quad \bx\in{\Bbb R}^d, \quad t\ge
0,\qquad l=-2,-1,1,0,1,2, \ee where for any time $t\ge0$, $\bx(t)$
satisfies the following second-order ODE system: \begin{align}
&\ddot\bx(t)+\Lambda \bx(t)=0,\\
\label{govern_eq3:sec5} &\bx(0) = \bx_0,  \quad\dot{\bx}(0) = \mathbf{a}^{(0)}.
 \end{align} In addition,
$\mathbf{a}(t)=(a_1(t), \ldots,a_d(t))^T$ and $b(t)$ satisfy \begin{align}
\label{eq:exact3:sec5} &\dot{\mathbf{a}}(t) =-\Lambda\,\bx(t), \quad \dot{b}(t)=-\frac{1}{2}|\mathbf{a}(t)|^2-\frac{1}{2}\bx(t)^T\,
\Lambda\,\bx(t),
\quad t>0,
 \end{align}
 with initial data $\mathbf{a}(0)=\mathbf{a}^{(0)}$, $b(0)=b^{(0)}$
 and $\Lambda=\text{diag}(\gamma_x^2)$ in 1D, $\Lambda=\text{diag}(\gamma_x^2,\gamma_y^2)$ in 2D and
 $\Lambda=\text{diag}(\gamma_x^2,\gamma_y^2,\gamma_z^2)$ in 3D.
\end{lemma}

\subsubsection{A numerical method}
To compute the dynamics of spin-2 system governed by the CGPEs \eqref{eq:cgpen:sec5} with prescribed initial data
\be
\Psi(\bx,0)=\Psi_0(\bx),
\ee
we adopt the time splitting technique \cite{Wang2011}. The key idea is to divide the evolution of the CGPEs \eqref{eq:cgpen:sec5}
into several subproblems.   Recently, Symes et al. \cite{SymesBlakie2017} introduced a time-splitting scheme where only two subproblems
are involved and the nonlinear subproblem is solved exactly and we will
sketch the procedure below.

After truncating the CGPEs \eqref{eq:cgpen:sec5}  onto a bounded domain with homogeneous Dirichlet boundary conditions or periodic boundary conditions,
we solve \eqref{eq:cgpen:sec5}  from time $t_n=n\tau$ to $t_{n+1}=t_n+\tau$ through the following subproblems. One first
solves \be \label{eq:ODE1:sec5} i\frac{\p\psi_l}{\p
t}=(-\frac{1}{2}\nabla^2+pl+ql^2)\psi_l, \qquad l=-2,-1,0,1,2, \ee
for the time step of length $\tau$, followed by solving \be
\label{eq:ODE3:sec5} i\partial_t\Psi=[V(\bx)+\beta_0|\Psi|^2+\beta_1(F_x^n\mathrm{f}_x+F_y^n\mathrm{f}_y+F_z^n\mathrm{f}_z)]\Psi+\beta_2A_{00}\mathbf{A}\overline{\Psi},
\ee for the same time step.
\eqref{eq:ODE1:sec5} can be integrated exactly in phase space.

For \eqref{eq:ODE3:sec5}, noticing $\mathrm{f}_\alpha$ ($\alpha=x,y,z$) are Hermitian matrices and $A_{00}(\Psi)=\Psi^T\mathbf{A}\Psi$, we find
$\partial_t|\Psi(t,\bx)|^2=0$ ($t\in(t_n,t_{n+1})$).  Similar to the spin-1 case,  the commutator relations $[\mathrm{f}_x,\mathrm{f}_y]=\mathrm{f}_x\mathrm{f}_y-\mathrm{f}_y\mathrm{f}_x=i\mathrm{f}_z$, $[\mathrm{f}_y,\mathrm{f}_z]=i\mathrm{f}_x$ and
$[\mathrm{f}_z,\mathrm{f}_x]=i\mathrm{f}_y$ hold and $\mathrm{f}_\alpha\mathbf{A}=-\mathbf{A}\mathrm{f}_\alpha$ ($\alpha=x,y,z$ ),  then we can compute that
\begin{align*}
\partial_t(\Psi^\ast f_\alpha\Psi)=&\text{Im}(\Psi^\ast f_\alpha[\beta_2A_{00}\mathbf{A}\overline{\Psi}])-\text{Im}(\Psi^T[\beta_2\bar{A}_{00}\mathbf{A}]f_\alpha\Psi)
\\
=&-\text{Im}(\beta_2A_{00}\Psi^\ast\mathbf{A}\mathrm{f}_\alpha\overline{\Psi})-\text{Im}(\Psi^T[\beta_2\bar{A}_{00}\mathbf{A}]f_\alpha\Psi)=0,
\end{align*}
which implies that the spin vector components $F_\alpha(\Psi(t,\bx))=F_\alpha(\Psi(\bx,t_n))$ ($t\in(t_n,t_{n+1})$) are independent of $t$. Similar computations show that
\begin{align*}
\partial_tA_{00}(\Psi)(t)=&\partial_t\Psi^T\mathbf{A}\Psi+\Psi^T\mathbf{A}\partial_t\Psi=-2i(V(\bx)+\beta_0|\Psi|^2+\beta_2|\Psi|^2)
A_{00},
\end{align*}
and thus
\be\label{eq:Ae}
A_{00}(\bx,t)=e^{-2i(t-t_n)\left(V(\bx)+(\beta_0+\beta_2)|\Psi(\bx,t_n)|^2\right)}A_{00}^n,\quad t_n\leq t\leq t_{n+1},
\ee
where $A_{00}^n=A_{00}(\Psi(\bx,t_n))$.
 Now, it is clear that \eqref{eq:ODE3:sec5} becomes a linear
ODE
\be\label{eq:odes1:sec5}
 i\partial_t\Psi=[V(\bx)+\beta_0|\Psi(\bx,t_n)|^2+\beta_1(F_x^n\mathrm{f}_x+F_y^n\mathrm{f}_y+F_z^n\mathrm{f}_z)]\Psi+\beta_2A_{00}(t)\mathbf{A}\overline{\Psi},
\ee
where the spin vector $\mathbf{F}^n=(F_x^n,F_y^n,F_z^n)^T$ is evaluated using $\Psi(\bx,t_n)$, i.e. $F_\alpha^n=F_\alpha(\Psi(\bx,t_n))$, and $A_{00}(t)$ is given in \eqref{eq:Ae}.
Introducing
\begin{align}
\widetilde{\Psi}^{(1)}(\bx,t)=&e^{i(t-t_n)\left(V(\bx)+(\beta_0+\beta_2)|\Psi(\bx,t_n)|^2\right)}\Psi(\bx,t),\quad t\in[t_n,t_{n+1}],\label{eq:sol1:sec5}\\
\widetilde{\Psi}^{(2)}(\bx,t)=&e^{i(t-t_n)\left(\beta_1 \mathbf{F}^n\cdot\mathbf{f}\right)}\widetilde{\Psi}^{(1)}(\bx,t),\quad t\in[t_n,t_{n+1}],\label{eq:sol2:sec5}
\end{align}
and plugging the above equalities into \eqref{eq:odes1:sec5}, noticing that $(\mathbf{F}^n\cdot\mathbf{f})\mathbf{A}=-\mathbf{A}(\mathbf{F}^n\cdot\mathbf{f})$,
which implies $\mathbf{A}\overline{e^{-i(t-t_n)\left(\beta_1 \mathbf{F}^n\cdot\mathbf{f}\right)}\Psi^{(1)}}=e^{-i(t-t_n)\left(\beta_1 \mathbf{F}^n\cdot\mathbf{f}\right)}\mathbf{A}\Psi^{(1)}$, we can derive that
\begin{align*}
i\partial_t\widetilde{\Psi}^{(1)}(\bx,t)=&\left[\beta_1\mathbf{F}^n\cdot\mathbf{f}-\beta_2|\Psi(\bx,t_n)|^2\right]\widetilde{\Psi}^{(1)}+\beta_2A_{00}^n
\mathbf{A}\overline{\widetilde{\Psi}^{(1)}},\\
i\partial_t\widetilde{\Psi}^{(2)}(\bx,t)=&-\beta_2|\Psi(\bx,t_n)|^2\widetilde{\Psi}^{(2)}+\beta_2A_{00}^n\mathbf{A}\overline{\widetilde{\Psi}^{(2)}},
\end{align*}
and
\begin{equation*}
\partial_{tt}\widetilde{\Psi}^{(2)}(\bx,t)=-\beta_2^2(|\Psi(\bx,t_n)|^4-|A_{00}^n|^2)\widetilde{\Psi}^{(2)},\quad t\in[t_n,t_{n+1}],
\end{equation*}
with solution given by
\begin{align}\label{eq:sol3:sec5}
\widetilde{\Psi}^{(2)}(\bx,t)=&\cos(\beta_2(t-t_n)\kappa^n(\bx))\Psi(\bx,t_n)\\
&+\frac{i}{\kappa^n(\bx)}\sin(\beta_2(t-t_n)\kappa^n(\bx))\left[|\Psi(\bx,t_n)|^2\Psi(\bx,t_n)-A_{00}^n\mathbf{A}\overline{\Psi(\bx,t_n)}\right],\nonumber
\end{align}
where $\kappa^n(\bx)=\sqrt{|\Psi(\bx,t_n)|^4-|A_{00}^n|^2}$.
Combining \eqref{eq:sol1:sec5}-\eqref{eq:sol3:sec5}, we find the solution to \eqref{eq:ODE3:sec5} as \cite{SymesBlakie2017}
\begin{align}
\Psi(\bx,t)=&e^{-i(t-t_n)\left(V(\bx)+(\beta_0+\beta_2)|\Psi(\bx,t_n)|^2\right)}e^{-i(t-t_n)\left(\beta_1 \mathbf{F}^n\cdot\mathbf{f}\right)}\bigg(\cos(\beta_2(t-t_n)\kappa^n(\bx))\Psi(\bx,t_n)\nonumber\\
&+\frac{i}{\kappa^n(\bx)}\sin(\beta_2(t-t_n)\kappa^n(\bx))\left[|\Psi(\bx,t_n)|^2\Psi(\bx,t_n)-A_{00}^n\mathbf{A}\overline{\Psi(\bx,t_n)}\right]\bigg),\label{eq:solode2:sec5}
\end{align}
where the exponential $e^{-it\left(\beta_1 \mathbf{F}^n\cdot\mathbf{f}\right)}$ is calculated as
\begin{align}
e^{-it\left(\beta_1 \mathbf{F}^n\cdot\mathbf{f}\right)}=&I_5+i\left(\frac{1}{6}\sin(2\beta_1|\mathbf{F}^n|t)-\frac{4}{3}\sin(\beta_1|\mathbf{F}^n|t)\right)\frac{\mathbf{F}^n\cdot\mathbf{f}}{|\mathbf{F}^n|}
\nonumber\\
&+\left(\frac{4}{3}\cos(\beta_1|\mathbf{F}^n|t)-\frac{1}{12}\cos(2\beta_1|\mathbf{F}^n|t)-\frac{5}{4}\right)\frac{(\mathbf{F}^n\cdot\mathbf{f})^2}{|\mathbf{F}^n|^2}\nonumber
\\&+i\left(\frac{1}{3}\sin(\beta_1|\mathbf{F}^n|t)-\frac{1}{6}\cos(2\beta_1|\mathbf{F}^n|t)\right)\frac{(\mathbf{F}^n\cdot\mathbf{f})^3}{|\mathbf{F}^n|^3}\nonumber\\
&+\left(\frac{1}{12}\cos(2\beta_1|\mathbf{F}^n|t)-\frac{1}{3}\cos(\beta_1|\mathbf{F}^n|t)+\frac{1}{4}\right)\frac{(\mathbf{F}^n\cdot\mathbf{f})^4}{|\mathbf{F}^n|^4},\nonumber
\end{align}
with $I_5$ being the 5-by-5 identity matrix.

Now, we are able to solve the two subproblems \eqref{eq:ODE1:sec5}-\eqref{eq:ODE3:sec5}, and a standard splitting procedure such as second order Strang splitting or fourth-order partition Runge-Kutta time splitting method \cite{Mc} can be applied to
construct a numerical scheme for solving the CGPEs \eqref{eq:cgpen:sec5}. Fourier/sine spectral discretizations can be used according to the periodic/homogeneous Dirichlet boundary conditions, and the details are omitted here for brevity.


\section{Summary and future perspectives}
\setcounter{equation}{0}
\setcounter{figure}{0}
In the previous sections, we have briefly reviewed the mathematical models, theories and numerical methods for the pseudo spin-1/2 system, spin-orbit coupled BEC, spin-1 and spin-2 systems.
When higher spin and/or other effects such as rotating frame, nonlocal dipole-dipole interactions and random potentials are relevant, more  complicated structure and interesting phenomenons would emerge and mathematical and numerical studies would be  quite challenging \cite{BaoCaiWang,Min,Bao2013,Baocai2013}. As examples, we will present the mean field models for spin-3 BEC and spinor dipolar BEC below.
\subsection{Spin-3 BEC and beyond}
For a spin-3 BEC system \cite{Pasq,Ueda,Ueda2014} at zero temperature, the condensate can be described by the  vector wave function
$\Psi:=\Psi(\bx,t)=(\psi_3,\psi_2,\psi_1,\psi_0,\psi_{-1},\psi_{-2},\psi_{-3})^T$ ($\psi_l$ for $m_F=l$ state, $l=-3,-2,-1,0,1,2,3$) satisfying the CGPEs as \cite{Ho1,Ueda,StamperUeda}:
\be\label{eq:cgpe:sec6}
i\hbar\partial_t\Psi=[\tilde{H}+c_0 \rho-\tilde{p} \mathrm{f}_z+\tilde{q} \mathrm{f}_z^2+c_1 \mathbf{F}\cdot\mathbf{f}]\Psi+c_2A_{00}\mathbf{A}\overline{\Psi}+c_3\sum_{l=-2}^2
A_{2l}\mathbf{A}_l\overline{\Psi},
\ee
where as described in section \ref{sec:mf:spin-1}, $\tilde{H}=-\frac{\hbar^2}{2m}\nabla^2+\tilde{V}(\bx)$ is the single particle Hamiltonian, $\tilde{p}$ and $\tilde{q}$ are the linear and quadratic Zeeman energy shifts, respectively,
$\rho=|\Psi|^2=\sum_{l=-3}^3|\psi_l|^2$ is the total density. $c_0=\frac{9g_4+2g_6}{11}$, $c_1=\frac{g_6-g_4}{11}$, $c_2=\frac{11g_0-21g_4+10g_6}{11}$ and $c_3=\frac{11g_2-18g_4+7g_6}{11}$  characterizes the spin-independent
interaction, spin-exchange interaction, spin-singlet interaction and spin-quintet interaction, respectively,
with $g_k=\frac{4\pi\hbar^2}{m}a_k$ ($k=0,2,4,6$) and $a_k$ being the s-wave scattering length for scattering channel of total hyperfine spin $k$.
The spin-3 matrices $\mathbf{f}=(\mathrm{f}_x,\mathrm{f}_y,\mathrm{f}_z)^T$ are given as
\be
\mathrm{f}_x=\begin{pmatrix}0&\sqrt{3/2}&0&0&0&0&0\\
\sqrt{3/2}&0&\sqrt{5/2}&0&0&0&0\\
0&\sqrt{5/2}&0&\sqrt{3}&0&0&0\\
0&0&\sqrt{3}&0&\sqrt{3}&0&0\\
0&0&0&\sqrt{3}&0&\sqrt{5/2}&0\\
0&0&0&0&\sqrt{5/2}&0&\sqrt{3/2}\\
0&0&0&0&0&\sqrt{3/2}&0\end{pmatrix},
\ee
\be
\mathrm{f}_y=\begin{pmatrix}0&i\sqrt{3/2}&0&0&0&0&0\\
-i\sqrt{3/2}&0&i\sqrt{5/2}&0&0&0&0\\
0&-i\sqrt{5/2}&0&i\sqrt{3}&0&0&0\\
0&0&-i\sqrt{3}&0&i\sqrt{3}&0&0\\
0&0&0&-i\sqrt{3}&0&i\sqrt{5/2}&0\\
0&0&0&0&-i\sqrt{5/2}&0&i\sqrt{3/2}\\
0&0&0&0&0&-i\sqrt{3/2}&0\end{pmatrix}
\ee
and
\be
\mathrm{f}_z=\text{diag}(3,2,1,0,-1,-2,-3).
\ee
The spin vector $\mathbf{F}=\mathbf{F}(\Psi)=(F_x,F_y,F_z)^T$ with $F_\alpha=\Psi^*\mathrm{f}_\alpha\Psi$ ($\alpha=x,y,z$) is given as
\begin{equation*}\begin{split}
&F_+=F_x+iF_y=\sqrt{6}\overline{\psi}_3\psi_2+\sqrt{10}\overline{\psi}_2\psi_1+2\sqrt{3}\overline{\psi}_1\psi_0+2\sqrt{3}\overline{\psi}_0\psi_{-1}+
\sqrt{10}\overline{\psi}_{-1}\psi_{-2}+\sqrt{6}\overline{\psi}_{-2}\psi_{-3},\\
&F_z=3|\psi_3|^2+2|\psi_2|^2+|\psi_1|^2-|\psi_{-1}|^2-2|\psi_{-2}|^2-3|\psi_{-3}|^2,
\end{split}
\end{equation*}
with $\mathbf{F}\cdot\mathbf{f}=F_x\mathrm{f}_x+F_y\mathrm{f}_y+F_z\mathrm{f}_z$. The matrices $\mathbf{A}$ and $\mathbf{A}_0$  are
defined as
\begin{equation*}
\mathbf{A}=\frac{1}{\sqrt{7}}\begin{pmatrix}
0&0&0&0&0&0&1\\
0&0&0&0&0&-1&0\\
0&0&0&0&1&0&0\\
0&0&0&-1&0&0&0\\
0&0&1&0&0&0&0\\
0&-1&0&0&0&0&0\\
1&0&0&0&0&0&0\end{pmatrix},
\mathbf{A}_0=\frac{1}{\sqrt{7}}\begin{pmatrix}
0&0&0&0&0&0&\frac{5}{2\sqrt{3}}\\
0&0&0&0&0&0&0\\
0&0&0&0&\frac{-\sqrt{3}}{2}&0&0\\
0&0&0&\sqrt{\frac{2}{3}}&0&0&0\\
0&0&\frac{-\sqrt{3}}{2}&0&0&0&0\\
0&0&0&0&0&0&0\\
\frac{5}{2\sqrt{3}}&0&0&0&0&0&0\end{pmatrix}.
\end{equation*}
$\mathbf{A}_{l}=(a_{l,jk})_{7\times7}$ ($l=\pm1,\pm2$)  and  $a_{l,jk}$ is zero except for those $j+k=8-l$;  for the simplicity of notations, we denote $\vec{a}_l=(a_{l,1(7-l)},
a_{l,2(6-l)},\ldots,a_{l,(7-l)1})^T\in\mathbb{R}^{7-l}$ for $l=1,2$ and $\vec{a}_l=(a_{l,(1-l)7},
a_{l,(-l)6},\ldots,a_{l,7(1-l)})^T\in\mathbb{R}^{7+l}$ for $l=-1,-2$ with
\begin{align*}
&\vec{a}_{\pm1}=\frac{1}{\sqrt{7}}\left(\frac{5}{2\sqrt{3}},-\frac{\sqrt{5}}{2},\frac{1}{\sqrt{6}},\frac{1}{\sqrt{6}},-\frac{\sqrt{5}}{2},\frac{5}{2\sqrt{3}}\right)^T,\\
&\vec{a}_{\pm2}=\frac{1}{\sqrt{7}}\left(\sqrt{\frac{5}{6}},-\sqrt{\frac{5}{3}},\sqrt{2},-\sqrt{\frac{5}{3}},\sqrt{\frac{5}{6}}\right)^T.
\end{align*}
 $A_{00}:=A_{00}(\Psi)=\Psi^T \mathbf{A}\Psi$ and $A_{2l}=A_{2l}(\Psi)=\Psi^T\mathbf{A}_l\Psi$ can be expressed as
\begin{align}\label{eq:A:sec6}
A_{00}=&\frac{1}{\sqrt{5}}(2\psi_3\psi_{-3}-2\psi_{2}\psi_{-2}+2\psi_1\psi_{-1}-\psi_0^2),\\
A_{20}=&\frac{1}{\sqrt{21}}(5\psi_3\psi_{-3}-3\psi_1\psi_{-1}+\sqrt{2}\psi_0^2),\\
A_{2\pm1}=&\frac{1}{\sqrt{21}}(5\psi_{\pm3}\psi_{\mp2}-\sqrt{15}\psi_{\pm2}\psi_{\mp1}+\sqrt{2}\psi_{\pm1}\psi_0),\\
A_{2\pm2}=&\frac{1}{\sqrt{21}}(\sqrt{10}\psi_{\pm3}\psi_{\mp1}-\sqrt{20}\psi_{\pm2}\psi_{0}+\sqrt{2}\psi_{\pm1}^2).
\end{align}

Similar to the spin-2 case, after nondimensionalization and proper dimension reduction,  the CGPEs \eqref{eq:cgpe:sec6} can be written as
\be\label{eq:cgpen:sec6}
i\partial_t\Psi=[-\frac{1}{2}\nabla^2+V(\bx)+\beta_0 \rho-p \mathrm{f}_z+q \mathrm{f}_z^2+\beta_1 \mathbf{F}\cdot\mathbf{f}]\Psi+\beta_2A_{00}\mathbf{A}\overline{\Psi}+\beta_3\sum_{l=-2}^2
A_{2l}\mathbf{A}_l\overline{\Psi},
\ee
where $\bx \in\mathbb{R}^d,\,d=1,2,3$, $\beta_{k}$ ($k=0,1,2,3$) are real constants. The CGPEs \eqref{eq:cgpen:sec6}  conserve the following three important quantities, i.e.
the {\sl mass} (or {\sl normalization})
\be\label{eq:norm:sec6}
N(\Psi(\cdot,t)):=\|\Psi(\cdot,t)\|^2=\int_{\Bbb R^d}\sum_{l=-3}^3|\psi_l(\bx,t)|^2\,d\bx=N(\Psi(\cdot,0))=1,
\ee
the {\sl magnetization} (with $M\in[-3,3]$)
\be\label{eq:mag:sec6}
M(\Psi(\cdot,t)):=\int_{\Bbb R^d}\sum_{l=-3}^3l|\psi_l(\bx,t)|^2\,d\bx
=M(\Psi(\cdot,0))=M,
\ee
and the {\sl energy per particle}
\begin{align}
E(\Psi(\cdot,t))=&\int_{\mathbb{R}^d
}\bigg\{\sum_{l=-3}^{3} \left(\frac{1}{2}|\nabla
\psi_l|^2+(V(\bx)-pl+ql^2)|\psi_l|^2\right)
+\frac{\beta_0}{2}|\Psi|^4+\frac{\beta_1}{2}|\mathbf{F}|^2+\frac{\beta_2}{2}|A_{00}|^2\nonumber\\
&\qquad\qquad+\frac{\beta_3}{2}\sum_{l=-2}^2|A_{2l}|^2\bigg\}\; d\bx \equiv
E(\Psi(\cdot,0)).\label{eq:energy:sec6}
\end{align}

For a spin-$F$ BEC system within the mean field regime, the order parameter has $2F+1$ components  as $\Psi=(\psi_{F},\psi_{F-1},\ldots,\psi_{-F})^T\in\mathbb{C}^{2F+1}$, and the corresponding CGPEs could be similarly derived \cite{Ueda}. Understanding  the ground state pattern and the dynamics of such higher spin BEC system requires extensive mathematical and numerical studies.

\subsection{Spinor dipolar BEC}
In a spinor condensate, the dipole-dipole interaction (DDI) due to the atomic spin or magnetization emerges and the DDI could affect the spin texture of the system for a
DDI strength comparable to the spin-dependent interactions.  For example, the DDI plays a crucial role in the spin-3  $^{52}$Cr system \cite{Pasq}. For a spinor dipolar BEC where the spin is not polarized by an external magnetic field and can vary in space, the
mean-field CGPEs will include a nonlocal term. Here, we present the  CGPEs for spin-1 dipolar BEC in 3D as \cite{Bar,Ueda,StamperUeda}
\be\label{eq:cgpe:sec7}
i\hbar\partial_t\Psi=[-\frac{\hbar^2}{2m}\nabla^2+\tilde{V}(\bx)+c_0 \rho-p_0 \mathrm{f}_z+q_0 \mathrm{f}_z^2+c_1 \mathbf{F}\cdot\mathbf{f}+c_{dd}\mathbf{V}(\Psi)\cdot\mathbf{f}]\Psi,
\ee
where $\Psi:=\Psi(t,\bx)=(\psi_1,\psi_0,\psi_{-1})^T$, $c_{dd}=\frac{\mu_0(g\mu_B)^2}{4\pi}$ is the DDI interaction strength with $\mu_0$ being the  magnetic permeability of vacuum, $\mu_B$ being the Bohr magneton and $g$ being the Land\'e $g$-factor for the particle, $\mathbf{V}:=\mathbf{V}(\Psi)=(V_x,V_y,V_z)^T$ is a  vector-valued function representing the DDI interaction induced potential, $\mathbf{V}\cdot\mathbf{f}=V_x\mathrm{f}_x+V_y\mathrm{f_y}+V_z\mathrm{f}_z$ and the rest parameters are the same as those in
the CGPEs \eqref{eq:cgpe:sec4}. In detail, for spin vector $\mathbf{F}=(F_x,F_y,F_z)^T$, the DDI interaction is given as
\be\label{eq:V:sec7}
V_\alpha=\int_{\Bbb R^3}\sum_{\alpha^\prime=x,y,z} U_{\alpha\alpha^\prime}(\bx-\bx^\prime)F_{\alpha^\prime}(\bx^\prime)\,d\bx^\prime,\quad \alpha=x,y,z,
\ee
and the DDI kernel $U=(U_{\alpha\alpha^\prime})$ \cite{Lahaye,Bar} is a $3\times3$ matrix with
\be
U_{\alpha\alpha^\prime}(\bx)=\frac{e_\alpha\cdot e_{\alpha^\prime}-3(e_\alpha\cdot\bx)(e_{\alpha^\prime}\cdot\bx)/|\bx|^2}{|\bx|^3},\quad \bx=(x,y,z)^T\in\mathbb{R}^3,\quad\alpha,\alpha^\prime=x,y,z,
\ee
where $e_x=(1,0,0)^T$, $e_y=(0,1,0)^T$ and $e_z=(0,0,1)^T$ are the corresponding unit vectors for $x$-, $y$- and $z$- axes, respectively.

After proper scaling, the dimensionless spin-1 dipolar Gross-Pitaevskii equations read as
\be\label{eq:cgpen:sec7}
i\partial_t\Psi(\bx,t)=[-\frac{1}{2}\nabla^2+V(\bx)+\beta_0 \rho-p \mathrm{f}_z+q \mathrm{f}_z^2+\beta_1 \mathbf{F}\cdot\mathbf{f}+\lambda\mathbf{V}(\Psi)\cdot\mathbf{f}]\Psi,\quad\bx\in\mathbb{R}^d,
\ee
where $V(\bx)$ is the real-valued trapping potential, $\beta_0$ denotes the spin-independent contact interaction, $\beta_1$ represents the spin-dependent interaction,
$\lambda$ is the DDI parameter,  the spin vector $\mathbf{F}$ and the DDI potential $\mathbf{V}(\Psi)$
are given in \eqref{eq:spin1v:sec4} and \eqref{eq:V:sec7}, respectively. The important conserved quantities of \eqref{eq:cgpen:sec7} include
the {\sl mass} (or {\sl normalization})
\be\label{eq:norm:sec7}
N(\Psi(\cdot,t)):=\|\Psi(\cdot,t)\|^2=\int_{\Bbb R^3}\sum_{l=-1}^1|\psi_l(\bx,t)|^2\,d\bx=N(\Psi(\cdot,0))=1,
\ee
the {\sl magnetization} (with $M\in[-1,1]$)
\be\label{eq:mag:sec7}
M(\Psi(\cdot,t)):=\int_{\Bbb R^3}\sum_{l=-1}^{1}l|\psi_l(\bx,t)|^2\,d\bx
=M(\Psi(\cdot,0))=M,
\ee
and the {\sl energy per particle}
\begin{align}
E(\Psi(\cdot,t))=&\int_{\mathbb{R}^3
}\bigg\{\sum_{l=-1}^{1} \left(\frac{1}{2}|\nabla
\psi_l|^2+(V(\bx)-pl+ql^2)|\psi_l|^2\right)
+\frac{\beta_0}{2}|\Psi|^4+\frac{\beta_1}{2}|\mathbf{F}|^2+\frac{\lambda}{2}\mathbf{V}(\Psi)\cdot\mathbf{F}\bigg\}\; d\bx\nonumber\\ \equiv&
E(\Psi(\cdot,0)).\label{eq:energy:sec7}
\end{align}

Similarly, the CGPEs with nonlocal DDI can be obtained for spin-2 (or higher spin-$F$) dipolar BEC by including the $c_{dd}$ (or $\lambda$) term (the same form as above, see \cite{Ueda}) into the spin-2 CGPEs \eqref{eq:cgpe:sec7} (or \eqref{eq:cgpen:sec7})  and we omit the details here.
For lower dimensions (1D and 2D), dimension reduction of the CGPEs with nonlocal DDI in 3D could be done following \cite{BaoBenCai,CaiRosen,Ro}. Similarly,
the efficient and accurate numerical methods for computing the ground state
and dynamics of single-component dipolar BEC \cite{BaoCaiWang,BaoTZ,Jiang,Tang}, especially the methods for handling the DDI,  can be extended directly for spinor dipolar BEC.
 In presence of the long range DDI, the ground state structure and dynamical properties of the spinor dipolar BEC would be very rich and complicated, which requires further mathematical and numerical studies.

\section*{Acknowledgements}
We acknowledge support from
the Ministry of Education of Singapore grant R-146-000-223-112 (W. Bao),
the Natural Science Foundation of China grant 91430103 (W. Bao),
the Natural Science Foundation of China  grants U1530401, 91630204 and 11771036 (Y. Cai).

\end{document}